\documentclass[a4paper,12pt]{article}
\pdfoutput=1 

\usepackage{a4wide}

\usepackage{cite}

\usepackage[T1]{fontenc} 

\usepackage[latin1]{inputenc}
\usepackage[english]{babel}

\usepackage{empheq}
\usepackage{amssymb}
\usepackage{amsmath}
\usepackage{amsfonts}
\usepackage{xcolor}

\numberwithin{equation}{section}

\usepackage[colorlinks]{hyperref}

\hypersetup{
 citecolor=blue,
 linkcolor=black,
 urlcolor=blue}

\newcommand{\ep}{\epsilon}

\newcommand{\be}{\begin{equation}}
\newcommand{\ee}{\end{equation}}
\newcommand{\bea}{\begin{eqnarray}}
\newcommand{\eea}{\end{eqnarray}}
\newcommand{\bes}{\begin{equation}\begin{split}}

\definecolor{darkblue}{rgb}{0,0,0.7}

\newcommand{\lp}{\left(}
\newcommand{\rp}{\right)}

\def \as {\alpha_s}

\def \asontwopimu {\frac{\alpha_s(\mu)}{2\pi}}
\def \ep {\epsilon}

\def \Li {\text{Li}}
\def \gsb {g_{s,b}}
\def \M {{\cal M}}

\def \Ca {C_A}
\def \Cf {C_F}
\def \kt {k_{\perp}}
\def \nkt {\kappa_{\perp}}
\def \Em {E_{\rm max}}
\def \d {{\rm d}}

\def \SS {S{\hspace{-5pt}}S}
\def \CC {C{\hspace{-6pt}C}}
\def \FLM {F_{LM}}
\def \FRV {F_{LRV}}
\def \FV {F_{LV}}

\def \FVsqF {F^{\rm fin}_{LV^2}}
\def \FVF {F^{\rm fin}_{LV}}
\def \FVV {F_{LVV}}
\def \FVVF {F^{\rm fin}_{LVV}}

\def \I {I}
\def \pref {\left[ \frac{1}{8\pi^2} \frac{(4\pi)^\ep}{\Gamma(1-\ep)}\right]}

\def \ONLO {\hat{\mathcal O}_{\rm NLO}}

\newcommand{\dg}[1]{[dg_{#1}]}

\def \PPOP {\hat{\mathcal P}_{qq}^{(-)}}

\def \PP {\mathcal P_{qq}}
\def \D {\mathcal D}
\def \Dt {\tilde{\mathcal D}}
\def \zb {\bar z}

\newcommand{\LM}[1]{L_{#1}}
\newcommand{\LMu}{\ln\lp\frac{\mu^2}{s}\rp}

\def \TC {T_C}

\def \NLO {NLO }
\def \NNLO {NNLO }

\def \PqqRRtwodelta {\mathcal P_{qq,R}}
\def \PqqRRtwodeltaEp {\mathcal P^{(\ep)}_{qq,R}}
\def \zm {z_{\rm min}}

\begin{document}
\vspace{-5.0cm}
\begin{flushright}
CERN-TH-2017-029, IPPP/17/10, TTP17-003
\end{flushright}

\vspace{2.0cm}

\begin{center}
{\Large \bf 
Nested soft-collinear subtractions in NNLO QCD computations
}\\
\end{center}

\vspace{0.5cm}

\begin{center}
Fabrizio Caola$^{1,2}$, Kirill Melnikov$^{3}$, Raoul R\"ontsch$^3$.\\
\vspace{.3cm}
{\it
{}$^1$CERN, Theoretical Physics Department, Geneva, Switzerland\\
{}$^2$IPPP, Durham University, Durham, UK\\
{}$^3$Institute for Theoretical Particle Physics, KIT, Karlsruhe, Germany
}

\vspace{1.3cm}

{\bf \large Abstract}
\end{center}
We discuss a modification of the next-to-next-to-leading order (NNLO) 
subtraction scheme based on the
residue-improved sector decomposition that reduces the number of
double-real emission sectors from five to four. In particular, a
sector where energies and angles of unresolved particles vanish in a
correlated fashion is redundant and can be discarded. This simple
observation allows us to formulate a transparent iterative subtraction
procedure for double-real emission contributions, to demonstrate the
cancellation of soft and collinear singularities in an explicit and
(almost) process-independent way and to write the result of a NNLO
calculation in terms of quantities that can be computed in four
space-time dimensions. We illustrate this procedure explicitly in the
simple case of $\mathcal O(\alpha_s^2)$ gluonic corrections to the
Drell-Yan process of $q \bar q$ annihilation into a lepton pair.  We
show that this framework leads to fast and numerically stable
computation of QCD corrections.

\thispagestyle{empty}

\clearpage
\tableofcontents
\thispagestyle{empty}
\clearpage

\pagenumbering{arabic}

\allowdisplaybreaks

\section{Introduction} 

One of the most important recent advances in perturbative QCD 
was the discovery of practical ways to perform fully
differential next-to-next-to-leading order (NNLO) QCD computations for hadron collider
processes. These methods, that include antenna~\cite{ant},
residue-improved sector-decomposition~\cite{czakonsub,czakonsub4d} 
(see also~\cite{Boughezal:2011jf}) and
projection to Born~\cite{Cacciari:2015jma} subtraction
schemes, as
well as $q_\perp$ \cite{qt1,qt2} and $N$-jettiness \cite{njet1,njet2}
slicing methods, were used to perform an impressive number of NNLO QCD
computations relevant for LHC
phenomenology~\cite{Cacciari:2015jma,qt1,njet2,nnloH,nnloDY,nnloAA,nnloVA,nnloWW,
nnloZZ,nnloWZ,nnloVH,nnloHH,nnloMCFM,nnloHbb,nnloTdec,nnlottbar,nnlostop,
nnloHj,nnloVJ,nnloJJ}.\footnote{We also note that recently the CoLoRFulNNLO scheme
  was fully worked out for $e^+e^-$ colliders~\cite{colorful}.}

However, in spite of these remarkable successes, it is important to
recognize that existing implementations of subtraction schemes are complex,
not transparent, and require significant CPU time to produce stable
results.  On the other hand, slicing methods, while conceptually simple, have to be
carefully controlled to avoid dependence of the final
result on the slicing parameter.  Given these shortcomings, it is
important to study whether improvements to existing methods are
possible.  In the context of the $N$-jettiness slicing method, 
there has been recent progress towards a better control of 
the soft-collinear region~\cite{Moult:2016fqy,Boughezal:2016zws}.

In this paper we study the residue-improved subtraction scheme
introduced in Refs.~\cite{czakonsub,czakonsub4d}. This scheme is
interesting because it is the only existing framework for NNLO QCD
computations that is fully local in multi-particle phase space. As
such, it should demonstrate exemplary numerical stability, at least in
theory.  Although this scheme is well-understood and was applied to a
large number of non-trivial problems, we will argue in this paper that
certain aspects of it are redundant.
Interestingly, once this redundancy is recognized and removed, the
residue-improved subtraction scheme becomes very transparent and
physical. In addition, the technical simplifications that occur become so
significant that the cancellation of the divergent terms can be
demonstrated independently of the hard matrix element and almost entirely
analytically, and the final finite result for the NNLO contribution
to (in principle) any process can be written in a compact form in
terms of {\it generic} four-dimensional matrix elements.\footnote{A
  different improvement of the original scheme~\cite{czakonsub} that
  also allows to deal with four-dimensional matrix elements was 
  presented in Ref.~\cite{czakonsub4d}.}

Although the improvements that we just described hold true for an
arbitrary complicated process, in this paper, for the sake of clarity, 
we  restrict our discussion  to the production 
of a colorless final state in 
$q \bar q$ annihilation. This allows us to discuss all the
relevant conceptual and technical aspects of the computational framework, without
cluttering the notation and limiting the bookkeeping to a
minimum. The generalization of the framework described here to
arbitrary processes is -- at least conceptually -- straightforward.

Admittedly, compared to NNLO QCD problems studied recently, the
production of a colorless final state in $q \bar q$ annihilation is
a very simple process, which has been discussed in the literature many times. However,
we believe that the simplicity of our approach and the structures that 
emerge justify revisiting it one more time.  Moreover, thanks to
the simplicity of this process, we will be able to describe our
approach in detail and demonstrate many intermediate steps of the
calculation. Hopefully this will allow us to make the rather technical
subject of NNLO subtractions accessible to a broader part of the
particle physics community.

The paper is organized as follows. We begin with preliminary remarks in Section~\ref{sec:intro},
where we also precisely define the problem that we plan to address. In
Section~\ref{sect:mynlo} we discuss the next-to-leading order (NLO) QCD computation as a
prototype of  the following NNLO QCD construction. In
Section~\ref{sect:gennnlo}, we describe how the NLO computation 
generalizes to the NNLO case.  We elaborate on this in Section~\ref{sect:uvpdf},
where we discuss ultraviolet and collinear renormalization, and in Sections
~\ref{sect:virt},~\ref{sect:rv} and ~\ref{sect:nnlorr}, where we study
two-loop virtual corrections,  one-loop corrections to single-real emission
process, and the double-real emission contributions, respectively.
In Section~\ref{sect:fres}, we combine the different
contributions and present the final result for the NNLO QCD
corrections to color singlet production in $q\bar q$ annihilation. 
In Section~\ref{sect:numerics},
 we show some numerical results and
a comparison with earlier analytic calculations. We conclude in Section~\ref{sect:conclusion}.
A collection of useful formulas is provided in the appendices.

\section{Preliminary remarks}
\label{sec:intro}

We consider the production of a colorless final state $V$ in the collision
of two protons
\be
P+P\to V + X.
\label{eq:procdef}
\ee
We are interested in computing the differential cross
section for the process in Eq.~\eqref{eq:procdef}
\be {\rm d} \sigma = \sum \limits_{ij}^{} \int {\rm d} x_1
{\rm d}x_2 f_i (x_1) f_j(x_2) {\rm d}\hat\sigma_{ij}(x_1, x_2), \ee where
${\rm d}\hat\sigma_{ij}$ is the finite partonic scattering cross section,
$f_{i,j}$ are parton distribution functions and $x_{1,2}$ are momenta
fractions of the incoming hadrons that are carried to a hard collision
by partons $i$ and $j$, respectively. The dependence on the renormalization
and factorization scales and all other parameters of the process is understood.
 The finite partonic scattering
cross section is obtained after the renormalization of the strong
coupling constant removes all ultraviolet divergences, all soft
and final state collinear divergences cancel in the sum of cross
sections with different partonic multiplicities, and the initial state
collinear divergences are subtracted  by redefining parton distribution 
functions. 

Since the process under consideration is driven by a conserved current
that is independent of $\alpha_s$, the ultraviolet renormalization
reduces to the following ($\overline{\rm MS}$) redefinition of the
strong coupling constant \be \frac{\alpha_{s,b}}{2\pi}
(\mu_0^2)^{\ep}S_{\ep}= \asontwopimu \mu^{2\ep}
\left[1-\frac{\beta_0}{\ep}\asontwopimu + \mathcal O\lp\as^2\rp \right],
\label{eq2.3}
\ee where $S_\ep= (4\pi)^\ep e^{-\ep\gamma_E},~\gamma_E\approx 0.577216$ 
is the Euler-Mascheroni constant, $\ep=(4-d)/2$ and 
\be \beta_0 =
\frac{11}{6}\Ca - \frac{2}{3} T_R n_f, ~~~~ \Ca = 3,~T_R = \frac{1}{2},
\label{eq2.4}
\ee 
is the leading-order (LO) QCD 
$\beta$-function.

Collinear divergences associated with initial state QCD radiation are
removed by a redefinition of parton distributions. In the
$\overline{\rm MS}$ scheme, this amounts to the replacement
 \bes f_{i,b} & \to  
\bigg[1 +
  \asontwopimu \frac{\hat P^{(0)}_{ij}}{\ep}  \\
&+ \lp\asontwopimu\rp^2\left[ \frac{1}{2\ep^2}\lp
    \hat P^{(0)}_{ik} \otimes \hat P^{(0)}_{kj} - \beta_0 \hat P^{(0)}_{ij} \rp +
    \frac{1}{2\ep} \hat P^{(1)}_{ij} \right] + \mathcal O(\as^3)
  \bigg]\otimes f_j (\mu),
\label{eq2.5}
\end{split}
\ee
where $\otimes$ stands for the convolution 
\be
g(z) = [ f_1 \otimes f_2 ](z) =  \int \limits_{0}^{1}  {\rm d} x {\rm d} y f_1(x) f_2(y) \delta(z-x y),
\label{eq2.6}
\ee
and $\hat P_{ij}^{(0,1)}$ are the Altarelli-Parisi splitting functions.

As we already mentioned, we focus on gluonic corrections to
the $q\bar q$ annihilation channel
\be
q+\bar q \to V + ng.
\label{eq2.7}
\ee
This allows us to present all the features of the  framework while
limiting the bookkeeping to a minimum and, therefore, to keep the  discussion
relatively concise. All other partonic channels relevant for the 
Drell-Yan process  can be obtained by a simple generalization of 
what we will describe. 

The collinear-renormalized partonic cross section for $q \bar q $ annihilation into a vector boson is 
expanded  in series of   $\alpha_s$. We write 
\be
{\rm d}\hat\sigma \equiv {\rm d}\hat\sigma_{q\bar q}= {\rm d} \hat\sigma^{\rm LO} 
+ {\rm d} \hat\sigma^{\rm NLO} + {\rm d} \hat\sigma^{\rm NNLO},
\label{eq2.8}
\ee
where 
\be
\begin{split} 
& {\rm d}\hat\sigma^{\rm NLO} = {\rm  d} \sigma^{\rm V} + {\rm d} \sigma^{\rm R}
+ \frac{\as(\mu)}{2 \pi \ep} \left ( 
\hat P_{qq}^{(0)} \otimes  {\rm d} {\hat\sigma^{\rm LO}}
+
  {\rm d} \hat\sigma^{\rm LO} \otimes \hat P_{qq}^{(0)} 
  \right ),  
\\
& {\rm d}\hat\sigma^{\rm NNLO} = {\rm d} \sigma^{\rm VV} 
+ {\rm d} \sigma^{\rm RV} + {\rm d} \sigma^{\rm RR}  + {\rm d}\sigma^{\rm ren} + {\rm d} \sigma^{\rm CV}.
\label{eq2.9}
\end{split}
\ee
Various contributions in Eq.~\eqref{eq2.9} refer to virtual and real corrections, as 
well as to contributions to cross sections that arise because of 
the ultraviolet and  collinear renormalizations. The latter are obtained with the procedure
just described and read
\bes
&{\rm d}\sigma^{\rm ren} = -\asontwopimu \frac{\beta_0}{\ep} {\rm d}\hat\sigma^{\rm NLO},\\
&{\rm d}\sigma^{\rm CV} = \asontwopimu
\bigg[
\Gamma_1 \otimes \d\hat\sigma^{\rm NLO} + \d\hat\sigma^{\rm NLO}\otimes\Gamma_1\bigg]\\
&
~~~~
-\lp\asontwopimu\rp^2
\bigg[
\Gamma_1\otimes\d\hat\sigma^{\rm LO}\otimes\Gamma_1
+\Gamma_2\otimes\d\hat\sigma^{\rm LO}
+\d\hat\sigma^{\rm LO}\otimes\Gamma_2
\bigg],
\label{eq:ren}
\end{split}
\ee
where
\bes
\Gamma_1 = \frac{\hat P_{qq}^{(0)}}{\ep},~~~~~~~~~~~~
\Gamma_2 = \frac{\hat P_{qq}^{(0)}\otimes \hat P_{qq}^{(0)}+\beta_0 \hat P_{qq}^{(0)}}{2\ep^2}
-\frac{\hat P^{(1)}_{qq}}{2\ep}
,
\label{eq:rensplit}
\end{split}
\ee
and the relevant splitting functions are provided in Appendix~\ref{sect:defs}.

The cross section ${\rm d} \hat\sigma^{\rm (N)NLO}$ is finite but all the
individual contributions in Eq.~\eqref{eq2.9} are divergent.  The
well-known problem is that these divergences are explicit in some of
the terms and implicit in the others. Indeed, soft and collinear
divergences appear as explicit $1/\ep$ poles in virtual corrections
but they only become evident in real corrections once integration
over gluon momenta is performed.  However, since we would like to keep
the kinematics of all the final state particles intact, we can not
integrate over momenta of any of the final state particles if it is
resolved. It is this point that makes extraction of implicit
singularities complicated and requires us to devise a procedure to 
do it.

Depending on how these
implicit singularities are extracted, it may or may not be
straightforward to recognize how they combine and
cancel, once all contributions to the physical cross section are put
together.  At NNLO, this was done for the antenna subtraction scheme
and, in a less transparent way,  for the residue-improved sector
decomposition. One thing we would like to do, therefore, is to combine
the individual terms that contribute to partonic cross sections,  and
cancel all the $1/\ep$ divergences explicitly, without any reference
to the matrix elements that contribute
to the different terms in Eq.~\eqref{eq2.9}.  In the next section we
show how to do that at next-to-leading order in the perturbative
expansion for the Drell-Yan process. This will allow us to set up the
formalism and the notation that will be used for the \NNLO
analysis of Sections~\ref{sect:gennnlo}-\ref{sect:nnlorr}.

\section{The \NLO calculation}
\label{sect:mynlo}

We will illustrate our approach by studying the production of a lepton pair in 
quark-antiquark annihilation at next-to-leading order in perturbative QCD. 
We note that, at this order,  the  method that we would like to 
describe is identical to the FKS 
subtraction scheme  introduced in Refs.~\cite{Frixione:1995ms,Frixione:1997np}. However, 
we formulate the FKS method in a way that makes its extension to 
next-to-next-to-leading order  as straightforward as possible.\footnote{
For earlier efforts, see Ref.~\cite{Frixione:2004is}.}
 One point 
that we found helpful, especially for bookkeeping, was to introduce soft and collinear 
subtraction operators, and we show how to use them in the NLO computation below. 

We are interested in the calculation  of the  finite partonic cross section 
$ {\rm d}\hat\sigma^{\rm NLO} $ defined in  Eq.~\eqref{eq2.9}. It receives contributions 
from the virtual and real corrections and the collinear subtraction term. 
We will start the  discussion with the  real emission contribution. 
It refers to the process 
\be
q(p_1) + \bar q(p_2) \to V + g(p_4),
\label{eq3.2}
\ee
where $V$ is a generic notation for all  colorless particles in the final state. We write 
the cross section for the process in Eq.~\eqref{eq3.2} as 
\be
{\rm d} \sigma^{\rm R} = \frac{1}{2 s}\int \dg4 F_{LM}(1,2,4),
\label{eq3.3}
\ee
where $s$ is the partonic center-of-mass energy,
\be
\dg4 = \frac{{\rm d}^{d-1} p_4}{(2\pi)^d 2 E_4} \theta(\Em-E_4),
\label{eq3.4}
\ee
and 
\be
F_{LM}(1,2,4) = {\rm d}{\rm Lips}_V\; |{\cal M}(1,2,4,V)|^2\; 
{\cal F}_{\rm kin}(1,2,4,V).
\label{eq3.5}
\ee 
In Eq.~\eqref{eq3.5}, ${\rm d}{\rm Lips}_V$ is the
Lorentz-invariant phase space for colorless particles, including 
the momentum-conserving $\delta^{(d)}(p_1+p_2-p_4-p_V)$, ${\cal
  M}(1,2,4,V)$ is the matrix element for the process in
Eq.~\eqref{eq3.2} and ${\cal F}_{\rm kin}(1,2,4,V)$ is an (infra-red
safe) observable that depends on kinematic variables of all particles
in the process.  Also, $\Em$ is an arbitrary auxiliary parameter
that has to be large enough to accommodate all possible kinematic
configurations for $q \bar q \to V+g$.
The need  to introduce such a parameter is a
consequence of our construction, as explained in detail below.

We would like to isolate and 
extract soft and collinear singularities that appear when the integration over 
$\dg4$ in  Eq.~\eqref{eq3.3} is attempted. To this end, we introduce 
two operators that define soft and collinear projections
\be
S_i A = \lim_{E_i \to 0} A,\;\;\;\;
C_{ij} A = \lim_{\rho_{ij} \to 0} A,
\label{eq3.6}
\ee
where $\rho_{ij} = 1 - n_{i} \cdot n_{j} $ and  $n_{i}$ is a unit vector that describes  
the direction of the momentum of the $i$-th particle in $(d-1)$-dimensional space. 
By definition, operators in Eq.~\eqref{eq3.6}   act on everything that appears to the right 
of them. The limit operations, on   the right hand side
of  Eq.~\eqref{eq3.6},  are to be understood 
in the sense of extracting the most singular contribution   provided that limits 
in the conventional sense do not exist. We will also  use the averaging 
sign $\langle .... \rangle $ to represent  integration over momenta 
of final state particles. 
This  integration  is supposed to be performed in 
the center-of-mass frame of  incoming partons. {\it We emphasize that this remark is important since 
our construction of the subtraction terms is frame-dependent and not Lorentz-invariant.} 

We rewrite Eq.~\eqref{eq3.3} in the following way 
\be
\begin{split} 
&  \int [ {\rm d} g_4] F_{LM}(1,2,4) 
= 
\langle F_{LM}(1,2,4) \rangle 
= \langle S_4 F_{LM}(1,2,4) \rangle 
+\langle ( I - S_4) F_{LM}(1,2,4) \rangle 
\\
& 
= 
\langle S_4 F_{LM}(1,2,4) \rangle
+ \langle (C_{41} + C_{42} )  ( I - S_4)  F_{LM}(1,2,4) \rangle 
+ \langle \hat O_{\rm  NLO}  F_{LM}(1,2,4) \rangle,  
\label{eq3.8}
\end{split} 
\ee
where $I$ is the identity operator and $\hat O_{\rm NLO}$ is a short-hand notation for a combination of soft and collinear 
projection operators
\be
\hat O_{\rm NLO} = ( I - C_{41} - C_{42} ) ( I - S_4).
\label{eq:onlo}
\ee
Note that in Eq.~\eqref{eq3.8} soft and collinear projection operators act on $\FLM(1,2,4)$ that, according 
to Eq.~\eqref{eq3.5},  contains  the energy-momentum conserving $\delta$-function; we stress that 
the soft and collinear limits must  be taken in that $\delta$-function as well. 

The reason for re-writing ${\rm d} \sigma^{\rm R}$ as in
Eq.~\eqref{eq3.8} is that the last term there is finite, thanks to the
nested structure of subtraction terms.  This term can, therefore, be
integrated numerically in four dimensions.  We emphasize again that 
the subtraction terms, as formulated here, are not
Lorentz-invariant. This means that all the three terms in
Eq.~\eqref{eq3.8} should be computed in the same reference
frame that, as already mentioned, is taken to be the center-of-mass
reference frame of the colliding partons.

We now consider the remaining two terms in Eq.~\eqref{eq3.8}.  Their
common feature is either complete or partial decoupling of the gluon
$g_4$ from the matrix element thanks to the fact that they
 contain either soft or collinear projection operators. Hence, those
terms can be re-written in such a way that all singularities are 
extracted and canceled, without specifying the matrix elements for the
hard process.

To see this explicitly, consider first two terms in Eq.~\eqref{eq3.8} 
and write  them as 
\be
\big\langle ( I - C_{41} - C_{42} ) S_4 F_{LM}(1,2,4) \big\rangle + 
\big\langle (C_{41} + C_{42} ) F_{LM}(1,2,4) \big\rangle.
\label{eq3.9}
\ee
It is easy to see that the first term in Eq.~\eqref{eq3.9} vanishes.\footnote{This feature 
is particular to the process under consideration.} Indeed, in the limit 
when the gluon $g_4$ becomes soft, we find 
\be
S_4 F_{\rm LM}(1,2,4)  =   \frac{2 \Cf g_{s,b}^2 }{E_4^2}  
\frac{\rho_{12}}{\rho_{14} \rho_{24} } F_{LM}(1,2),
\ee
where $g_{s,b}$ is the bare QCD coupling, $\Cf=4/3$ is the QCD color factor, and $F_{LM}(1,2)$ is closely related to the LO cross section
\be
 \langle F_{LM}(1,2) \rangle  =2s \cdot {\rm d} \hat\sigma^{\rm LO} = 
 \int {\rm d}{\rm Lips}_V\; |{\cal M}(1,2,V)|^2\; 
{\cal F}_{\rm kin}(1,2,V).
\label{eq:lo}
\ee  
The action of the collinear operators on $\rho$'s gives 
\be
C_{41} \frac{\rho_{12}}{\rho_{14}\rho_{24}} = \frac{1}{\rho_{14}},~~~~
C_{42} \frac{\rho_{12}}{\rho_{14}\rho_{24}} = \frac{1}{\rho_{24}}.
\ee
Since for head-on collision $\rho_{12}=2$, $\rho_{24}=2-\rho_{14}$, we find
\be
\frac{\rho_{12}}{\rho_{14}\rho_{24}} = \frac{1}{\rho_{14}} + \frac{1}{\rho_{24}};
\ee
this implies $(\I-C_{41}-C_{42})S_4\FLM(1,2,4) = 0$.

Hence, the only term that we need to consider in Eq.~\eqref{eq3.9} is the collinear subtraction
\be
\big\langle (C_{41} + C_{42} ) F_{LM}(1,2,4) \big\rangle.
\ee
We will consider the action of the operator $C_{41}$  on $F_{LM}(1,2,4)$ and then infer the result 
for the operator $C_{42}$. First, we find the collinear limit 
\be
C_{41} F_{LM}(1,2,4) = \frac{g_{s,b}^2}{E_4^2 \rho_{41}} (1-z) P_{qq}(z) \frac{F_{LM}( z \cdot 1, 2 )}{z}.
\label{eq3.14}
\ee
We note that a new variable $z = 1 - E_4/E_1$  is introduced in Eq.~\eqref{eq3.14}. 
The notation $F_{LM}( z \cdot 1, 2 )$
 implies   that in the computation of $F_{LM}(1,2)$, c.f. Eq.~\eqref{eq:lo},  
the momentum $p_1$ is replaced with $z p_1$ everywhere, including 
the energy-momentum conserving $\delta$-function.  We also used $P_{qq}(z)$ to denote  the splitting function 
\be
P_{qq}(z) = C_F \left [ \frac{1+z^2}{1-z} - \ep (1-z) \right] = P_{qq}^{(0)}(z) + \ep P_{qq}^{(\ep)}(z).
\label{eq3.15}
\ee

To simplify $\langle C_{41} F_{LM}(1,2,4) \rangle$, we integrate over the emission angle of the 
gluon $g_4$, rewrite the integration over its energy as an integral over $z$ and express 
$g_{s,b}$ in terms of the renormalized coupling $\as(\mu)$.
After straightforward 
manipulations we find 
\be
\langle C_{41} F_{LM}(1,2,4) \rangle 
= 
-\frac{[\alpha_s]}{\ep} \frac{\Gamma^2(1-\ep)}{\Gamma(1-2\ep)}(2 E_1)^{-2 \ep}  
\int \limits_{\zm}^{1} \frac{{\rm d} z }{(1-z)^{2\ep}}   P_{qq}(z) \frac{F_{LM}( z \cdot 1, 2 )}{z},
\label{eq3.20}
\ee
where $\zm = 1-\Em/E_1$ and we introduced the short-hand notation
\be
[\as]\equiv \asontwopimu \frac{\mu^{2\ep} e^{\ep\gamma_E}}{\Gamma(1-\ep)}.
\ee
We note that in  Eq.~\eqref{eq3.20}  integration over $z$ leads to divergences caused by the 
soft $ z \to 1$ 
singularity in the  splitting functions. These singularities are regulated  dimensionally in 
Eq.~\eqref{eq3.20}.  
On the other hand, this equation has the form of a  convolution of a hard matrix element 
with a splitting function, so  we expect that 
 divergences present there will  cancel against the collinear subtraction terms. 
However,  collinear subtractions employ  regularization of soft singularities that is 
based on the plus-prescription. Our goal, therefore,  is to rewrite 
Eq.~\eqref{eq3.20} in such a way that 
all  soft singularities  are regulated by the plus-prescription; once this is done, combining this 
contribution with virtual corrections and collinear subtractions becomes straightforward.    

To simplify the notation, we  denote  
$
F_{LM}( z \cdot 1, 2 )/z = G(z)
$
and split $P_{qq}(z)$ into a piece that is 
singular at $z = 1$ and a regular piece
\be
 P_{qq}(z) = \frac{2\Cf}{(1-z)} + P^{\rm reg}_{qq}(z).
\ee
We also note that we can extend the integral over $z$ in Eq.~\eqref{eq3.20}
 to $z = 0$ since if $E_4 > E_{\rm max}$, $F_{LM}(z \cdot 1,2)$ 
will vanish because there is not enough energy to produce the final state. 
We will use this fact frequently in our \NNLO analysis.
We write 
\be 
\begin{split} 
& \int \limits_{\zm}^{1} \frac{{\rm d} z }{(1-z)^{2\ep}}   P_{qq}(z) \frac{F_{LM}( z \cdot 1, 2 )}{z} 
= \int \limits_{0}^{1} {\rm d} z\;
\left [ \frac{2 C_F }{(1-z)^{1+2\ep} } + (1-z)^{-2\ep}\; P^{\rm reg}_{qq}(z) 
\right ] G(z)
\\
& = -\frac{C_F}{\ep} G(1) + 
\int \limits_{0}^{1} {\rm d} z\;
\left [ \frac{2 C_F }{(1-z)^{1+2\ep} } \left ( G(z) - G(1) \right )  + (1-z)^{-2\ep}\; P^{\rm reg}_{qq}(z) G(z)
\right ].
\label{eq3.21}
\end{split} 
\ee
The expression in Eq.~\eqref{eq3.21} can be expanded in a power series in 
$\ep$ to the required order and the plus-distributions 
can be used to write the result in a compact form. Indeed, the following 
equation holds 
\be
\frac{ G(z) - G(1)  }{(1-z)^{1+2\ep} } 
=  \left [ \sum \limits_{n=0}^{\infty} \frac{(-1)^n (2\ep)^n}{n!} {\cal D}_n(z) \right ]\;  G(z), 
\ee
where ${\cal D}_n(z) = [\ln^n(1-z)/(1-z)]_+$. 
It is now straightforward to rewrite Eq.~\eqref{eq3.20} in such a way that all soft, $z\to 1$,
singularities 
are regulated using the plus-prescription. We use the fact that we are in the 
center-of-mass frame of the incoming $q\bar q$ pair, so that $2 E_1 = 2 E_2 = \sqrt{s}$. 
We find 
\be
\begin{split}
& \langle C_{41} F_{LM}(1,2,4) \rangle 
= 
-\frac{[\alpha_s] s^{- \ep}   }{\ep} \frac{\Gamma^2(1-\ep)}{\Gamma(1-2\ep)}
\\
& \times  \left [ - \left ( \frac{C_F}{\ep} + \frac{3 C_F }{2}  \right ) \bigl<F_{LM}(1,2) \bigr>
+  \int \limits_{0}^{1}   {\rm d} z \PqqRRtwodelta (z) 
\left\langle
\frac{F_{LM}( z \cdot 1, 2 )}{z}\right\rangle
\right ].
\end{split}
\label{eq3.23}
\ee
The splitting function in Eq.~\eqref{eq3.23} reads\footnote{
The $\mathcal O(\ep^2)$ contribution to $\PqqRRtwodelta$, relevant for NNLO contributions,
is reported in Appendix~\ref{sect:defs}.
}
\be
\PqqRRtwodelta(z)  = {\hat P}_{qq}^{(0)}(z) + \ep \PqqRRtwodeltaEp(z) + \mathcal O(\ep^2), 
\ee
where   ${\hat P}_{qq}^{(0)}(z)$ is the LO Altarelli-Parisi splitting kernel, see 
Eq.~\eqref{eq:Pqq_AP}, 
and $\PqqRRtwodeltaEp$ is defined as
\be
\PqqRRtwodeltaEp(z) = \Cf\bigg[2 (1+z) \ln(1-z) - (1-z) - 4 {\cal D}_1(z)\bigg].
\ee

The result for $\langle C_{42} F_{LM}(1,2,4) \rangle$ is obtained 
by a simple replacement  $\FLM(z \cdot 1, 2) \to \FLM(1, z\cdot 2)$
in Eq.~\eqref{eq3.23}.  Putting everything together, we find the following result 
for the real emission cross section 
\be
\begin{split}
& 2 s\cdot{\rm d}\sigma^{\rm R} = 
2 [\alpha_s] s^{-\ep} \left ( \frac{C_F}{\ep^2} + \frac{3C_F}{2\ep} \right ) 
\frac{\Gamma^2(1-\ep)}{\Gamma(1-2\ep)}
\big\langle F_{LM}(1,2) \big\rangle  +  \big\langle \hat O_{\rm NLO} F_{LM}(1,2,4) \big\rangle 
\\
& 
-\frac{[\alpha_s] s^{-\ep} }{\ep} \frac{\Gamma^2(1-\ep)}{\Gamma(1-2\ep)} 
\int \limits_{0}^{1}   {\rm d} z \PqqRRtwodelta(z)
\left \langle \frac{F_{LM}( z \cdot 1, 2 )}{z}   +  \frac{F_{LM}( 1, z \cdot 2 )}{z} \right \rangle.
\end{split}
\label{eq3.27}
\ee
We note that in Eq.~\eqref{eq3.27}  all singularities of the real-emission contribution 
are explicit and a straightforward expansion in $\ep$ is, in principle, possible. 
However, such an expansion is inconvenient since it involves higher-order $\ep$ 
terms of lower-multiplicity amplitude. To avoid these contributions, it is useful to combine 
Eq.~\eqref{eq3.27} with virtual corrections and collinear counterterms. 

For the virtual corrections, all divergent parts can be separated using 
Catani's representation of  renormalized one-loop scattering amplitudes \cite{Catani:1998bh}. 
We obtain 
\be
\begin{split}
2 s \cdot {\rm d}\sigma^{\rm V} &= 
\big\langle F_{LV}(1,2) \big\rangle  = 
\int {\rm d}{\rm Lips}_V\; 2 {\rm Re} \left\{ {\cal M}(1,2) {\cal M}^*_{{\rm 1-loop}}(1,2) \right\}\; 
{\cal F}_{\rm kin}(1,2,V) \\
&= -2[\as]\cos(\ep\pi)
\lp\frac{\Cf}{\ep^2}+\frac{3}{2}\frac{\Cf}{\ep}\rp
s^{-\ep}
\big\langle \FLM(1,2) \big\rangle 
+ 
\big\langle \FVF(1,2)\big\rangle,
\label{eq:FVF}
\end{split}
\ee
where $\FVF(1,2)$ is free of singularities and $\mu$-independent. 

To arrive at the final result,  we add 
virtual, real and collinear subtraction terms, c.f. Eq.~\eqref{eq2.9}, 
and obtain 
\be
\begin{split}
& 2 s\cdot {\rm d}\hat\sigma^{\rm NLO} = \bigg.
[\as]\big\langle \mathcal S(1,2)\FLM(1,2)\big\rangle 
+  \langle \hat O_{\rm NLO} F_{LM}(1,2,4) \rangle
+  \left\langle \FVF(1,2)\right\rangle
\\
& 
-\frac{[\alpha_s] s^{-\ep}}{\ep} \frac{\Gamma^2(1-\ep)}{\Gamma(1-2\ep)} 
\int \limits_{0}^{1}   {\rm d} z \PqqRRtwodelta(z)
\left \langle \frac{F_{LM}( z \cdot 1, 2 )}{z}   +  \frac{F_{LM}( 1, z \cdot 2 )}{z} \right \rangle
\\
& + \frac{\as(\mu)}{2\pi\ep}
\int \limits_{0}^{1} {\rm d} z \; \hat P_{qq}^{(0)}(z)
\left \langle \frac{F_{LM}( z \cdot 1, 2 )}{z}   +  \frac{F_{LM}( 1, z \cdot 2 )}{z} \right \rangle, 
\end{split}
\label{eq3.30}
\ee
where 
\be
\mathcal S(1,2) = 
2 s^{-\ep}\lp\frac{\Cf}{\ep^2}+\frac{3}{2}\frac{\Cf}{\ep}\rp 
\left[\frac{\Gamma^2(1-\ep)}{\Gamma(1-2\ep)}-\cos(\pi\ep)\right],
\ee
and the extra $z$ terms in the denominator of the last line of Eq.~\eqref{eq3.30} 
appear because of the 
$z$-dependent flux factor in the collinear counterterm cross section. 
Taking  the limit $\ep \to 0$ in Eq.~\eqref{eq3.30}, 
we find the final formula for the NLO QCD contribution to the scattering 
cross section for $q(p_1) + \bar q(p_2) \to V +X$. It reads
\be
\begin{split}
& 2 s\cdot{\rm d} \hat\sigma^{\rm NLO} = 
\left\langle \FVF(1,2) + \asontwopimu \left[ \frac{2}{3}\pi^2 \Cf \FLM(1,2)\right]\right\rangle
+  \big\langle \hat O_{\rm NLO} F_{LM}(1,2,4) \big\rangle+
\\
& 
+\asontwopimu
\int \limits_{0}^{1} {\rm d} z 
\left [ \ln \lp\frac{s}{\mu^2}\rp  \hat P_{qq}^{(0)}(z) - \PqqRRtwodeltaEp(z) \right ] 
\left \langle \frac{F_{LM}( z \cdot 1, 2 )}{z}   +  \frac{F_{LM}( 1, z \cdot 2 )}{z} \right \rangle. 
\end{split}
\ee
It follows that the NLO cross section is computed as a sum of low-multiplicity 
terms, including those where $\FLM(z \cdot 1, 2)$ or $\FLM(1,z \cdot 2)$ are 
convoluted with particular splitting functions, and the subtracted real emission 
term described by $\langle {\cal O}_{\rm NLO} \FLM(1,2,4) \rangle $. We note that 
terms that involve matrix elements of different multiplicities, as well as 
terms that involve different types of convolutions, are separately finite. 
We will use this observation at NNLO, to check  for the cancellation of $1/\ep$ divergences
in an efficient way.

\section{ The \NNLO computation: general considerations} 
\label{sect:gennnlo}

We would like to extend the above framework 
to NNLO in QCD.
Apart from the UV and collinear renormalization discussed in Sec.~\ref{sec:intro},
the NNLO cross section receives contributions from two-loop virtual 
corrections to $q \bar q \to V$ (double virtual), from
one-loop corrections to the 
process with an additional gluon in the final state $ q \bar q  \to V+g$  (real-virtual),  
and from the tree-level process 
with two additional gluons in the final state $q \bar q \to V+gg$ (double real).  

The double-virtual corrections can be dealt with in a straightforward
way since all the divergences of the two-loop matrix elements are explicit, 
universal and well-understood \cite{Catani:1998bh}. For our purposes,
we only need to write them in a convenient form.  The real-virtual
corrections are more tricky, but do not require new conceptual developments. Indeed, the kinematic regions that lead to
singularities in one-loop amplitudes with an additional gluon in the
final state are identical to those appearing in the NLO
computations and, furthermore, the limiting behavior of one-loop amplitudes with
one additional parton is well-understood
\cite{Bern:1999ry,Kosower:1999rx,Catani:2000pi}.  Hence, we can deal with the 
real-virtual contribution by a simple generalization of what we did at
NLO.

The qualitatively new element of the NNLO computation is the double-real emission 
process $q \bar q \to V+gg$. The methods that are applicable at next-to-leading 
order  need to be adjusted  to become  useful 
in the NNLO  case. However, somewhat surprisingly, these adjustments appear to 
be {\it relatively  minor} although, of course, the bookkeeping becomes much more complex.

We  begin by discussing the ultraviolet and PDF renormalizations at NNLO,  as well as
the double-virtual and the real-virtual contributions. 
We then move on to a more involved analysis of the double-real emission 
contribution to ${\rm d} \hat\sigma^{\rm NNLO}$.

\section{The \NNLO computation: ultraviolet and PDF renormalization}
\label{sect:uvpdf}

In this section we study the 
contributions to ${\rm d} \hat\sigma^{\rm NNLO}$  
coming from  the ultraviolet and the collinear renormalization, beginning with the former.
As mentioned previously, because the process 
$ q \bar q \to V$ is driven by a conserved current which is independent of $\alpha_s$, the ultraviolet
renormalization contribution is very simple. Combining Eq.~\eqref{eq:ren}
and the first two lines of Eq.~\eqref{eq3.30},
it is straightforward to obtain
\bes
& 2s\cdot{\rm d}\sigma^{\rm ren} = 
-\frac{\beta_0}{\ep}\asontwopimu\bigg[\big\langle \ONLO \FLM(1,2,4)\big\rangle 
+ \big\langle \FVF(1,2)\big\rangle\bigg]\\
&-\frac{\beta_0}{\ep}\lp\asontwopimu\rp^2\frac{\Gamma(1-\ep)e^{\ep\gamma_E}}{\Gamma(1-2\ep)}
\lp\frac{\mu^2}{s}\rp^{\ep}
\Bigg\{
\Cf
\left[\frac{2}{\ep^2}+\frac{3}{\ep}\right]
\times\left[ 1 -\cos(\pi\ep)\frac{\Gamma(1-2\ep)}{\Gamma^2(1-\ep)}\right]\\
&
\times\big\langle\FLM(1,2)\big\rangle
-\frac{1}{\ep}\int_0^1 \d z \;\PqqRRtwodelta(z)
\left\langle\frac{\FLM(z\cdot 1,2)}{z}+\frac{\FLM(1,z\cdot2)}{z}\right\rangle
\Bigg\}.
\label{eq5.1}
\end{split}
\ee

We proceed to the collinear subtraction. Rewriting Eq.~\eqref{eq:ren}
to make the convolutions explicit, we obtain
\be
\begin{split}
&2s\cdot\d\sigma^{\rm CV} = 2s\cdot
\lp\asontwopimu\rp\frac{1}{\ep}\lp
\hat P_{qq}^{(0)} \otimes  \big[\d\sigma^{\rm R}+\d\sigma^{\rm V}\big]
+\big[\d\sigma^{\rm R}+\d\sigma^{\rm V}\big] \otimes \hat P_{qq}^{(0)} 
\rp\\
&+\lp\asontwopimu\rp^2\frac{1}{\ep^2}
\int\limits_0^1 \d z \; \d\bar z \; \hat P^{(0)}_{qq}(z)
\times\left\langle
\frac{\FLM(z\cdot 1,\bar z\cdot 2)}{z\bar z} 
\right\rangle
\times
\hat P^{(0)}_{qq}(\bar z)\\
&+\lp\asontwopimu\rp^2
\int\limits_0^1 \d z \left[\frac{\big[\hat P^{(0)}_{qq}\otimes \hat P^{(0)}_{qq}\big](z)
-\beta_0 \hat P^{(0)}_{qq}(z)}{2\ep^2} + \frac{\hat P^{(1)}_{qq}(z)}{2\ep}\right]
\Bigg\langle\frac{\FLM(z\cdot 1,2)}{z}\Bigg\rangle\\
&+\lp\asontwopimu\rp^2
\int\limits_0^1 \d z \left[\frac{\big[\hat P^{(0)}_{qq}\otimes \hat P^{(0)}_{qq}\big](z)
-\beta_0 \hat P^{(0)}_{qq}(z)}{2\ep^2} + \frac{\hat P^{(1)}_{qq}(z)}{2\ep}\right]
\Bigg\langle
\frac{\FLM(1,z\cdot 2)}{z}\Bigg\rangle.
\end{split}
\label{eq:pdfren_flm}
\ee
Terms that involve convolutions of the various splitting functions with 
$\FLM$ are, in principle, straightforward to deal with. These terms 
 are fully regulated and can be expanded in powers of $\ep$ 
without further ado.  In practice, we combine those terms with other  
contributions in order to cancel the singularities prior to integration 
over $z$.   

It is less straightforward to re-write  $\hat P\otimes \d\sigma^{\rm R+V}$
and $ \d\sigma^{\rm R+V} \otimes \hat P$ 
in a form convenient for combining them with other contributions to 
${\rm d} \hat\sigma^{\rm NNLO}$.
We focus on  $\hat P\otimes \d\sigma^{\rm R+V}$, and consider 
the effect of the convolution on the first two lines of Eq.~\eqref{eq3.30}.

First, we consider the term proportional to $\langle \mathcal S(1,2) \FLM(1,2) \rangle$
in Eq.~\eqref{eq3.30}.  It receives contributions from the divergent
part of virtual corrections and from the soft regularization of collinear
subtraction terms. These terms scale differently with $z$. The virtual correction depends on $s^{-\ep}$ which becomes $(zs)^{-\ep}$ once 
the momentum $p_1$ is changed to  $z p_1$. On the other hand, the soft
remainders of the collinear subtracted terms scale as $E_i^{-2\ep}$,
with $i = 1,2$. Hence, in the calculation of ${\rm d}\sigma^{\rm
  R+V}(z \cdot 1, 2)$, the corresponding contribution scales with $z$ either as
$\sim z^{-2\ep}$ or as $\sim 1$.  Therefore, we have to compute
\be
\int\limits_{0}^{1} \d z\; \hat P^{(0)}_{qq}(z) \mathcal
S(z\cdot 1,2) \times \bigg\langle\frac{\FLM(z\cdot 1,2)}{z}\bigg\rangle,
\label{eq41}
\ee
where 
\bes
&\mathcal S(z\cdot 1,2) = s^{-\ep}
\lp\frac{\Cf}{\ep^2}+\frac{3}{2}\frac{\Cf}{\ep}\rp
\left[
\frac{\Gamma^2(1-\ep)}{\Gamma(1-2\ep)}
\big[ z^{-2\ep}+1\big]-2 \cos(\pi\ep)z^{-\ep}
\right]=
\\
& =s^{-\ep}\Cf \bigg[
\frac{2}{3}\pi^2 + \ln^2 z + 
\lp
\pi^2-\frac{2}{3}\pi^2\ln z + \frac{3}{2}\ln^2 z - \ln^3 z - 4\zeta_3
\rp\ep
\bigg] + \mathcal O(\ep^2).
\end{split}
\ee
For future convenience, we re-write Eq.~\eqref{eq41} as
\be
\int \limits_{0}^{1} \d z\; \hat P^{(0)}_{qq}(z) \mathcal S(z\cdot 1,2) 
\times\bigg\langle \frac{\FLM(z\cdot 1,2)}{z}\bigg\rangle
= s^{-\ep}
\int \limits_{0}^{1} \d z\; \mathcal P_{qq,{\rm NLO_{CV}}}(z)
\bigg\langle 
\frac{\FLM(z\cdot 1,2)}{z}
\bigg\rangle,
\ee
where the splitting function $\mathcal P_{qq,{\rm NLO_{CV}}}(z)$ is defined in Eq.~\eqref{eq:PqqNLOCV}.

The other two terms that we need involve convolutions
of splitting function $\PqqRRtwodelta$ with $\FLM(1,2)$, c.f. Eq.~\eqref{eq3.30}.
The first term can be written as a double convolution 
\bes
\int \limits_0^1 \d x \;\hat P^{(0)}_{qq}(x) x^{-2\ep} s^{-\ep} \int \limits_0^1 \d y \;
\PqqRRtwodelta(y)\;
\Bigg\langle
\frac{\FLM(xy\cdot 1,2)}{xy}
\Bigg\rangle = 
\\
 = s^{-\ep}
\int \limits_0^{1} \d z \; \big[\mathcal P_{qq}\otimes\mathcal P_{qq}\big]_{\rm NLO_{CV}}(z)
\;\Bigg\langle 
\frac{\FLM(z \cdot 1,2)}{z}\Bigg\rangle,
\end{split}
\ee
where the splitting function $\big[\mathcal P_{qq}\otimes\mathcal P_{qq}\big]_{\rm NLO_{CV}}$
is defined in Eq.~\eqref{eq:pqqOpqqNLOcv}. The second term is the left-right convolution
\be
\begin{split} 
s^{-\ep} 
\int \limits_{0}^{1} \d x\; \hat P^{(0)}_{qq}(x)
\int \limits_{0}^{1} \d y\; \PqqRRtwodelta(y)
\;
\Bigg\langle
\frac{\FLM(x\cdot1,y\cdot2)}{xy}
\Bigg\rangle.
\end{split}
\ee
Combining 
all these terms we find 
\bes
& 2s\cdot \bigg[\hat P^{(0)}_{qq}\otimes \d\sigma_{qq}^{\rm R+V}+
\d\sigma^{\rm R+V}\otimes \hat P^{(0)}_{qq}\bigg]
= -\frac{[\as] s^{-\ep} }{\ep}\frac{\Gamma^2(1-\ep)}{\Gamma(1-2\ep)} \times 
\Bigg \{
\\
&
 \int\limits_0^1 \d z\; \d\bar z~
\PqqRRtwodelta(z)  \Bigg\langle \frac{\FLM(z\cdot 1,\bar z\cdot 2)
+\FLM(\bar z\cdot 1,z\cdot 2)}{z\bar z} \Bigg\rangle
\hat P^{(0)}_{qq}(\bar z)
\\
& +  \int \limits_0^1 \d z 
\big[\mathcal P_{qq}\otimes\mathcal P_{qq}\big]_{\rm NLO_{CV}}(z)
 \Bigg\langle \frac{\FLM(z\cdot 1,2)+\FLM(1,z\cdot 2)}{z}\Bigg\rangle
\Bigg \} \\
&+[\as] s^{-\ep}
\int \limits_{0}^{1} \d z\; \mathcal P_{qq,{\rm NLO_{CV}}}(z)
 \Bigg\langle \frac{\FLM(z\cdot 1,2)+\FLM(1,z\cdot 2)}{z}\Bigg\rangle\\
&+\int\limits_0^1 \d z~ \hat P^{(0)}_{qq}(z)  \Bigg\langle \ONLO 
  \frac{\FLM(z\cdot 1,2,4)+\FLM(1,z\cdot 2,4)}{z}\Bigg\rangle
\\
&+ \int \limits_{0}^{1} \d z \;\hat P^{(0)}_{qq}(z) 
\Bigg\langle \frac{\FVF(z\cdot 1,2)+\FVF(1,z\cdot 2)}{z}\Bigg\rangle.
\end{split}
\label{eq:final_pnlo}
\ee
Each term  that appears on the right hand side of Eq.~\eqref{eq:final_pnlo} is 
regularized and can be expanded in powers of $\ep$ independently of the other terms 
in that equation.

\section{The \NNLO computation: double-virtual corrections} 
\label{sect:virt}

The calculation of double-virtual corrections proceeds in the standard way. We start 
with  the scattering amplitude for $q \bar q \to V$ and write it as an expansion 
in the renormalized strong coupling constant 
\bes
\mathcal M = \mathcal M_{\rm tree} + 
\asontwopimu \mathcal M_{\rm 1-loop} + 
\left[\asontwopimu\right]^2 \mathcal M_{\rm 2-loop} + 
\mathcal O(\as^3),
\end{split}
\ee
where the dependence of the scattering amplitudes on the momenta of the external particles is suppressed. 
By analogy with what was done in Section~\ref{sect:mynlo}, we write
\bes
\!
2s\cdot 
{\rm d} \sigma^{\rm VV} &= 
\left[\asontwopimu\right]^2
\int {\rm dLips}_{12 \to V}  \left [ 
2 {\rm Re} \{{\cal M}_{\rm 2-loop}
 {\cal M}_{\rm tree}^*\}  + |{\cal M}_{\rm 1-loop}|^2
\right ]  \mathcal F_{\rm kin}(1,2,V) \\
&+\frac{\beta_0}{\ep}
\left[\asontwopimu\right]^2
\int {\rm dLips}_{12 \to V}
2 {\rm Re} \{{\cal M}_{\rm 1-loop}
 {\cal M}_{\rm tree}^*\} \mathcal F_{\rm kin}(1,2,V) = \\
&
\equiv \big\langle \FVV(1,2) \big\rangle,
\end{split} 
\ee 
where in the second line we removed the renormalization contribution that is
already accounted for in Eq.~\eqref{eq5.1}.

$\FVV$ can be directly expanded in a Laurent series in $\ep$ and
integrated over the phase space of the final state particles since
this integration does not introduce  soft or collinear divergences.
Before doing that, it is convenient to explicitly extract the 
$1/\ep$ poles from $\langle \FVV  \rangle$. 
Soft and collinear singularities of a generic scattering 
amplitude are given in Ref.~\cite{Catani:1998bh}. Using these results, 
we rewrite $\langle \FVV \rangle $ as
\bes
\big\langle \FVV(1,2) \big\rangle  &= 
\left[\asontwopimu\right]^2 \lp\frac{\mu^2}{s}\rp^{2\ep}
\Bigg\{
\frac{e^{2\ep \gamma_E}}{\Gamma^2(1-\ep)}\cos^2(\ep\pi)
\Cf^2\left[\frac{2}{\ep^4}+\frac{6}{\ep^3}+\frac{9}{2\ep^2}
\right]\\
&+\frac{e^{\ep \gamma_E}}{\Gamma(1-\ep)}\cos(2\ep\pi)
\bigg[
\frac{\Cf^2}{\ep}\lp-\frac{3}{8}+\frac{\pi^2}{2}-6\zeta_3\rp+
\Ca\Cf\times\\
&\times\lp-\frac{11}{12\ep^3}-\frac{83}{18\ep^2}
+\frac{\pi^2}{12\ep^2}
-\frac{961}{216\ep}-\frac{11\pi^2}{48\ep}
+\frac{13\zeta_3}{2\ep}\rp
\bigg]
\Bigg\} \big\langle \FLM(1,2) \big\rangle \\
&+\left[\asontwopimu\right]
\bigg[
\frac{e^{\ep\gamma_E}}{\Gamma(1-\ep)}
\cos(\ep\pi)\Cf
\lp\frac{\mu^2}{s}\rp^\ep
\lp\frac{2}{\ep^2}+\frac{3}{\ep}\rp
 + \frac{\beta_0}{\ep}\bigg] 
\big\langle \FVF(1,2) \big\rangle   \\
&+ \big\langle \FVsqF(1,2) \big\rangle
+ \big\langle \FVVF(1,2) \big\rangle.
\end{split}
\label{eq4v.3}
\ee 
The sum of the last two  terms in Eq.~\eqref{eq4v.3}  is a finite remainder of the 
${\cal O}(\alpha_s^2)$ contribution to the virtual corrections once its 
divergent part is written in a form suggested in Ref.~\cite{Catani:1998bh}.
More specifically, $\FVsqF$ is the finite remainder of the one-loop amplitude squared,
while $\FVVF$ is the genuine two-loop finite remainder.
Note that, contrary to $\FVF(1,2)$ and $\FVsqF$, $\FVVF$ is scale-dependent;
the scale-dependent contribution reads
\be
\FVVF(\mu^2,s) - \FVVF(s,s) =
\frac{44}{3} \Cf \Ca \log\left(\frac{s}{\mu^2}\right) \FLM .
\ee
As follows from Eq.~\eqref{eq4v.3}, the singularities of the double-virtual corrections 
are proportional to the leading order contribution $\FLM(1,2)$ and 
to the finite part of the virtual corrections  to the NLO cross section 
 $\FV^{\rm fin}(1,2)$, given in Section~\ref{sect:mynlo}.
Our goal is 
to rewrite the real-virtual and the double-real emission 
contributions in a way that allows explicit cancellation of the divergences 
in Eq.~\eqref{eq4v.3} without specifying hard matrix elements. 

\section{The \NNLO computation: real-virtual corrections} 
\label{sect:rv}

The kinematics of the real-virtual corrections is identical to the NLO case 
described in Section~\ref{sect:mynlo}.
The procedure for making these corrections expandable in $\ep$
is, therefore, the same.  We write
\bes
& 2s\cdot\d\sigma^{\rm RV} \equiv \big\langle \FRV(1,2,4)\big\rangle = 
 \big\langle S_4 \FRV(1,2,4)\big\rangle + \\
&+ \left\langle \big(\I-S_4 \big)\big(C_{41}+C_{42}\big) \FRV(1,2,4) \right\rangle+
 \big\langle {\cal O}_{NLO} \FRV(1,2,4) \big\rangle.
\end{split}
\label{eq7.1}
\ee
We remind the reader that, according to our notation, soft- and
collinear-projection operators in Eq.~\eqref{eq7.1} do not act on the phase space of the gluon $g_4$. 
It remains to compute the corresponding limits in Eq.~\eqref{eq7.1} and to re-write them, 
where appropriate,  as  convolutions of the hard matrix elements 
with splitting functions. 

The soft limit of a  general one-loop amplitude  is discussed in 
Refs.~\cite{Bern:1999ry,Catani:2000pi}.
Adapting those results to our case, we find 
\bes
E_4^2 S_4 \FRV(1,2,4) = 2\Cf \gsb^2 \bigg[ \frac{\rho_{12}}{\rho_{14}\rho_{24}}
\FV(1,2)  \\
-\Ca[\as]\frac{1}{\ep^2}\frac{\Gamma^5(1-\ep)\Gamma^3(1+\ep)}
{\Gamma^2(1-2\ep)\Gamma(1+2\ep)}
\lp\frac{\rho_{12}}{\rho_{14}\rho_{24}}\rp^{1+\ep}
E_4^{-2\ep} 
2^{-\ep}\FLM(1,2)\bigg].
\end{split}
\label{eq5.5}
\ee
We need to integrate Eq.~\eqref{eq5.5} over the phase space of the
gluon $g_4$. This can be easily done, with the result
\bes
\big\langle S_4 \FRV(1,2,4)\big\rangle = 2\Cf [\as]\bigg[
\frac{\lp 4\Em^2\rp^{-\ep}}{\ep^2} \big\langle\FV(1,2)\big\rangle\\
-\Ca \frac{[\as]}{\ep^4}
\frac{\Gamma^5(1-\ep)\Gamma^3(1+\ep)}
{\Gamma^2(1-2\ep)\Gamma(1+2\ep)}
\frac{\lp 4\Em^2\rp^{-2\ep}}{4}
 \big\langle\FLM(1,2)\big\rangle\bigg].
\end{split}
\label{eq5.8}
\ee
Note that in order to obtain a meaningful result, it is crucial that the
integration over gluon energy  is bounded from above; as we already explained 
in the context of the NLO computations, we use a parameter 
$\Em$ for this purpose, c.f.  Eq.~\eqref{eq3.4}.

The second term that we need to consider is the soft-regulated collinear subtraction term
\be
\left\langle\big(\I-S_4\big) \big( C_{41} + C_{42}\big)  \FRV(1,2,4)\right\rangle.
\ee
We will only discuss   the  collinear projection operator $C_{41}$; the contribution of $C_{42}$ is 
obtained  along the same lines.
Collinear limits of loop amplitudes were studied in Refs.~\cite{Kosower:1999rx,Bern:1999ry}. 
Using these results and adapting 
them to our case, we find
\bes
C_{41} \FRV(1,2,4) = \frac{\gsb^2}{E_4^2\rho_{41}}
\Bigg[
(1-z)P_{qq}(z) \frac{\FV(z\cdot 1,2)}{z}  \\
+[\as]\frac{\Gamma^3(1-\ep)\Gamma(1+\ep)}{\Gamma(1-2\ep)}
\frac{ 2^{-\ep} E_1^{-2\ep}}{\rho_{41}^{\ep}}
P_{qq}^{\rm loop,i}(z) \frac{\FLM(z\cdot 1,2)}{z}\Bigg],
\label{eq:coll_rv_qqi}
\end{split}
\ee
where  the splitting function 
$P_{qq}^{\rm loop,i}$ is given in Eq.~\eqref{eq:pqqloopi}.
The soft-collinear limit is easily obtained by taking the collinear approximation 
in Eq.~\eqref{eq5.5}.  We find 
\bes
S_4 C_{41} \FRV(1,2,4) = 
\frac{\gsb^2}{E_4^2\rho_{41}}\Bigg[
2\Cf \FV(1,2)\\
-2\Ca\Cf\frac{[\as]}{\ep^2}2^{-\ep}
\frac{\Gamma^5(1-\ep)\Gamma^3(1+\ep)}
{\Gamma^2(1-2\ep)\Gamma(1+2\ep)}
E_4^{-2\ep}\rho_{41}^{-\ep}\FLM(1,2)\Bigg].
\end{split}
\ee
Integrating over emission angles of the gluon $g_4$ and rewriting the result through 
plus-distributions, following the discussion in Section~\ref{sect:mynlo}, 
we obtain a convenient representation  for the soft-regulated collinear subtraction term.
It reads
\bes
\left\langle\big(\I-S_4\big)C_{41} \FRV(1,2,4)\right\rangle = 
\frac{[\as] s^{-\ep}}{\ep}\frac{\Gamma^2(1-\ep)}{\Gamma(1-2\ep)} 
\int \limits_{0}^{1} \d z\;\mathcal P_{qq,RV,1}(z) 
\left\langle\frac{\FV(z\cdot 1,2)}{z}\right\rangle\\
+\;\frac{[\as]^2s^{-2\ep} }{\ep}
\frac{\Gamma^4(1-\ep)\Gamma(1+\ep)}{\Gamma(1-3\ep)}
\int \limits_{0}^{1} \d z\; \mathcal P_{qq,RV,2}(z) 
\left\langle\frac{\FLM(z\cdot 1,2)}{z}\right\rangle,
\label{eq5.12}
\end{split}
\ee
where the two splitting functions are defined in Eqs.~(\ref{eq:PqqRV_1},\ref{eq:PqqRV_2}).
We replace $\FLM(z\cdot 1,2)$ with 
$\FLM(1,z\cdot 2)$ in Eq.~\eqref{eq5.12}  to obtain 
the result for $\left\langle\big(\I-S_4\big)C_{42} \FRV(1,2,4)\right\rangle$.

We are now in position to present the final result for the real-virtual part. 
Collecting results shown in Eqs.~(\ref{eq5.8},\ref{eq5.12}), we obtain 
\be
\begin{split}
&\big\langle \FV(1,2,4)\big\rangle = \\
&=\big\langle
\mathcal O_{\rm NLO} \FV(1,2,4)
\big\rangle+2\Cf [\as] s^{-\ep} \Bigg[
\frac{1}{\ep^2} 
\lp\frac{4\Em^2}{s}\rp^{-\ep}
\big\langle\FV(1,2)\big\rangle  \\
&
-\Ca \frac{[\as] s^{-\ep}}{4\ep^4}
\frac{\Gamma^5(1-\ep)\Gamma^3(1+\ep)}
{\Gamma^2(1-2\ep)\Gamma(1+2\ep)}
\lp\frac{4\Em^2}{s}\rp^{-2\ep}
\big\langle\FLM(1,2)\big\rangle\Bigg]\\
&
+\frac{[\as] s^{-\ep} }{\ep}\frac{\Gamma^2(1-\ep)}{\Gamma(1-2\ep)} 
\int \limits_{0}^{1} \d z~ \mathcal P_{qq,RV_1}(z) 
\bigg\langle 
\frac{\FV(z\cdot 1,2)+\FV(1,z\cdot 2)}
{z}\bigg\rangle\\
&
+\frac{[\as]^2 s^{-2\ep}}{\ep}
\frac{\Gamma^4(1-\ep)\Gamma(1+\ep)}{\Gamma(1-3\ep)}
\int\limits_0^1 \d z\; \mathcal P_{qq,RV_2}(z) 
\bigg\langle
\frac{\FLM(z\cdot 1,2)+\FLM(1,z\cdot 2)}{z}
\bigg\rangle. 
\end{split}\label{eq:final_rv}
\ee
We stress that each term on the r.h.s.
in Eq.~\eqref{eq:final_rv} can 
be expanded in powers of $\ep$; we will make full use of this 
to cancel $1/\ep$ singularities when  combining  Eq.~\eqref{eq:final_rv}
with other
contributions   to ${\rm d}\hat\sigma^{\rm NNLO}$. To this
end, it is useful to make all the $1/\ep$ singularities explicit in Eq.~\eqref{eq:final_rv} 
by writing
$\FV(1,2)$ in terms of $\FVF(1,2)$ and $\FLM(1,2)$, c.f.  Eq.~\eqref{eq:FVF},
 and $\FV(1,2,4)$ as 
\bes
 \FV(1,2,4) &=
 [\as]\Bigg[\lp\frac{1}{\ep^2}+\frac{3}{2\ep}\rp \cos(\ep\pi)
(\Ca-2\Cf)s_{12}^{-\ep}\\
&-\lp\frac{\Ca}{\ep^2}+\frac{3\Ca}{4\ep}\rp(s_{14}^{-\ep}+s_{24}^{-\ep})\Bigg]\FLM(1,2,4)
+\FVF(1,2,4).
\end{split}
\ee
We used $s_{ij} = 2p_i\cdot p_j$ and denoted  a finite remainder which  does not depend 
on the scale $\mu$ by $\FVF(1,2,4)$.

\section{The \NNLO computation: double-real emission}
\label{sect:nnlorr}

\subsection{General considerations}

In this section, we discuss the double-real emission contributions to
${\rm d}\hat\sigma^{\rm NNLO}$.  Similar to the NLO case, we need to
determine all kinematic configurations that may lead to singularities
and understand the factorization of the matrix element that describes $q\bar q \to V+ gg$ in these regions.  In the case of the two-gluon emission in $q \bar q$ annihilation into a colorless final state, the singular regions
correspond to soft and/or collinear emissions, with collinear
directions being the collision axis and the direction of either one of
the two gluons.

The difficulty of the NNLO case is that 
each of these kinematic limits can be 
approached in  several different ways and all of them have to be identified 
and regularized separately.  To do so, we introduce  several 
 soft and collinear projection  operators. They are defined as follows. Consider 
a quantity $A$ that depends on the four-momenta of some or all of the particles in the 
process. The action of  operators $\SS, S_{4,5}, \CC_{1,2}, C_{14},C_{15},C_{24},C_{25},C_{45}$  
on $A$ is described by the following formulas
 \bes
\SS A &= \lim_{E_4,E_5\to 0} A, {\rm~ at~ fixed~ }E_5/E_4,\;\;\;\; S_i A = \lim_{E_i \to 0} A,\\
\CC_i A &= \lim_{\rho_{4i},\rho_{5i}\to 0} A, {\rm~with~non~vanishing~}
\rho_{4i}/\rho_{5i},\rho_{45}/\rho_{4i},\rho_{45}/\rho_{5i},
\;\;\;\;
 C_{ij} A = \lim_{\rho_{ij}\to 0} A.
\end{split}
\ee

To make use of these projection operators, we need to re-write the two-gluon phase space 
in a way that allows these limits to be taken. It is convenient to order  gluon 
energies as the first step.   We write   
\begin{align}
& 2s\cdot \d\sigma^{\rm RR}=\int
\frac{1}{2!} [dg_4][dg_5] |\M(1,2,4,5,V)|^2 \d{\rm Lips}_{12-45\to V} \mathcal F_{\rm kin}(1,2,4,5,V)=
\nonumber
\\
& = 
\int
[dg_4][dg_5] \theta(E_4-E_5)|\M(1,2,4,5,V)|^2\d{\rm Lips}_{12-45\to V} \mathcal F_{\rm kin}(1,2,4,5,V)
=
\\
&=
\int
[dg_4][dg_5] \theta(E_4-E_5)\FLM(1,2,4,5) = 
\big\langle F_{LM}(1,2,4,5) \big\rangle,
\nonumber
\end{align}
where as in Section~\ref{sect:mynlo} $\d{\rm Lips}$ is the phase space for the final state $V$, 
including momentum conserving delta-function.
The gluon phase space elements $[{\rm d} g_{4,5}]$ are defined as in Eq.~\eqref{eq3.4}
\be
[dg] = \frac{d^{d-1}p_g}{(2\pi)^{d-1}2E_g} \theta(\Em-E_g).
\label{eq:emax}
\ee
As we already saw when considering 
the  real-virtual contribution, 
it is necessary to introduce the  $\theta$-function in order   to define integrals 
over gluon energies in the soft limits.  We work in the center-of-mass frame of 
the colliding quark and antiquark; 
it is in this frame that all the energies in the above formulas are defined. 

We recall that, similar to the NLO case, soft and collinear projection
operators act on everything that appears to the right of
them. However, in the NNLO case we will find it convenient, occasionally, to
also simplify the phase space in the collinear limits. If so, we will
explicitly show the corresponding part of the phase space to the
right of the operator.  For example 
\bes \big\langle \mathcal O \dg4
\FLM(1,2,4,5) \big\rangle \equiv \int_{E_4>E_5} \dg5 \mathcal O \dg4
\FLM(1,2,4,5),
\end{split}
\ee
implies that the operator ${\cal O}$ acts on $\FLM(1,2,4,5)$ {\it} and on the phase space element 
 $\dg4$. 

We begin by extracting soft singularities from the double-real process, largely  
repeating what we have 
done  at next-to-leading order.\footnote{The very possibility of regulating soft singularities 
independently of collinear ones follows directly from QCD color coherence. It is the primary reason
why we don't need to consider sectors where soft and collinear singularities are
entangled. Note that this feature does not apply to individual diagrams but to on-shell
QCD scattering amplitudes as a whole.} We write
\bes
\big\langle F_{LM}(1,2,4,5) \big\rangle 
&= \big\langle \SS    \FLM(1,2,4,5) \big\rangle 
 + \big\langle ( I - \SS) \FLM(1,2,4,5) \big\rangle  
\\
& = 
\big\langle \SS    \FLM(1,2,4,5) \big\rangle  +  \big\langle S_5 ( I - \SS) \FLM(1,2,4,5) \big\rangle 
\\
&+
\big\langle (I - S_5) ( I - \SS) \FLM(1,2,4,5) \big\rangle.
\end{split}
\label{eq8.4}
\ee
In Eq.~\eqref{eq8.4},  the last  term is soft-regulated, in  the second term 
gluon $g_5$ is soft 
and soft singularities associated with $g_4$ are regulated, and
in the  first term  {\it both}  $g_4$ and $g_5$ are soft. 

All of these terms contain collinear singularities. Regulating these is more difficult  since 
collinear singularities overlap.  For this 
reason,  we first need to split the phase space into mutually-exclusive 
partitions that, ideally, select a single kinematic configuration that leads 
to singularities. We write 
\be
1 = \sum_i w^i, 
\ee
where the label  $i$ runs through the elements 
of the set $i \in \{14,15;~ 24,25;~ 14,25;~ 15,24\}$. We refer to the first two 
elements of the set as triple-collinear and to the last two elements of the set as 
double-collinear partitions. 
We construct the weights $w^i$ in such a way that when they
multiply the matrix element ${\cal M}(1,2,4,5)$ squared, the
resulting expression is only singular in a well-defined subset of
limits.  For example, in the partition $14,15$ collinear singularities
in $w^{14,15}|{\cal M}|^2$ only occur when gluons $4$ and/or $5$ are
emitted along the direction of the incoming quark $q(p_1)$ or when
their momenta are parallel to each other.  Similarly, in the partition
$24,25$, the singularities occur when momenta of $g_4$ and/or $g_5$ are
parallel to $p_2$ or to each other. In the partitions $14,25$ and $15,24$, singularities only occur when momenta of $g_4$ and $g_5$ are
collinear to $p_1$ and $p_2$ or $p_2$ and $p_1$, respectively. Apart
from these requirements, the specific form of $w^i$ is arbitrary.
Weights used in this calculation are given in 
Appendix~\ref{sect:phsp}.
In the following, we assume that weights  do not depend on gluon energies  
and, therefore, commute with soft operators. 

The triple-collinear partitions require further splitting to factorize all the relevant singularities. 
The purpose of this splitting is to establish a well-defined hierarchy for 
the parameters $\rho_{4i},\rho_{5i},\rho_{45}$, 
since  different orderings correspond
to different limiting behavior.
This splitting is not unique; a possible choice consistent with the phase space 
parametrization that we employ  later (c.f. Appendix~\ref{sect:phsp}) reads
\bes
1 &= 
\theta\lp\eta_{51} < \frac{\eta_{41}}{2}\rp + 
\theta\lp\frac{\eta_{41}}{2} < \eta_{51} < \eta_{41}\rp  \\
& + 
\theta\lp\eta_{41} < \frac{\eta_{51}}{2}\rp + 
\theta\lp\frac{\eta_{51}}{2} < \eta_{41} < \eta_{51}\rp 
 \equiv \theta^{(a)} + \theta^{(b)} + \theta^{(c)} + \theta^{(d)}, 
\end{split}\label{eq:tcsectors}
\ee
where, as usual,  $\eta_{ij} = \rho_{ij}/2 = (1-\cos\theta_{ij})/2$. We will refer 
to the four contributions shown in Eq.~\eqref{eq:tcsectors} as sectors  $a,b,c$ and $d$. 
We note that only two of the sectors are qualitatively
different, since the other two are just obtained by the $4\leftrightarrow 5$ replacement. 
However, because of  the energy ordering $E_5 < E_4$, 
we no longer have the $4\leftrightarrow 5$ symmetry, and have to consider all the four sectors 
separately.

A suitable parametrization of all angular variables that supports splitting 
of the angular phase space as shown in Eq.~\eqref{eq:tcsectors} and allows 
factorization of singularities in hard amplitudes was provided  in  
Ref.~\cite{czakonsub} and is reviewed in Appendix~\ref{sect:phsp}. 
We will use this parametrization to carry out integrations 
over sectors $\theta^{(a)},\theta^{(b)},\theta^{(c)},\theta^{(d)}$ explicitly in what 
follows. 

Having introduced  partitions and sectors as a tool to identify singularities that may 
appear in the course of integrating over the angles of the final state gluons, 
we are now in position to write the result for
 the double-real emission cross section as a sum of terms that  either can be integrated 
in four dimensions, 
or that depend on hard matrix elements of lower multiplicity. The latter contributions still 
diverge, either explicitly or implicitly, and we will have to combine them with double-virtual 
and real-virtual contributions to arrive at the finite result. 
  
We can thus rewrite the double-real emission cross section as 
\be
\begin{split} 
\big\langle \FLM  (1,2,4,5)\big\rangle & = \left\langle \SS \FLM(1,2,4,5)\right\rangle 
+\left\langle \big[\I-\SS\big] S_5 \FLM(1,2,4,5)\right\rangle 
\\
& + \big\langle \FLM^{s_rc_s}(1,2,4,5) \big\rangle  
+ \big\langle \FLM^{s_rc_t}(1,2,4,5) \big\rangle  
+ \big\langle \FLM^{s_rc_r}(1,2,4,5) \big\rangle,  
\label{eq4.9}
\end{split}
\ee
where the soft-regulated, single-collinear term $\langle \FLM^{s_rc_s} \rangle$ reads
\bes
& \langle \FLM^{s_rc_s} \rangle =
\sum_{(ij)\in dc}
\left\langle
\big[\I-\SS\big]\big[\I-S_5\big]
\bigg[ C_{4i} \dg4 + C_{5j}\dg5 \bigg] w^{i4,j5}\FLM(1,2,4,5)
\right\rangle\\
&\quad\quad
+\sum_{i\in tc} 
\bigg\langle
\big[\I-\SS\big]\big[\I-S_5\big]
\bigg[
\theta^{(a)} C_{5i} + \theta^{(b)} C_{45} + \theta^{(c)} C_{4i} + \theta^{(d)} C_{45}
\bigg]\\
&\quad\quad\quad\quad\quad
\times\dg4 \dg5 w^{i4,i5}\FLM(1,2,4,5)
\bigg\rangle,
\label{eq4.10}
\end{split} 
\ee
the soft-regulated, triple-collinear terms $ \langle \FLM^{s_rc_t} \rangle $ reads 
\bes
& \langle \FLM^{s_rc_t} \rangle =
-\sum_{(ij)\in dc}\bigg\langle
\big[\I-\SS\big]\big[\I-S_5\big]
C_{4i}C_{5j}\dg4\dg5 w^{i4,j5}\FLM(1,2,4,5)
\bigg\rangle\\
&\quad\quad
+\sum_{i\in tc} 
\bigg\langle
\big[\I-\SS\big]\big[\I-S_5\big]
\bigg[
\theta^{(a)} \CC_i\big[\I-C_{5i}\big] + \theta^{(b)} \CC_i\big[\I-C_{45}\big]  \\
&\quad\quad\quad\quad~~
 + \theta^{(c)} \CC_i\big[\I-C_{4i}\big]+ \theta^{(d)} \CC_i\big[\I-C_{45}\big]
\bigg]\dg4 \dg5 w^{i4,i5}\FLM(1,2,4,5)
\bigg\rangle, 
\label{eq4.11}
\end{split} 
\ee
and the fully regulated term $\langle \FLM^{s_rc_r} \rangle $ reads  
\bes
& \langle \FLM^{s_rc_r} \rangle  = 
\sum_{(ij)\in dc}\bigg\langle
\big[\I-\SS\big]\big[\I-S_5\big]
\bigg[(\I- C_{5j})(\I-C_{4i})\bigg]\\
&\quad\quad\quad\quad\quad\quad
\times\dg4\dg5 w^{i4,j5}\FLM(1,2,4,5)
\bigg\rangle\\
&\quad\quad
+\sum_{i\in tc} 
\bigg\langle
\big[\I-\SS\big]\big[\I-S_5\big]
\bigg[
\theta^{(a)} \big[\I-\CC_i\big]\big[\I-C_{5i}\big] + 
\theta^{(b)} \big[\I-\CC_i\big]\big[\I-C_{45}\big]  \\
&\quad\quad\quad\quad~~
 + \theta^{(c)} \big[\I-\CC_i\big]\big[\I-C_{4i}\big]+ 
\theta^{(d)} \big[\I-\CC_i\big]\big[\I-C_{45}\big]
\bigg]\\
&\quad\quad\quad\quad~~
\times\dg4 \dg5 w^{i4,i5}\FLM(1,2,4,5)
\bigg\rangle .
\end{split}
\label{eq4.12}
\ee
For the process under consideration, $dc=\{(1,2),(2,1)\}$ and
$tc = \{1,2\}$. 
The above results are obtained by combining  soft-regulated expression 
for $\FLM(1,2,4,5)$ with multiple partitions of unity for the angular projections, 
and the understanding of which collinear divergences can appear in each partition 
and sector.  One can easily check, starting from Eq.~\eqref{eq4.9}, that the  
collinear projection operators add up to an identity operator for each partition and each sector.

For example, the contribution of the  double-collinear sector $i4,j5$ follows 
from an expansion of an identity operator written in the following form 
\be
I =  (I - C_{4i} + C_{4i}) ( I - C_{5j}+C_{5j}) = C_{4i} + C_{5j} - C_{4i}C_{5j} 
+(I - C_{4i}) (I - C_{5j} ).
\ee
The reason we restrict ourselves to the subtraction of the $C_{4i}$ and $C_{5j}$ collinear projection 
operators is that in the partition $i4,j5$ no other collinear singularities appear,
thanks to the damping factor $w^{i4,j5}$.
Similarly, taking e.g. the sector $a$ of the triple-collinear partition $w^{i4,i5}$, we 
write 
\be
I =  (I - \CC_i + \CC_i) ( I - C_{5i}+C_{5i}) = C_{5i} + \CC_i(\I -C_{5i}) 
 + (I - \CC_i) (I - C_{5i} ), 
\ee
because in this case a singularity can only occur either in a triple-collinear limit 
$\eta_{4i} \sim \eta_{5i} \to 0$ or if $\eta_{5i} \to 0$  at fixed $\eta_{4i}$. 

It is worth pointing out a few things in connection with Eq.~\eqref{eq4.9}.
\begin{itemize}

\item The procedure that we used to write Eq.~\eqref{eq4.9}
is, in principle,  process- and phase space parametrization-independent. 
We will use a particular parametrization of the phase space 
to perform the required computation but  one should keep in mind that the freedom 
of changing the parametrization exists and, perhaps, it is worth exploring it in the future. 

\item The first term in Eq.~\eqref{eq4.9}, $\langle\SS \FLM(1,2,4,5)\rangle$,  is the 
double-soft subtraction term. It contains unregulated soft and collinear
singularities and  cannot be directly expanded in $\ep$. However,
it only involves the tree-level matrix element $\FLM(1,2)$ which means that emitted gluons 
decouple from  both the hard matrix element and the phase space constraints.
When integrated over gluon energies and angles, this term gives rise to $1/\ep^n$ poles, $ n \le 4$. 

\item The second term in Eq.~\eqref{eq4.9}, $\left\langle
  \big[\I-\SS\big] S_5 \FLM(1,2,4,5)\right\rangle $, is the
  double-soft regulated, single-soft subtraction term.  It contains
  $\FLM(1,2,4)$ and matrix elements of {\it lower} multiplicity.
  This term still contains unregulated singularities that occur when
  the momentum of the gluon $g_4$ becomes collinear to the collision
  axis or to the direction of $g_5$.  When integrated over gluon
  energies and angles, this term gives rise to $1/\ep^n$ poles, with
  $n \le 3$.

\item  The term $\langle \FLM^{s_rc_s} \rangle $ in Eq.~\eqref{eq4.10} 
is the soft-regulated single-collinear subtraction term. Note
that thanks to the damping and $\theta$ factors only one kind of collinear singularity
per term is present. $\langle \FLM^{s_rc_s} \rangle $ involves $\FLM(1,2,4(5))$, depending on 
the partition and the sector and, therefore,   contains  unregulated
collinear singularities related to gluon emissions along the collision axis. 
It gives rise to  $1/\ep^2$ and $1/\ep$ poles. 

\item The term $\langle \FLM^{s_rc_t} \rangle $ in Eq.~\eqref{eq4.11} is  the 
triple-collinear subtraction, where all other singularities are 
regulated. As we will see, contributions of 
the double-collinear partitions  to   $\langle \FLM^{s_rc_t} \rangle $
have  the ``double-convolution'' structure. 
The $\langle \FLM^{s_rc_t} \rangle $ term  contains  $1/\ep$ poles in
contributions  of triple-collinear partitions, and  $1/\ep^2$ poles in contributions 
of  double-collinear partitions. 

\item  The term $\langle \FLM^{s_rc_r} \rangle $ in Eq.~\eqref{eq4.12}
is completely  regulated, thanks to the nested subtractions.
It can be evaluated in four dimensions. It is the only term that involves  the
full hard matrix element for the process $q(p_1) + \bar q(p_2) \to V + g(p_4) + g(p_5)$. 
\end{itemize}

Following our general strategy, we 
need to study the first four terms 
in Eq.~\eqref{eq4.9},  which  involve  matrix elements of reduced multiplicity, and 
re-write them in terms of integrable 
quantities that admit straightforward expansions in the dimensional regularization 
parameter $\ep$.  We will  discuss how to do this in the following subsections. 

\subsection{The double-soft subtraction term}
We begin with the discussion of the first term in Eq.~\eqref{eq4.9},  
$\left\langle \SS \FLM(1,2,4,5)\right\rangle$. It  corresponds to the kinematic 
situation where momenta of both gluons vanish  at a comparable rate. The corresponding 
limit for the amplitude squared  is given in 
Refs.~\cite{Berends:1988zn,Catani:1999ss} and allows us to write
\be
\left\langle \SS \FLM(1,2,4,5)\right\rangle = g_{s,b}^2\;
\int_{E_4 > E_5} \dg4 \dg5 \; {\rm Eik}(1,2,4,5) \;
\big\langle \FLM(1,2 )\big\rangle,
\ee
where 
\bes
{\rm Eik}(1,2,4,5) =  
4 \Cf^2 S_{12}(4) S_{12}(5) 
+\Ca\Cf \big[ 2 S_{12}(4,5)-S_{11}(4,5)-S_{22}(4,5)\big],
\end{split}
\ee
and~\cite{Catani:1999ss}
\bes
S_{ij}(q) \equiv & \frac{p_i\cdot p_j}{(p_i\cdot q)(p_j\cdot q)} = 2\frac{s_{ij}}{s_{iq}s_{jq}}, \\
S_{ij}(q_1,q_2) =& S_{ij}^{so}(q_1,q_2) 
+ \frac{p_i\cdot q_1 p_j\cdot q_2+p_i\cdot q_2 p_j\cdot q_1}{p_i\cdot q_{12} p_j\cdot q_{12}}
\left[\frac{(1-\ep)}{(q_1\cdot q_2)^2}-\frac{1}{2} S_{ij}^{so}(q_1,q_2)\right]  \\
&-2\frac{p_i\cdot p_j}{q_1\cdot q_2 p_i\cdot q_{12} p_j \cdot q_{12}},
\end{split}
\ee
with
\be
 S_{ij}^{so}(q_1,q_2) = 
\frac{p_i\cdot p_j}{q_1\cdot q_2}
\lp
\frac{1}{p_i\cdot q_1 p_j\cdot q_2} + \frac{1}{p_i\cdot q_2 p_j\cdot q_1}\rp
-\frac{(p_i\cdot p_j)^2}{p_i\cdot q_1 p_j\cdot q_1 p_i\cdot q_2 p_j\cdot q_2}.
\ee
At this point, we stress again that the hard matrix element $\FLM(1,2)$ corresponds 
to a tree-level process and that the emitted gluons 
have no impact on the kinematic properties of the final state $V$ because 
they decouple from the energy-momentum conserving $\delta$-function.

The goal now is to integrate the eikonal factor over the momenta 
of the two gluons. We note that,    at this point, unless put in by hand,  
the integration over gluon energies becomes  unconstrained 
since the 
energy-momentum conserving $\delta$-function becomes independent 
of the gluon momenta after the double-soft limit is taken. It is for this 
reason that we need to introduce $E_{\rm max}$ as in Eq.~\eqref{eq:emax}.

 To satisfy constraints on gluon energies, we 
parametrize them as 
\be
E_4 = E_{\rm max} ~x_1,\;\;\;\;\;E_5 = E_4 x_2 = E_{\rm max}\; x_1 x_2.
\label{eq8.19}
\ee
Written in these variables, the eikonal factor becomes
\be
{\rm Eik}(1,2,4,5) = E_{\rm max}^{-4} x_1^{-4} x_2^{-2} {\cal E} (x_2, n_1,  n_2,  n_4,   n_5 ).
\ee
The important point is that the dependence on the overall energy scale $x_1$ factorizes 
and that the remaining (complicated) function  ${\cal E}$ is independent of energies of 
the incoming partons. 
We also use the parametrization of energies Eq.~\eqref{eq8.19}
in the phase space to obtain 
\begin{equation}
\int \dg4\dg5 {\rm Eik}(1,2,4,5) = 
-\frac{E_{\mathrm{max}}^{-4\epsilon}}{4\epsilon}   
\int \limits_{0}^{1}\frac{\d x_2}{x_2^{1+2\epsilon}} 
\frac{{\rm d} \Omega_4}{2 (2\pi)^{d-1}}
\frac{{\rm d} \Omega_5}{2 (2\pi)^{d-1}}
{\cal E} (x_2, n_1,  n_2,  n_4,   n_5 ).
\label{eq7.21}
\end{equation}
For the case of a color-singlet final state, this integral is just a 
constant.\footnote{In a more general 
NNLO problem, this integral is  a function of the scalar product of the three-momenta 
of the two hard partons.} 
The abelian contribution is simple to obtain since it is just the product of
\NLO structures. 
In principle it should be possible to compute the non-abelian contribution 
analytically along the lines of e.g. Refs.~\cite{Matsuura:1988sm,deFlorian:2012za,Czakon:2013hxa}. 
However, it is also 
straightforward to obtain it numerically. 
To do this, we partition the phase space as in 
residue-improved sector decomposition~\cite{czakonsub}.  The corresponding formulas 
for the angular phase space are given in Appendix~\ref{sect:phsp}.
Performing the required  
decomposition and integrating Eq.~\eqref{eq7.21} numerically, we obtain
the   $\ep$-expansion 
of the double-soft subtraction term
\begin{equation}
\bigl< \SS F_{LM}(1,2,4,5) \bigr> =
 [\alpha_s]^2  
\big\langle E_{\mathrm{max}}^{-4\epsilon} F_{LM}(1,2)\big\rangle
 \left( \frac{c^{\SS}_{4}}{\epsilon^4}
+\frac{c^{\SS}_{3}}{\epsilon^3}+\frac{c^{\SS}_{2}}{\epsilon^2}
+\frac{c^{\SS}_{1}}{\epsilon}+c^{\SS}_{0} \right),
\label{eq4.19}
\end{equation}
where $c^{\SS}_i = \Cf^2\cdot c^{\SS}_{i,\Cf^2}  + \Ca\Cf \cdot c^{\SS}_{i,\Ca\Cf}$.
Numerical values of  the coefficients $c^{\SS}$ are shown  in Table~\ref{tab:ds}.  
There we also report numerical results for the abelian contribution, which are in
perfect agreement with the analytic calculation. 
The result for the double-soft subtraction $\bigl< \SS F_{LM}(1,2,4,5) \bigr> $ 
does not require any further regularization; we will later combine it with 
other contributions with tree-level  kinematics to cancel the $1/\ep$ singularities 
explicitly.

\begin{table}[t]
\begin{center}
\scalebox{0.85}{
\begin{tabular}{|c|c|c|c|c|}
\hline
$c^{\SS}_{4}$ & $c^{\SS}_{3}$ & $c^{\SS}_{2}$ & $c^{\SS}_{1}$ & $c^{\SS}_{0}$ \\
\hline
$5.55554(2)$ & $-11.73653(7)$ & $-7.3253(7)$ & $ -20.796(5)$&  $-54.65(7)$\\
\hline
$1.999995(8)~\Cf^2$ &
$-5.54530(5)~\Cf^2$ &
$1.1077(3)~\Cf^2$&
$1.522(1)~\Cf^2$&
$1.961(4)~\Cf^2$
\\
\hline
$0.499999(2)~\Ca\Cf$&
$-0.46960(1)~\Ca\Cf$&
$-2.3236(1)~\Ca\Cf$&
$-5.876(1)~\Ca\Cf$&
$-14.52(1)~\Ca\Cf$
\\
\hline
\end{tabular}}
\end{center}
\caption{Coefficients of the $\ep$ expansion of the double-soft projected 
real emission contribution. Full results are given 
in the first row.   Results for individual color factors are given in the  
second and third rows. Numerical errors are such that their contribution
to the final result is below the  per mille level.}
\label{tab:ds}
\end{table}

\subsection{The single-soft term}
We now consider the second term that contributes to  Eq.~\eqref{eq4.9}. It is 
a double-soft regulated, single-soft singular expression that reads 
\be
\left\langle \big[\I-\SS\big] S_5 \FLM(1,2,4,5)\right\rangle.
\ee
Note that this contribution implicitly depends on 
$\FLM(1,2,4)$ and $\FLM(1,2)$, and
the hard matrix element that appears in $\FLM(1,2,4)$ still contains collinear 
singularities that arise  when the momentum of gluon $g_4$ becomes parallel  to the momenta 
of the incoming partons or to the direction of $g_5$. 
These divergences will have to  be extracted and regulated. 

We start by  computing   the soft limit for the gluon $g_5$. We find  
\be
S_5 \FLM(1,2,4,5) = 
\frac{\gsb^2}{E_5^2} \bigg[
4(2\Cf-\Ca)\frac{\rho_{12}}{\rho_{15}\rho_{45}} 
+ \Ca\lp\frac{\rho_{14}}{\rho_{15}\rho_{45}}+\frac{\rho_{24}}{\rho_{25}\rho_{45}}\rp
\bigg] \FLM(1,2,4).
\ee
Since the gluon $g_5$ decouples from the hard matrix element, we can integrate over its 
momentum.  We find
\be
\langle( I - \SS)  S_5\FLM(1,2,4,5) \rangle 
= \langle J_{124}\; ( I - S_4)  \FLM(1,2,4) \rangle, 
\label{eq4.25}
\ee
where  
\bes
J_{124} = 
\frac{ [\as] E_4^{-2\ep}}{\ep^2}
\bigg[
(2\Cf-\Ca)(2\rho_{12})^{-\ep}K_{12} + 
\Ca \big[(2\rho_{14})^{-\ep} K_{14} + (2\rho_{24})^{-\ep} K_{24}\big]
\bigg], 
\end{split}
\ee
and
\be
 K_{ij} = \frac{\Gamma^2(1-\ep)}{\Gamma(1-2\ep)}
\eta_{ij}^{1+\ep}F_{21}(1,1,1-\ep,1-\eta_{ij}) = 
1 + \left[ \Li_2(1-\eta_{ij}) - \frac{\pi^2}{6}\right] \ep^2 +\mathcal O(\ep^3).
\ee

We need to simplify  Eq.~\eqref{eq4.25} because it still contains 
 collinear singularities that appear when the momentum of 
gluon $g_4$ becomes parallel to the collision axis.   To extract them, we write 
\bes
\big\langle
J_{124}\big[\I-S_4\big]\FLM(1,2,4)
\big\rangle  &= \big\langle
\big[\I - C_{41} - C_{42}\big]J_{124}\big[\I-S_4\big]\FLM(1,2,4)
\big\rangle\\
&+ 
\big\langle
\big[ C_{41} + C_{42}\big]J_{124}\big[\I-S_4\big]\FLM(1,2,4)
\big\rangle .
\label{eq4.27}
\end{split}
\ee
We reiterate that according to our notational conventions, the collinear projection operators 
do not act on the phase space element of the gluon $g_4$ in Eq.~\eqref{eq4.27}.
The first term in Eq.~\eqref{eq4.27} is explicitly regulated 
and can be expanded in   powers of $\epsilon$; for this reason, 
we will only be concerned with  the second term. 
We  focus  on  the projection operator   $C_{41}$; 
the contribution of the projection operator $C_{42}$ is then 
obtained  by analogy. 

First, we consider how $C_{41}$ acts on $J_{124}$. Using $\eta_{12}=1$, 
$C_{41}\rho_{24}=\rho_{12}$ and taking
the $\rho_{41}\to 0$ limit on $K_{14}$, $K_{24}$ we obtain
\bes
C_{41} J_{124} = \frac{[\as]}{\ep^2}E_4^{-2\ep}
\frac{\Gamma^2(1-\ep)}{\Gamma(1-2\ep)}
\bigg[ 2^{1-2\ep}\Cf + \Ca \Gamma(1+\ep)\Gamma(1-\ep)(2\rho_{14})^{-\ep} \bigg].
\end{split}
\ee
Integrating over the energy and angle of the gluon $g_4$ we arrive at
\bes
\big\langle
C_{41}&J_{124}\big[\I-S_4\big]\FLM(1,2,4) \big\rangle = 
-\frac{[\as]^2 s^{-2\ep}}{\ep^3}
\bigg[
2\Cf  \frac{\Gamma^4(1-\ep)}{\Gamma^2(1-2\ep)}\\
& + \Ca \frac{\Gamma^4(1-\ep)\Gamma(1+\ep)}{2 \Gamma(1-3\ep)} 
\bigg] 
\int \limits_{z_{\rm min}}^1 \frac{\d z}{(1-z)^{1+4\ep}}
  \PPOP 
\bigl<\FLM(z\cdot 1,2)\bigr>,
\label{eq4.34}
\end{split}
\ee
where $\zm = 1-\Em/E_4$. 
The splitting function operator   $\PPOP$ is defined by means of the following 
equation 
\be
\PPOP f(z) = {\cal P}_{qq}(z) f(z) - 2 C_F f(1),
\label{eq.ppop}
\ee
where  
\be
{\cal P}_{qq}(z) = (1-z) \frac{P_{qq}(z)}{z},
\label{eq.capqq}
\ee
and the  splitting function $P_{qq}(z)$ is given in Eq.~\eqref{eq3.15}.

Note that $ \big\langle
C_{41} J_{124}\big[\I-S_4\big]\FLM(1,2,4) \big\rangle  $   in 
Eq.~\eqref{eq4.34} can be directly expanded in powers of $\ep$ since all the 
singularities are regulated. The only problem that needs to be addressed 
is the fact that the integration over $z$ does not start at $z=0$, as is 
the case for the convolutions. The lower integration boundary 
$z_{\rm min}$  must be kept in  Eq.~\eqref{eq4.34} because of  the subtraction term 
$ 2 C_F \FLM(1,2)$. Indeed, if $z_{\rm min}$ is replaced with zero, 
the integration over the gluon energy for this term extends to the region $E_4 > E_{\rm max}$, 
in contradiction the with original phase space parametrization.  The extension of the integration 
region in Eq.~\eqref{eq4.34} is accomplished following  steps discussed in the context 
of the NLO QCD computation in Section~\ref{sect:mynlo}. 
Effectively, this leads to a redefinition of the splitting function 
\be
\int \limits_{z_{\rm min}}^{1} \frac{{\rm d} z }{(1-z)^{1+4\ep}}
\PPOP  \FLM(z\cdot 1,2) 
\equiv  \int \limits_{0}^{1} {\rm d} z {\cal P}_{qq,RR_1}(z) \frac{\FLM(z \cdot 1, 2)}{z},
\ee
where ${\cal P}_{qq,RR_1}(z)$ is given in  Eq.~\eqref{eq:pqqrrI_b}.
We note that the contribution of the collinear operator $C_{42}$ 
to the second term in Eq.~\eqref{eq4.27} is computed 
in a  similar way;  the computation  leads to the same result as 
in Eq.~\eqref{eq4.34}  up to 
an obvious replacement $\FLM(z \cdot 1, 2) \to \FLM(1, z \cdot 2)$. 

Putting everything together, we find the final result for the 
double-soft regulated single-soft singular contribution to $\langle \FLM(1,2,4,5) \rangle$
\be
\begin{split}
& \big\langle\big[\I-\SS\big] S_5 \FLM(1,2,4,5)\big\rangle = 
\big\langle \big[\I - C_{41} - C_{42}\big]\big[\I-S_4\big] J_{124} \FLM(1,2,4) \big \rangle \bigg. \\
& -\frac{[\as]^2 s^{-2\ep}}{\ep^3}
\left[ 2\Cf \frac{\Gamma^4(1-\ep)}{\Gamma^2(1-2\ep)}
+\frac{\Ca}{2}\frac{\Gamma^4(1-\ep)\Gamma(1+\ep)}{\Gamma(1-3\ep)} \right]
\\
& \;\;\;\;\;\;\;\;\;\;\;\;\;\;\;\;\;\;\;\;\;\times
\int \limits_0^1 \d z\;\mathcal P_{qq,RR_1}(z) 
\Bigg\langle \frac{\FLM(z\cdot 1,2) + \FLM(1,z\cdot 2) }{z} 
\Bigg\rangle.
\end{split}
\label{eq:final_second}
\ee

\subsection{The single-collinear term}

Next, we consider  the soft-regulated, 
single-collinear contribution to $\langle \FLM(1,2,4,5) \rangle $
\begin{align}
& \langle \FLM^{s_rc_s} \rangle 
= \sum_{(ij)\in dc}
\left\langle
\big[\I-\SS\big]\big[\I-S_5\big]
\bigg[ C_{4i} \dg4 + C_{5j}\dg5 \bigg] w^{i4,j5}\FLM(1,2,4,5)
\right\rangle\nonumber\\
& \quad\quad
+\sum_{i\in tc} 
\bigg\langle
\big[\I-\SS\big]\big[\I-S_5\big]
\bigg[
\theta^{(a)} C_{5i} + \theta^{(b)} C_{45} + \theta^{(c)} C_{4i} + \theta^{(d)} C_{45}
\bigg]
\label{eq4.36}
\\
& \quad\quad\quad\quad\quad
\times\dg4 \dg5 w^{i4,i5}\FLM(1,2,4,5)
\bigg\rangle. \nonumber
\end{align}
We need to rewrite Eq.~\eqref{eq4.36}  in such a way that extraction 
of the remaining collinear singularities becomes straightforward.
We note that $\langle \FLM^{s_rc_s} \rangle $
contains contributions from  double- and triple-collinear partitions, 
which we will treat  separately. We will start with the double-collinear 
partitions since they are  somewhat simpler. 

\subsubsection{The double-collinear  partitions}
In this subsection, we will consider the contribution of the double-collinear 
partitions to $\langle \FLM^{s_rc_s} \rangle $.  We begin 
with the partition $14,25$.  For the first term, we need to compute 
\be
\big[\I-\SS\big] \big[\I-S_5\big]C_{41} w^{14,25}\FLM(1,2,4,5) = 
\tilde w_{4||1}^{14,25}  \big[\I-\SS\big] \big[\I-S_5\big]C_{41}\FLM(1,2,4,5), 
\label{eq4.37}
\ee
where $\tilde w_{4||1}^{14,25} = 
\lim_{\rho_{41}\to 0} w^{14,25}$ does not depend on the momentum of gluon $g_4$  anymore.  
To further simplify Eq.~\eqref{eq4.37}, note that collinear and soft projection 
operators commute with each other and that 
\be
\SS (\I-S_5) C_{41}\FLM(1,2,4,5) \sim \SS \FLM(1-4,2,5) - \SS S_5 \FLM(1-4,2,5) = 0.
\ee
This implies that we can  drop the $\SS$ term in Eq.~\eqref{eq4.37}.  We use 
the collinear limit for $C_{41} \FLM(1,2,4,5)$  obtained by a straightforward 
generalization of Eq.~\eqref{eq3.14}.  We define  $z = 1-E_4/E_1$
and obtain 
\be
C_{41}(\I-S_5)\FLM(1,2,4,5) = 
\frac{\gsb^2}{E_4^2 \rho_{41}} {\cal P}_{qq}(z)
\big[\I-S_5\big] \FLM(z\cdot 1,2,5).
\ee
The function  ${\cal P}_{qq}(z)$ was  introduced in Eq.~\eqref{eq.capqq}. 

We now consider the phase space. According to Eq.~\eqref{eq4.36}, $C_{41}$ acts
on the phase space element $\dg4$. We introduce
$x_3 = (1-\cos\theta_{41})/2$ to get
\be
\gsb^2 \dg4 \theta(E_4-E_5) = [\as] s^{-\ep} E_4^2 \rho_{14} \frac{\d z}{(1-z)^{1+2\ep}}
\big[x_3(1-x_3)\big]^{-\ep}\frac{\d x_3}{x_3}\times \theta(z_{\rm max}(E_5)-z),
\label{eq8.38}
\ee
with
\be
z_{\rm max}(E_5)  = 1- \frac{E_5}{E_1} = 1 - \frac{2E_5}{\sqrt{s}} .
\ee 
In this parametrization, the action of $C_{41}$ implies 
replacing $[ x_3 ( 1- x_3) ]^{-\ep}$ with $x_3^{-\ep}$.
Putting everything together, we obtain 
\bes
&\big\langle \big[\I-\SS\big] \big[\I-S_5\big]C_{41} w^{14,25} \dg4 \FLM(1,2,4,5)\big\rangle =
\\
&=-\frac{[\as] s^{-\ep} }{\ep} 
\int\limits_{\zm}^{z_{\rm max}(E_5)}
\frac{\d z}{(1-z)^{1+2\ep} }
{\cal P}_{qq}(z)
\big\langle
\tilde w_{4||1}^{14,25}
\big[\I-S_5\big]\FLM(z\cdot 1,2,5)\big\rangle.
\label{eq4.41}
\end{split}
\ee
Note that, similar to the NLO case, the lower boundary 
$\zm$ is not important when 
integrating  $\FLM(z\cdot 1,2,4)$ since for $z<\zm$ there is no 
sufficient energy in the incoming partons to produce the required final state. 

Next we consider the action of the $C_{52}$ projection operator.
Following the preceding discussion and  using $z = 1-E_5/E_2 = 1-2E_5/\sqrt{s}$, we obtain
\bes
&\left\langle \big[\I-\SS\big] \big[\I-S_5\big]C_{52} w^{14,25} \dg5 \FLM(1,2,4,5)\right\rangle=\\
&=-\frac{[\as] s^{-\ep} }{\ep} 
\int \limits_{z_{\rm min}(E_4)}^{1}
\frac{\d z}{ ( 1-z)^{1+2\ep} }
\PPOP 
\big\langle
\tilde w_{5||2}^{14,25} \; 
\FLM(1,2 \cdot z,4)\big\rangle.
\label{eq4.46}
\end{split}
\ee
The operator  $\PPOP$ was introduced in Eq.~\eqref{eq.ppop} and
\be
\zm(E_4) = 1-\frac{E_4}{E_2}=1-\frac{2E_4}{\sqrt{s}}.
\ee
The sum of Eq.~\eqref{eq4.41} and Eq.~\eqref{eq4.46}  gives the required result 
for the collinear sector $14,25$. 

The partition $15,24$  
is  obtained from the results for $14,25$ after a few obvious replacements. 
We find 
\bes
&\left\langle \big[\I-\SS\big] \big[\I-S_5\big]C_{42} 
w^{15,24} \dg4 \FLM(1,2,4,5)\right\rangle =\\
&=-\frac{[\as] s^{-\ep}}{\ep} 
 \int \limits_{z_{\rm min}}^{z_{\rm max}(E_5)} 
\frac{\d z}{ ( 1-z)^{1+2\ep} }
{\cal P}_{qq}(z)
\big\langle
\tilde w_{4||2}^{15,24}
\big[\I-S_5\big]\FLM(1,z\cdot 2,5)\big\rangle,
\end{split}
\ee
and
\bes
\left\langle\big[\I-\SS\big] \big[\I-S_5\big]C_{51} w^{15,24} \dg5 \FLM(1,2,4,5)\right\rangle=\\
=-\frac{[\as] s^{-\ep}}{\ep} 
\int \limits_{z_{\rm min}(E_4)}^{1} \frac{\d z}{(1-z)^{1+2\ep}}
\PPOP 
\big\langle
\tilde w_{5||1}^{15,24} \;
\FLM(1\cdot z,2,4)\big\rangle.
\end{split}
\ee

We can now combine the contributions of the two double-collinear partitions. In doing so, it is 
convenient to always denote the ``resolved'' (i.e. the non-collinear)  gluon by $g_4$.  
Out of the four terms that we need to combine, two correspond to the 
collinear emission along the direction of the incoming quark $p_1$ and two along 
the direction of the incoming antiquark $p_2$. 
We consider terms that belong to the former category first.

When combining results, it is important to realise that 
$z_{\rm min}(E_4) = z_{\rm max}(E_4) = 1 - 2E_4/\sqrt{s}$. We will denote it 
by $z_4 > z_{\rm min} = 1-2 E_{\rm max}/\sqrt{s}$. After straightforward
manipulations we find
\begin{align}
&\left\langle\big[\I-\SS\big]\big[\I-S_5\big]
\left[ C_{41} w^{14,25} +C_{51} w^{15,24}\right]
\dg4 \dg5 \FLM(1,2,4,5)\right\rangle =
\nonumber
\\
&=
-\frac{[\as s^{-\ep}]}{\ep}
 \int\limits_0^1 \frac{\d z}{(1-z)^{1+2\ep}}
\Bigg\langle
\tilde w_{5||1}^{15,24}
\bigg(
\PPOP\big[\I-S_4\big]\FLM(z\cdot 1,2,4)
\label{eq4.51}
\\
&\quad\quad+
\theta\lp z_4 - z \rp
2\Cf \big[\I-S_4\big]\FLM(1,2,4) 
+\theta\lp z - z_4 \rp 
\PPOP S_4 \FLM(z\cdot 1,2,4)\bigg)\Bigg\rangle.
\nonumber
\end{align}
Note that the lower  integration boundary  in this formula should be $z = z_{\rm min}$ 
but we can  extend the integration region to  $z = 0$, without making a mistake. This is 
so because  every time $\FLM(z\cdot 1,...)$ appears in the integrand,  the $z>z_{min}$
condition is automatically enforced by the requirement that the initial state should have 
enough energy to produce the final state. 
On the other hand, if $\FLM(z\cdot 1,...)$ does not appear, 
$\theta$-functions require that  $z > z_4 > z_{\rm min}$.  
We also note that, thanks to explicit  subtractions and 
constraints due to $\theta$-functions,  each term 
in Eq.~\eqref{eq4.51} vanishes if  $z\to 1$ or $E_4\to 0$.
Finally, we stress that $\PPOP$ and $S_4$ commute since
they act on different variables. 

We can write a similar equation for the sum of the two terms where the 
collinear gluon is emitted along the direction of the antiquark 
$\bar q(p_2)$.  Finally, putting everything together, we obtain 
the contribution of the double-collinear partitions to $\langle \FLM^{s_r c_s} \rangle$. 
We find 
\bes
\bigg\langle
&\big[\I-\SS\big]\big[\I-S_5\big]
\left[ (C_{41} + C_{52}) w^{14,25} + ( C_{51} + C_{42} ) w^{24,15}
\right] \\
& 
\times \dg4 \dg5 \FLM(1,2,4,5)\bigg\rangle=
-\frac{[\as] s^{-\ep}}{\ep}
 \int\limits_0^1 \frac{\d z}{(1-z)^{1+2\ep}}
\\
&
 \times 
\bigg\langle
\tilde w_{5||1}^{15,24}
\Bigg\{
\PPOP\big[\I-S_4\big] \big[ \FLM(z\cdot 1,2,4)+ \FLM(1,z \cdot 2,4) \big]
\\
& ~~~~~~~~~~~~~~~
+ \theta\lp z_4 - z \rp
4\Cf \big[\I-S_4\big]\FLM(1,2,4) 
\\
& ~~~~~~~~~~~~~~~
+\theta\lp z - z_4 \rp \PPOP S_4 \big[ 
\FLM(z\cdot 1,2,4) + \FLM(1,z\cdot 2,4)
\big]
\bigg\}\Bigg\rangle.
\end{split}
\ee
Note that the  second term in the curly bracket only 
depends on $z$ through the $\theta$-function and so the $z$-integration of this term 
can be performed 
explicitly.

\subsubsection{The triple-collinear partition $14,15$}

We consider the triple-collinear partition $14,15$ and study the contribution of single-collinear limits in Eq.~\eqref{eq4.36}. We begin with sector $(a)$. The 
relevant expression reads 
\be
\left\langle
\big[\I-\SS\big]\big[\I-S_5\big]\theta^{(a)} 
C_{51} \dg5 w^{14,15} \FLM(1,2,4,5)\right\rangle.
\ee
The calculation is identical to the case of the double-collinear partition 
except that we need to account for the constraint that defines sector $(a)$ when integrating 
over the angle of gluon $g_5$. Writing $\rho_{14} = 2x_{3}$ and $\rho_{15} = 2x_4$, 
and taking $\theta_a = \theta(\rho_{14}/2 - \rho_{15})$,
we find 
\be
\int\limits_0^1\theta^{(a)} \frac{\d x_4}{x_4^{1+\ep}} =
\int\limits_0^{x_3/2} \frac{\d x_4}{x_4^{1+\ep}} =
-\frac{(x_3/2)^{-\ep}}{\ep} = -\frac{(\rho_{14}/4)^{-\ep}}{\ep}.
\ee
Using this result, we obtain
\bes
&\left\langle\big[\I-\SS\big] \big[\I-S_5\big]\theta^{(a)}
C_{51} w^{14,15} \dg5 \FLM(1,2,4,5)\right\rangle=\\
&=-\frac{[\as] s^{-\ep}}{\ep} 
 \int \limits_{0}^{1} \frac{\d z}{(1-z)^{1+2\ep}} 
\left\langle
\tilde w_{5||1}^{14,15} \lp\frac{\rho_{14}}{4}\rp^{-\ep}
\theta(z - z_4)
\PPOP \FLM(1,2,4)\right\rangle. 
\end{split}
\ee
A similar calculation for the sector $(c)$ gives 
\bes
&\left\langle\big[\I-\SS\big] \big[\I-S_5\big]\theta^{(c)}
C_{41} w^{14,15} \dg4 \FLM(1,2,4,5)\right\rangle=
-\frac{[\as]s^{-\ep}}{\ep} \\
&\times
\int \limits_{0}^{1} \frac{\d z}{(1-z)^{1+2\ep} }
\bigg\langle
\tilde w_{4||1}^{14,15} \lp\frac{\rho_{15}}{4}\rp^{-\ep} 
\theta(z_5 - z) 
\times {\cal P}_{qq}(z)  \big[\I-S_5\big]\FLM(z\cdot 1,2,5)\bigg\rangle.
\end{split}
\ee

In parallel to the case of the double-collinear partitions, it is again convenient to always 
call the resolved gluon $g_4$. We  combine contributions of  sectors $(a)$ and $(c)$, 
renaming $g_5 \to g_4$ where appropriate,  and obtain 
\bes
&\left\langle\big[\I-\SS\big] \big[\I-S_5\big]\lp\theta^{(a)}C_{51}+\theta^{(c)}C_{41}\rp
w^{14,15} \dg4\dg5 \FLM(1,2,4,5)\right\rangle=\\
&=-\frac{[\as] s^{-\ep} }{\ep} 
\int \limits_{0}^{1} \frac{\d z}{(1-z)^{1+2\ep}} 
\Bigg\langle
\tilde w_{5||1}^{14,15} \lp\frac{\rho_{14}}{4}\rp^{-\ep} 
 \bigg\{
\left[\I- \theta(z_4 - z) S_4\right]
\\
&\times
 \PPOP\FLM(z\cdot 1,2,4)
+ \theta(z_4 - z) 2\Cf \big[\I-S_4\big] \FLM(1,2,4)\bigg\}\Bigg\rangle.
\end{split}
\label{eq4.57}
\ee

\be\widetilde{~~~~~}\nonumber\ee

We now turn to sectors $(b)$ and $(d)$. These sectors are different from the other triple-collinear sectors
and from the double-collinear partitions. Indeed, 
the single-collinear limits that we consider in sectors $(b)$ and $(d)$ 
correspond to gluons $g_4$ and $g_5$ becoming collinear to each other.  We consider 
\be
\left\langle\big[\I-\SS\big] \big[\I-S_5\big]\theta^{(b,d)}
C_{45} w^{14,15} \dg5 \FLM(1,2,4,5)\right\rangle, 
\ee
and start with the discussion of how the collinear projection operator $C_{45}$  acts on  $\FLM$. 
We find  
\bes
C_{45}&\FLM(1,2,4,5) = \frac{\gsb^2}{E_5^2\rho_{45}}
\frac{E_5}{E_4} P_{gg,\mu \nu}(z)\FLM^{\mu\nu}(1,2,4+5)\\
&=\frac{\gsb^2}{E_5^2\rho_{45}}\frac{z}{1-z}\bigg[
P_{gg}^{(0)}(z)\FLM(1,2,45) + 
P_{gg}^\perp(z)\kappa_{\perp,\mu}\kappa_{\perp,\nu} \FLM^{\mu\nu}(1,2,45)
\bigg],
\end{split}\
\label{eq:coll_45}
\ee
where $p_{4+5} = p_{45} = (E_4+E_5)/E_4 \cdot p_4$, i.e. the hard matrix element must be taken in the 
collinear limit. The splitting functions are 
\bes
&P_{gg}^{(0)}(z) = 2\Ca \lp \frac{z}{1-z}+\frac{1-z}{z}\rp,\;\;\;\;\;\;P^\perp_{gg}(z) = 4\Ca (1-\ep)z(1-z),
\end{split}
\ee
and  $z$ is the fraction of the total momentum $p_{45} = p_4 + p_5$ 
carried by gluon $g_5$,
\be
z = E_5/( E_4 + E_5).
\label{eq:defzc45}
\ee
The vector  $\kappa_\perp$ is a normalized transverse vector 
\be
\kappa = k_\perp/\sqrt{-k_\perp^2}, 
\ee
defined by the following decomposition 
\be
p_5  = \alpha p_4 + \beta {\bar p}_4 + \kt,
\ee
where ${\bar p}_4 = (p_4^{(0)}, - \vec p_4)$ and $\kt \cdot p_4 = \kt \cdot {\bar p}_4 = 0$.

We now construct these vectors explicitly.  For this, we need the parametrization of 
the angular phase space of the two gluons valid for sectors $(b)$ and $(d)$; it is given
in Appendix~\ref{sect:phsp}. Here we repeat the relevant formulas and discuss simplifications 
that occur in the limit where the momenta of $g_4$ and $g_5$ become collinear. 
We write the four-momenta of $g_4$ and $g_5$ as
\be
\begin{split} 
& p_4^\mu = E_4 \big( t^\mu + \cos \theta_{41} e_3^{\mu} + \sin \theta_{41} b^\mu\big),
\\  
& p_5 = E_5 \big( t^\mu + \cos \theta_{51} e_3^\mu + \sin \theta_{51} \left ( \cos \varphi_{45} b^\mu 
+ \sin \varphi_{45} a^{\mu} \right )\big), 
\end{split}
\label{eq166}
\ee
where  $t^\mu = (1,\vec 0)$,  $e_3^\mu = (0,0,0,1;0...)$, 
$b \cdot t = b \cdot e_3 = 0$ and  $a \cdot t = a \cdot e_3 = a \cdot b = 0$. 
Our goal is to parametrize the phase space in such a way that explicit averaging 
over directions of $k_\perp$ can be performed. 
The phase space parametrization for sectors $(b)$ and $(d)$ can be written as
\bes
\dg4 \dg5 = \lp E_4^{1-2\ep} \d E_4\rp
\lp   E_5^{1-2\ep} \d E_5 \rp
\theta(\Em-E_4)
\theta(E_4-E_5)
\d\Omega^{(b,d)}_{45},
\end{split}
\ee
with
\bes
C_{45}\left[\frac{\d\Omega_{45}^{(b,d)}}{\eta_{45}}\right] &= N_{\ep}^{(b,d)}
\frac{\d\Omega_{g_4}}{(2\pi)^{d-1}}
\pref
\left[\frac{\d\Omega_{d-3,a}}{\Omega_{d-3}}\right]
\frac{\d x_3}{x_3^{1+\ep}(1-x_3)^{-\ep}}
\frac{\d x_4}{x_4^{1+2\ep}}\d\Lambda,
\label{eq8.60}
\end{split}
\ee
where 
\be
\d\Lambda \equiv
\frac{\Gamma(1+\ep)\Gamma(1-\ep)}{\Gamma(1+2\ep)\Gamma(1-2\ep)}
\frac{\lambda^{-1/2+\ep}(1-\lambda)^{-1/2-\ep}}{\pi} \d\lambda,
\;\;\;
N_\ep^{(b,d)} = \left[
\frac{\Gamma(1-\ep)\Gamma(1+2\ep)}{\Gamma(1+\ep)}
\right].
\ee
Here, $x_4\to 0$ corresponds to the $4||5$
collinear limit,
 $\eta_{45} = (1-\cos \theta_{45} )/2$, $x_3 = \rho_{41}/2$ and
$\lambda$ is related to the azimuthal angle $\varphi_{45}$.
 Further details about the parametrization 
as well as expressions of scalar products in terms of $x_{3,4}$ and $\lambda$ can be found 
in Appendix~\ref{sect:phsp}.
In this parametrization, the vector $\kappa_\perp$ reads\footnote{
This expression is valid for sector $(b)$. For sector $(d)$, one should replace
$r\to-\tilde r$ with $\tilde r = \sin\theta_{51} \; e_3 - \cos\theta_{51}\;b$ 
in Eq.~\eqref{eq172}.}
\be
\kappa_\perp = a \sqrt{1-\lambda} + r \sqrt{\lambda},\;\;\;
r = \sin \theta_{41} \;  e_3 - \cos \theta_{41} \;  b.
\label{eq172}
\ee

Using this expression in Eq.~\eqref{eq:coll_45} together with 
momenta parametrization Eq.~\eqref{eq166} and 
the phase space limit Eq.~\eqref{eq8.60}, we observe that integrations  over 
$\lambda$ and the directions of the vector $a^\mu$ can be performed since neither $\lambda$ nor $a^\mu$
appear in the hard matrix element.   We define 
\be
\big\langle \nkt^\mu \nkt^\nu\big\rangle  \equiv
\int \d\Lambda  \frac{\d\Omega_{d-3,a}}{\Omega_{d-3}} 
\nkt^\mu\nkt^\nu.
\ee
Using 
\be
\int \frac{\d\Omega_{d-3,a}}{\Omega_{d-3}}\; a^\mu = 0,
\;\;\;\;\;
\int \frac{\d\Omega_{d-3,a}}{\Omega_{d-3}}\; a^\mu a^\nu = -\frac{g_{\perp,(d-3)_a}^{\mu \nu}}{d-3},
\ee
and
\be
\int \d\Lambda = 1,~~~ \int \lambda \d\Lambda = \frac{1+2\ep}{2}
,~~~ \int (1-\lambda) \d\Lambda = \frac{1-2\ep}{2},
\ee
we find 
\bes
\big\langle \nkt^\mu \nkt^\nu\big\rangle  &= 
-\frac{g^{\mu\nu}_{\perp,(d-3)_a}}{2} + 
\frac{1+2\ep}{2} r^\mu r^\nu\\
&=\frac{1}{2}\bigg[
-g^{\mu\nu}_{\perp,(d-3)_a}+r^\mu r^\nu\bigg] 
+\ep r^\mu r^\nu =-\frac{g_{\perp,(d-2)}}{2} + \ep r^\mu r^\nu.
\end{split}
\ee
It follows  that averaging over transverse directions introduces an $\ep$-dependent 
leftover, as a consequence of the chosen phase space parametrization~\cite{czakonsub}.

To write the result of the integration over unresolved phase space variables, 
it is convenient to define an additional splitting function 
\be
P_{gg}(z,\ep) = P_{gg}^{(0)}(z) + \frac{P_{gg}^\perp(z)}{2}=
2\Ca \lp \frac{1-z}{z} + \frac{z}{1-z} + z(1-z)(1-\ep)\rp.
\ee
Combining  the results discussed above, we  write an expression for the contribution of the 
$C_{45}$ collinear projection operator in sector $(b)$.  We obtain 
\bes
&\big\langle \big[\I-\SS\big] \big[\I-S_5\big]\theta^{(b)}
C_{45} w^{14,15} \dg5 \FLM(1,2,4,5)\big\rangle\bigg.\\
&=-\frac{[\as]}{2\ep} N_\ep^{(b)} 
\int_{E_4>E_5} \dg4
\tilde w_{4||5}^{14,15}
x_3^{-\ep}(1-x_3)^{\ep}
\frac{\d E_5}{E_{5}^{1+2\ep}}
\big[\I-\SS\big] \big[\I-S_5\big] {\cal P}_{45}(1,2,4,5),
\end{split}
\label{eq7.82}
\ee
where 
\bes
\mathcal P_{45}(1,2,4,5) = 
\frac{E_5}{E_4}
\Bigg[&P_{gg}\left[\frac{E_5}{E_4+E_5},\ep\right]
\FLM(1,2,4+5)  \\
&+
\ep P_{gg}^\perp\left[\frac{E_5}{E_4+E_5} \right] r_\mu r_\nu \FLM^{\mu\nu}(1,2,4+5)
\Bigg].
\end{split}
\ee 

It follows from Eq.~\eqref{eq7.82} that we need to know how an operator 
$(I - \SS)(I - S_5)$ acts on  $\mathcal P_{45}(1,2,4,5)$.  We recall that the action of 
$\SS$ on energy variables 
implies that $E_4,E_5 \to 0$ at fixed  $E_4/E_5$.  Computing 
the soft limits is simple and standard  except for the spin-correlated 
part that we address below. 
In principle, we need to know  three soft limits $\SS, S_5$ and $\SS S_5$. However, 
since  
\be
\lim_{E_5 \to 0}  P^\perp_{gg}(E_5/(E_4+E_5)) = 0,
\ee
we only need to consider 
$ \SS r_\mu r_\nu \FLM^{\mu\nu}(1,2,4+5)$.  We find it using the known soft limits
for amplitudes and the explicit form of the vector $r$ given in Eq.~\eqref{eq172}.  Indeed,  since 
$r \cdot p_4 = 0$ and $r^2 =-1$, $r^\mu$  
is a valid polarization vector of the gluon with momentum $4+5$, 
in the collinear $4||5$ approximation.
For this reason,  the soft limit of $r_\mu r_\nu \FLM^{\mu \nu}$ follows from the standard soft limit of the amplitude for 
$q \bar q \to V + g$, not averaged over gluon polarizations. 
 We obtain 
\bes
\SS r_\mu r_\nu &\FLM^{\mu\nu}(1,2,4) = 
\frac{\Cf}{E_4^2} \lp
\frac{n_2\cdot r}{\rho_{24}}-
\frac{n_1\cdot r}{\rho_{14}}\rp^2\FLM(1,2) \\
&=\frac{2\Cf}{E_4^2} \frac{2\sin^2\theta_{14}}{\rho_{14}^2 \rho_{24}^2} \FLM(1,2)= 
\frac{2\Cf}{E_4^2} \frac{\rho_{12}}{\rho_{14} \rho_{24}} \FLM(1,2) \\
& = S_4 \FLM(1,2,4).
\label{eq7.84}
\end{split}
\ee
Collecting all the soft limits, we find 
\bes
\big[\I - \SS \big] &\big[\I - S_5\big] \mathcal P_{45}(1,2,4,5) =\frac{z}{1-z} \bigg[
-g_{\mu \nu} P_{gg}(z,\ep) + \ep P^\perp_{gg}(z) r_\mu r_\nu
\bigg] 
\\
& \times
\big[\I-S_{45}\big]\FLM^{\mu\nu}(1,2,45) 
-2\Ca \big[\I-S_4\big] \FLM(1,2,4),
\label{eq:p45}
\end{split}
\ee
where $z$ is defined in Eq.~\eqref{eq:defzc45}. This implies
\be
E_5 = z E_{45},~~ E_4 = (1-z) E_{45},\;\;\;\;  E_{45} = E_4 + E_5.
\ee

We can now use Eq.~\eqref{eq:p45} in Eq.~\eqref{eq7.82} 
and  integrate over all variables 
that are not present in the hard matrix elements.  This requires 
different variable transformations in the first and the second terms 
in Eq.~\eqref{eq:p45}.  To integrate the  first term, 
we change the integration variables from $E_{4,5}$ to 
$E_{45}$ and $z$. 
\begin{figure}[t]
\centering
\includegraphics[width=0.45\textwidth]{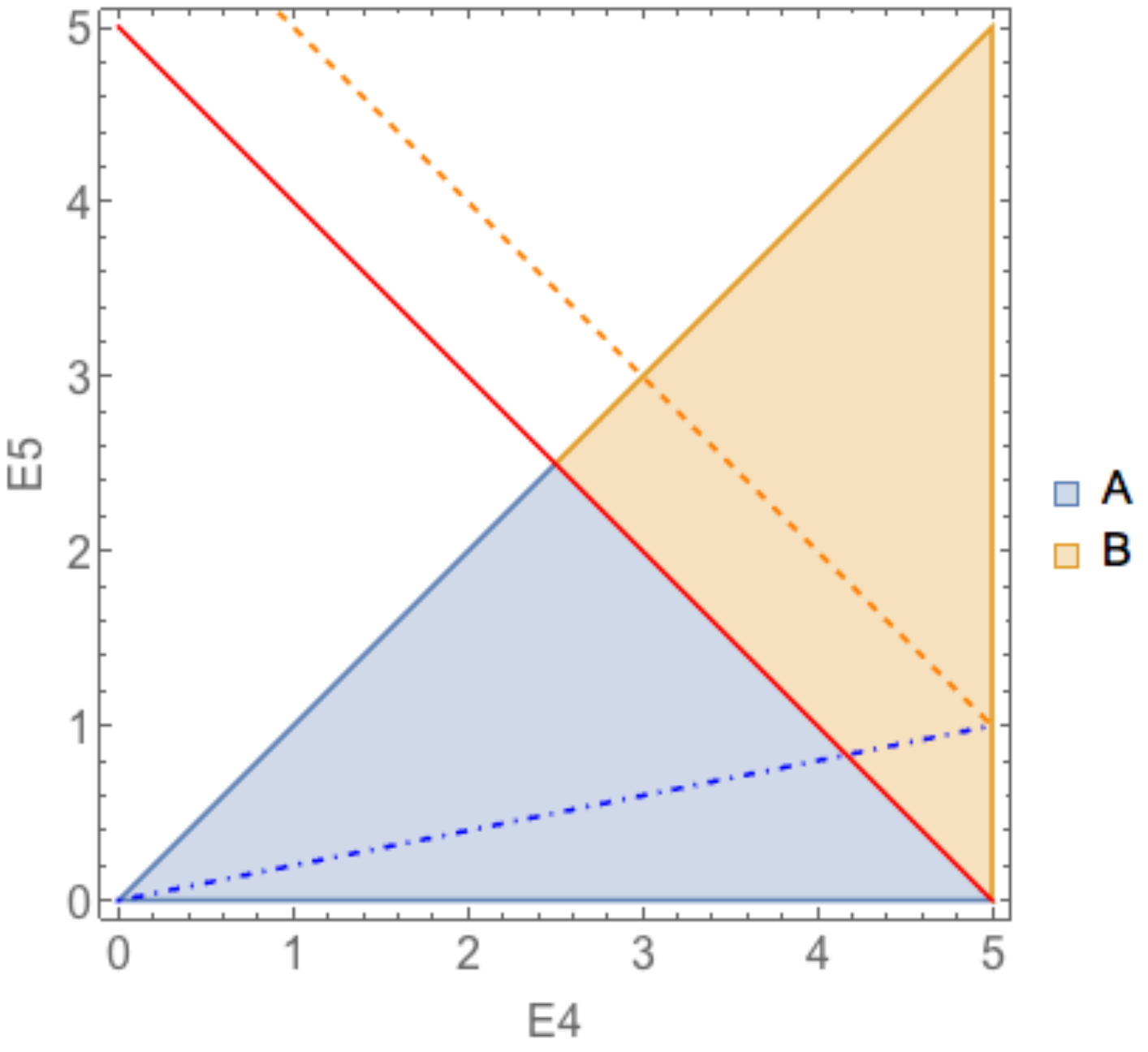}
\caption{Integration region for the $(E_4,E_5)\to (E_{45},z)$ change
of variables.
The colored triangle is the allowed $0<E_5<E_4<\Em$ region. The blue
region ``A'' is the ``physical'' one, i.e. the one which is not removed
by a phase space $\theta$-function inside $\FLM(1,2,45)$. The orange region
``B'' only contributes to the soft limit, since there no $\theta$-function from
$\FLM$ is preventing it. Lines of fixed $E_{45}$ are shown in solid red
(for $E_{45}=\Em$) and dashed orange (for $E_{45}>\Em$). In dot-dashed blue, lines of constant
$z$ are shown. In the ``physical'' region, only the $z<1/2$ condition is 
relevant. In the ``B'' region, we also have to impose $z>1-\Em/E_{45}$, to 
prevent the $E_{45}$ integration to go outside the triangle (see the
intersection of blue and orange lines).
}\label{fig:int_reg_b}
\end{figure}
We find that the integration region splits into two regions
(c.f. Fig.~\ref{fig:int_reg_b})
\be
\int \limits_{0}^{\Em} {\rm d}E_4 
\int \limits_{0}^{E_4} {\rm d} E_5  
= 
\int \limits_{0}^{\Em} E_{45}\;{\rm d} E_{45} 
  \int \limits_{0}^{1/2} {\rm d} z 
+ 
\int \limits_{\Em}^{2 \Em} E_{45}\; {\rm d} E_{45} 
  \int \limits_{1-\Em/E_{45}}^{1/2} {\rm d} z.
\ee
Following this separation, we split the integral into two parts
\bes
&\big\langle\big[\I-\SS\big] \big[\I-S_5\big]\theta^{(b)}
C_{45} w^{14,15} \dg5 \FLM(1,2,4,5)\big\rangle=I_{A}+I_{B},
\end{split}
\ee
where the integral $I_A$ corresponds to the region $E_{45}<\Em$ and 
the integral $I_B$ to the region 
$E_{45}>\Em$, see Fig.~\ref{fig:int_reg_b}.
We obtain an integral representation for $I_A$ 
starting from Eq.~\eqref{eq:p45}, changing  variables  
$(E_4,E_5)\to (E_{45},z)$ in the first term, and  $E_5=z E_4$
in the second term in Eq.~\eqref{eq:p45}  and, finally,  renaming $E_{45}\to E_4$. 
We obtain 
\begin{align}
&I_A=-\frac{[\as]}{2\ep}N_\ep^{(b)} \int\dg4 
\tilde w_{4||5}^{14,15} 
x_3^{-\ep}(1-x_3)^{\ep}
E_4^{-2\ep}
\int\limits_0^1\frac{\d z}{z^{1+2\ep}} \nonumber\\
&\times\bigg\{z(1-z)^{-2\ep}
\bigg[
-g_{\mu \nu} P_{gg}(z,\ep)  + \ep P^\perp_{gg}(z) r_\mu r_\nu
\bigg]
\big[\I-S_{4}\big]\FLM^{\mu\nu}(1,2,4)\theta(1/2 - z )
\label{eq188}
\\
&-2\Ca \big[\I-S_4\big]\FLM(1,2,4)\bigg\}.\nonumber
\end{align}
To compute  $I_B$, we notice that only the soft $S_{45}\FLM^{\mu\nu}(1,2,45)$ 
term of Eq.~\eqref{eq:p45} contributes.  Again renaming $E_{45} \to E_4$ and 
using Eq.~\eqref{eq7.84}, we obtain 
\bes
&I_B=\frac{[\as]^2}{2\ep}N_\ep^{(b)} 
\int\limits_{\Em}^{2\Em} \frac{\d E_4}{E_4^{1+4\ep}}
\int\limits_0^1 \frac{2 \d x_3}{\big[4x_3(1-x_3)\big]^\ep}
\tilde w_{4||5}^{14,15} 
x_3^{-\ep}(1-x_3)^{\ep}\\
&\times\int \limits_{z_4}^{1/2} \frac{\d z}{z^{1+2\ep}} 
z(1-z)^{-2\ep}
\bigg[
P_{gg}(z,\ep) + \ep P_{gg}(z,\kt)
\bigg]  \frac{2\Cf \rho_{12}}{\rho_{14}\rho_{24}}\FLM(1,2),
\end{split}
\label{eq189}
\ee
where $z_4 = 1-E_4/\Em$ and 
we expressed $\dg4$ as an integral over energy $E_4$ and 
the angular integration variable $x_3 = \rho_{14}/2$. We have 
also integrated over $d-3$ angular variables that do not appear in  
the hard matrix element and in the splitting function.

Finally, we consider sector ${(d)}$. We need to compute 
\bes
&\big\langle\big[\I-\SS\big] \big[\I-S_5\big]\theta^{(d)}
C_{45} w^{14,15} \dg4 \FLM(1,2,4,5)\big\rangle.
\end{split}
\ee
The calculation is similar to what we just described for sector $(b)$, 
apart from the following modifications of the integration boundaries 
\bes
& I_A:\;\;\;\;\theta(1/2-z) \to \theta(1-z) \;\theta(z-1/2),
\\
& I_B:\;\;\;\;\theta( z - z_4) \;\theta(1/2-z) \to \theta(z-1/2) \; \theta(1-z_4- z).
\end{split}
\ee
Incorporating these changes in Eqs.~(\ref{eq188},\ref{eq189})  provides us with the 
result for sector $(d)$.

Significant simplifications occur if the results for the two sectors are added; this happens 
because the $z$-integration boundaries in sectors $(b)$ and $(d)$  complement each other. 
Also, for both $I_A$ and $I_B$ the $z$-integration decouples from the rest 
and can be performed independently 
of the hard matrix element. In $I_A$, it yields
\bes
I_A^{(b)+(d)}
&=\frac{[\as]}{\ep}
\bigg\langle
\tilde w_{4||5}^{14,15} 
\lp\frac{\rho_{14}}{2}\rp^{-\ep}\lp 1-\frac{\rho_{14}}{2}\rp^{\ep}
E_4^{-2\ep} 
\times\\
&\times
\bigg[
\tilde\gamma_g(\ep) \big[\I-S_4\big]\FLM(1,2,4) + 
\ep \tilde\gamma_g(\ep,k_\perp) \big[\I-S_4\big]r_\mu r_\nu \FLM^{\mu\nu}(1,2,4)
\bigg]\bigg\rangle,
\end{split}
\ee
where we used $x_3 = \rho_{14}/2$ and the constants $\tilde\gamma_g(\ep)$, 
$\tilde\gamma_g(\ep,k_\perp)$ are
defined in Eq.~\eqref{eq:gamma}.
In the integral  $I_B$ the hard matrix element is that 
of the leading order process which implies that 
integration over all variables related to radiated gluons can be explicitly performed. 
We find 
\bes
I_B^{(b)+(d)}=\frac{[\as]^2}{\ep}\Em^{-4\ep}
\int\limits_0^1\frac{2\d x_3}{\big[4x_3(1-x_3)\big]^\ep}
\tilde w_{4||5}^{14,15} x_3^{-\ep}(1-x_3)^{\ep}
\delta_g(\ep)
\frac{ 2\Cf \rho_{12}}{\rho_{14}\rho_{24}}
\big\langle\FLM(1,2)\big\rangle,
\end{split}
\ee
with $\delta_g$ defined in Eq.~\eqref{eq:delta}.

\subsubsection{Summing double- and triple-collinear partitions}

Summing up the above results, we obtain an  intermediate representation of 
$\langle \FLM^{s_rc_s} \rangle $. We write it as a sum of four terms
\be
\begin{split}
& \langle \FLM^{s_rc_s} \rangle 
=
\Bigg\{
\sum_{(ij)\in dc}
\left\langle
\big[\I-\SS\big]\big[\I-S_5\big]
\bigg[ C_{4i} \dg4 + C_{5j}\dg5 \bigg] w^{i4,j5}\FLM(1,2,4,5)
\right\rangle\\
&\quad\quad
+\sum_{i\in tc} 
\bigg\langle
\big[\I-\SS\big]\big[\I-S_5\big]
\bigg[
\theta^{(a)} C_{5i} + \theta^{(b)} C_{45} + \theta^{(c)} C_{4i} + \theta^{(d)} C_{45}
\bigg]\\
&\quad\quad\quad\quad\quad
\times\dg4 \dg5 w^{i4,i5}\FLM(1,2,4,5)
\bigg\rangle\Bigg\}\\
& 
\quad\quad
=\big\langle \mathcal C_1(z\cdot 1,2,4)\big\rangle+
\big\langle \mathcal C_2(1,z\cdot 2,4)\big\rangle+
\big\langle \mathcal C_3(1,2,4)\big\rangle+
\big\langle 
\mathcal C_4(1,2,4)\big\rangle.
\end{split}
\ee
These terms are defined as 
\be
\begin{split}
& \big\langle \mathcal C_1(z\cdot 1,2,4)\big\rangle = 
-\frac{[\as] s^{-\ep}}{\ep}
\int \limits_{0}^{1}\frac{\d z}{(1-z)^{1+2\ep}}
\Bigg\langle
\lp \tilde w^{14,15}_{5||1}\lp \frac{\rho_{41}}{4}\rp^{-\ep} + \tilde w^{24,15}_{5||1}\rp\times
  \\
&
\times \bigg(
\left[\I- \theta( z_4 - z) S_4 \right] \PPOP\FLM(z\cdot 1,2,4) 
+2\Cf \theta( z_4 - z) \big[\I-  S_4\big] \FLM(1,2,4)\bigg)
\Bigg\rangle, 
\end{split}
\ee
\be
\begin{split}
& \bigg\langle \mathcal C_2(1,z \cdot 2,4)\bigg\rangle = 
-\frac{[\as] s^{-\ep}}{\ep}
\int \limits_{0}^{1} \frac{\d z}{(1-z)^{1+2\ep}}
\Bigg\langle
\lp \tilde w^{24,25}_{5||2}\lp \frac{\rho_{42}}{4}\rp^{-\ep} + \tilde w^{14,25}_{5||2}\rp\times
  \\
&
\times \bigg(
\left[\I- \theta( z_4 - z) S_4\right] \PPOP\FLM(1,z\cdot 2,4) 
+2\Cf \theta(z_4 - z) \big[\I-S_4\big]  \FLM(1,2,4)\bigg)
\Bigg\rangle, 
\end{split} 
\ee

\be
\begin{split} 
& \bigg\langle \mathcal C_3(1,  2,4)\bigg\rangle = 
\frac{[\as] }{\ep}\Bigg\langle
\left[\tilde w_{4||5}^{14,15} 
\lp\frac{\rho_{14}}{2}\rp^{-\ep}\lp 1-\frac{\rho_{14}}{2}\rp^{\ep}
+
\tilde w_{4||5}^{24,25} 
\lp\frac{\rho_{24}}{2}\rp^{-\ep}\lp 1-\frac{\rho_{24}}{2}\rp^{\ep}
\right] \\
&\quad\quad\quad\quad\quad
\times E_4^{-2\ep} \big[\I-S_4\big]
\bigg[
\tilde\gamma_g(\ep) \FLM(1,2,4) 
+\ep \tilde\gamma_g(\ep,\kt) r_\mu r_\nu \FLM^{\mu\nu}(1,2,4)
\bigg]\Bigg\rangle, 
\end{split}
\ee
\be
\begin{split} 
\bigg\langle 
\mathcal C_4(1,2,4)\bigg\rangle &=
\frac{[\as]^2 \Em^{-4\ep}  }{\ep}
\;\delta_g(\ep)
\int
\frac{\d\Omega_{(d-1),4}}{\Omega_{d-2}}
\bigg[\tilde w_{4||5}^{14,15} 
\lp\frac{\rho_{14}}{2}\rp^{-\ep}\lp 1-\frac{\rho_{14}}{2}\rp^{\ep}\\
&
+
\tilde w_{4||5}^{24,25} 
\lp\frac{\rho_{24}}{2}\rp^{-\ep}
\lp 1-\frac{\rho_{24}}{2}\rp^{\ep}\bigg]
 \left[2\Cf \frac{\rho_{12}}{\rho_{14}\rho_{24}}\right]
\big\langle\FLM(1,2)\big\rangle,
\end{split} 
\ee
with 
\be
\frac{\d\Omega_{d-1,4}}{\Omega_{d-2}}=\d\cos\theta \lp\sin^2\theta\rp^{-\ep}=
\frac{2\d x_3}{\big[4x_3(1-x_3)\big]^\ep}.
\ee

\subsection{The single-collinear term: extracting the last singularities }
\label{sec:coll_reg}

The four contributions to $\langle \FLM^{s_rc_s} \rangle $ described at the end 
of the previous section  require further manipulations because they cannot be expanded 
in series of $\ep$ as they are. Indeed, all of them 
exhibit  collinear singularities in the limits $4 || 1$ 
and $4 ||2$ that need to be extracted before expansion in $\ep$ becomes possible. 
To  deal with  this issue,  we again 
rewrite  the identity operator through collinear projections. For example, we write
\be
\big\langle \mathcal C_{1}(z \cdot 1, 2, 4) \big\rangle = 
\big\langle ( C_{41} + C_{42} ) \mathcal C_{1}(z \cdot 1, 2, 4) \big\rangle
+\big\langle ( I - C_{41} - C_{42} ) \mathcal C_{1}(z \cdot 1, 2, 4) \big\rangle.
\ee
The first two terms can be further simplified by considering respective collinear 
limits; the last term is regulated and can be Taylor-expanded in $\ep$. 
The single-collinear subtraction term can be analyzed in the same way as all 
the other collinear limits discussed previously. The only new element here is the action 
of the collinear projection operators on the spin-correlated part. Using 
the explicit expression for the vector $r$  in Eq.~\eqref{eq172} we find 
\be
E_4^2\rho_{41} C_{41}r_\mu r_\nu \FLM^{\mu\nu}(1,2,4) = \gsb^2 \Cf \frac{(1+z)^2}{2z}\FLM(z\cdot 1,2),
\ee
where $z$ is defined in the usual way $z = 1- E_4/E_1$. Taking this into account,
after tedious but straightforward calculations we arrive at
\begin{align}
& \langle \FLM^{s_rc_s} \rangle = 
\Bigg\{
\frac{[\as]}{\ep}\Bigg\langle
E_4^{-2\ep} \ONLO \Delta_{4||5}\bigg[
-g_{\mu\nu} \tilde\gamma_g   + \ep \tilde \gamma_g^\perp
r_\mu r_\nu\bigg] \FLM^{\mu\nu}(1,2,4)\Bigg\rangle
\nonumber
\\
&~~~
-\frac{[\as]}{\ep}\Bigg\langle
s^{-\ep}\left[\frac{(E_4/E_1)^{-2\ep}-1}{2\ep}\right]2\Cf
\ONLO\big[\Delta_{5||1}+\Delta_{5||2}\big] \FLM(1,2,4)\Bigg\rangle
\nonumber
\\
&~~~
-\frac{[\as]s^{-\ep}}{\ep}\int\limits_0^1 \d z\; \mathcal P_{qq,RR_2}(z)\Bigg\langle 
\ONLO
\frac{\Delta_{5||1} \FLM(z\cdot 1,2,4)  
+ \Delta_{5||2}\FLM(1,z\cdot 2,4)
}
{z}
\Bigg\rangle
\nonumber
\\
&~~~
-\frac{[\as]^2}{\ep}2\Cf  \left[\frac{s}{2}\right]^{-2\ep}
\int\limits_0^1 \d z\; \mathcal P_{qq,RR_3}(z)
\int[d\rho_{14}] \hat {\cal O}_C S_{12,\rho}^{(4)}
\Delta_{5||1}
\left\langle\frac{\FLM(z\cdot 1,2)}{z}\right\rangle
\nonumber
\\
&~~~
-\frac{[\as]^2}{\ep}2\Cf  \left[\frac{s}{2}\right]^{-2\ep}
\int\limits_0^1 \d z\; \mathcal P_{qq,RR_3}(z)
\int[d\rho_{24}]
\hat {\cal O}_C S_{12,\rho}^{(4)}\Delta_{5||2}
\left\langle\frac{\FLM(1,z\cdot 2)}{z} \right\rangle
\Bigg\}+
\nonumber
\\
&
+\Bigg\{\frac{[\as]^2 s^{-2\ep}}{\ep^2}\int\limits_0^1 \d z\Bigg(
\frac{1}{2^{1-\ep}} \frac{\Gamma(1-2\ep)\Gamma(1-\ep)}{\Gamma(1-3\ep)}
\big[\mathcal P_{qq}\otimes \mathcal P_{qq}\big]_{RR}(z)
\nonumber
\\
&~~~~~~~
+\frac{\Gamma^2(1-\ep)}{\Gamma(1-2\ep)}\Cf
\left[\frac{(\Em/E_1)^{-2\ep}-1}{\ep}\right]
\mathcal P_{qq,RR_2}(z)
\nonumber
\\
&~~~~~~~
+\left[
2^{\ep} \frac{\Gamma(1-2\ep)\Gamma(1-\ep)}{\Gamma(1-3\ep)}
+2\frac{\Gamma^2(1-\ep)}{\Gamma(1-2\ep)}\right]\Cf
 \mathcal P_{qq,RR_{3+4}}(z) 
\label{eq:final_third}
\\ 
&~~~~~~~
-\frac{1}{2^{1-2\ep}}\frac{\Gamma(1-2\ep)\Gamma(1-\ep)}{\Gamma(1-3\ep)}
\bigg[\tilde\gamma_g \mathcal P_{qq,RR_1}(z) +
\ep \tilde\gamma_g^\perp \mathcal P_{qq,RR_5}(z)\bigg]\Bigg)
\bigg\langle \FLM^z(1,2) \bigg\rangle
\nonumber
\\
&~~~
+2 \frac{[\as]^2 s^{-2\ep} }{\ep^2}\frac{\Gamma^2(1-\ep)}{\Gamma(1-2\ep)}
\int\limits_0^1 \d z~ \d\bar z ~\mathcal P_{qq,RR_2}(z) \mathcal P_{qq,RR_2}(\bar z)
\left\langle\frac{\FLM(z\cdot 1,\bar z\cdot 2)}{z\bar z}\right\rangle 
\nonumber
\\
&~~~
+[\as]^2\Cf\delta_g(\ep) \bigg[
\frac{1}{\ep}
\int [d\rho_{14}] \hat {\cal O}_C S_{12,\rho}^{(4)}\Delta_{4||5}
+ 
\frac{1}{\ep}
\int [d\rho_{24}] \hat {\cal O}_C S_{12,\rho}^{(4)}\Delta_{4||5}
\nonumber
\\
&~~~~~~~
-\frac{2^{1-2\ep}}{\ep^2}\frac{\Gamma(1-2\ep)\Gamma(1-\ep)}
{\Gamma(1-3\ep)}\bigg]
\big\langle
\Em^{-4\ep}\FLM(1,2)\big\rangle
\Bigg\}.
\nonumber
\end{align}
In the above equation, we used  the following notation 
\be
\begin{split}
& [d\rho_{ij}] = \d\cos\theta_{ij}\; (\sin^2\theta_{ij})^{-\ep},~~~~~\cos\theta_{ij} = 1 - \rho_{ij}.
\\
& \Delta_{5||1} = 
\lp \tilde w^{14,15}_{5||1}\lp \frac{\rho_{41}}{4}\rp^{-\ep} + \tilde w^{24,15}_{5||1}\rp
= 1 + \mathcal O(\ep),
\\
& \Delta_{5||2} = 
\lp \tilde w^{24,25}_{5||2}\lp \frac{\rho_{42}}{4}\rp^{-\ep} + \tilde w^{14,25}_{5||2}\rp
= 1 + \mathcal O(\ep),
\\
& \Delta_{4||5} = 
\left[\tilde w_{4||5}^{14,15} 
\lp\frac{\rho_{14}}{2}\rp^{-\ep}\lp 1-\frac{\rho_{14}}{2}\rp^{\ep}
+
\tilde w_{4||5}^{24,25} 
\lp\frac{\rho_{24}}{2}\rp^{-\ep}\lp 1-\frac{\rho_{24}}{2}\rp^{\ep}
\right]
= 1 + \mathcal O(\ep).
\end{split}
\ee
The relevant  splitting functions are defined in Appendix~\ref{sect:defs}.
Also, 
\bes
&~~~~~~~~~~~ \FLM^z(1,2) \equiv \frac{ \FLM(z\cdot 1,2)+\FLM(1,z\cdot 2)}{z},
\\
& \hat {\cal O}_C \equiv  I - C_{41} - C_{42},
\quad\quad\quad\quad
\ONLO \equiv \big[\I-S_4\big]\big[\I-C_{41}-C_{42}\big], 
\end{split}
\ee
and
\be
S_{12,\rho}^{(4)}\equiv\frac{\rho_{12}}{\rho_{14}\rho_{24}}.
\ee

We note that in  Eq.~\eqref{eq:final_third}  the first curly bracket is fully regulated,
while the second contains subtraction terms. Note also that since 
$\Delta_{i||j} = 1+\mathcal O(\ep)$, if we are only interested in the $1/\ep$ poles,  we can
substitute $\Delta_{i||j}\to 1$ in Eq.~\eqref{eq:final_third}. 
We also note that, for the process of interest,  
terms that  contain ${\cal O}_C S_{12,\rho}^{(4)}$ can be easily integrated 
over the relative angles of the gluon $g_4$ with respect to the collision axis. 
We find 
\be
\begin{split}
& \int [\d\rho_{41}] {\cal O}_C S_{12,\rho}^{(4)}\Delta_{5||1}=
\int [\d\rho_{42}] {\cal O}_C S_{12,\rho}^{(4)}\Delta_{5||2}=
(1+\ln 2)\ep + \mathcal O(\ep^2),
\\
& \int [\d\rho_{41}] {\cal O}_C S_{12,\rho}^{(4)}\Delta_{4||5}=
\int [\d\rho_{41}] {\cal O}_C S_{12,\rho}^{(4)}\Delta_{4||5}=
\lp 2-\frac{\pi^2}{3}\rp \ep + \mathcal O(\ep^2).
\end{split}
\ee
We will use these results when presenting the final formula for the double-real contribution.

\subsection{The double-unresolved collinear limit: double collinear}
\label{sec:double_unr_dc}

We now turn to the term $\langle \FLM^{s_rc_t} \rangle $ and begin 
by considering the contribution 
of the double-collinear partitions. It reads 
\bes
-\sum_{(ij)\in dc}\bigg\langle
\big[\I-\SS\big]\big[\I-S_5\big]
C_{4i}C_{5j}\dg4\dg5 w^{i4,j5}\FLM(1,2,4,5)
\bigg\rangle=\\
=-\bigg\langle
\big[\I-\SS\big]\big[\I-S_5\big]
\big[C_{41}C_{52}+C_{51}C_{42}\big]\dg4\dg5\FLM(1,2,4,5)
\bigg\rangle.
\label{eq7.132}
\end{split}
\ee
Note that,  following our notational convention, the collinear projection 
operators  act on the phase space elements $\dg4$ and $\dg5$. 

We begin with  the $C_{41}C_{52}$ term. Introducing
\be
E_4 = (1-z)E_1,~~~ E_5=(1-\zb)E_2,
\ee
and calculating collinear limits  we obtain 
\bes
E_4^2 E_5^2 \rho_{14} \rho_{25} C_{41}C_{52}\FLM(1,2,4,5) = \gsb^4
\PP(z)\PP(\zb) \FLM(z\cdot 1,\zb \cdot 2).
\end{split}
\ee
Since the momenta of gluons $g_{4}$ and $g_5$  decouple from each other, we find 
\bes
\SS C_{41}C_{52} \FLM(1,2,4,5) = 
\SS S_5  C_{41}C_{52} \FLM(1,2,4,5) = 
\frac{4\gsb^4\Cf^2}{E_4^2 E_5^2 \rho_{14} 
\rho_{15} } \FLM(1,2).
\end{split}
\ee
As the result, the original expression simplifies 
\be
\big[\I-\SS\big]\big[\I-S_5\big]
C_{41}C_{52}\FLM(1,2,4,5)=\big[\I-S_5\big]C_{41}C_{52}\FLM(1,2,4,5).
\ee
Performing the angular integrations and accounting for the 
hierarchy of energies $E_4 > E_5$, we obtain 
\bes
& -\big\langle
\big[\I-\SS\big]\big[\I-S_5\big]
C_{41}C_{52}\dg4\dg5\FLM(1,2,4,5)
\big\rangle= 
\\
& -\frac{[\as]^2 s^{-2\ep}}{\ep^2}
\int \limits_{0}^{1} \frac{\d z}{(1-z)^{1+2\ep}} \int \limits_{z}^{1} 
\frac{\d\zb}{(1-\zb)^{1+2\ep}}
 \\
& \times \big\langle
 \PP(z)\PP(\zb) \FLM(z\cdot 1,\zb \cdot 2)
-2\Cf \PP(z)\FLM(z\cdot 1,2)
\big\rangle, 
\end{split}
\ee
where, as usual,  the $z$ integrals do not need a lower cut-off whenever $z$ is
present in  $\FLM$.

The term with collinear operators $C_{51}C_{42}$ 
in Eq.~\eqref{eq7.132}  can be simplified  in a similar way. 
Combining the two contributions, we obtain 
\bes
-\big\langle
\big[\I-\SS\big]\big[\I-S_5\big]
\big[C_{41}C_{52}+C_{51}C_{42}\big]\dg4\dg5\FLM(1,2,4,5)
\big\rangle = \\
-\frac{[\as]^2 s^{-2\ep} }{\ep^2}
\Bigg\{\int \limits_{0}^{1} \frac{\d z}{(1-z)^{1+2\ep}} \frac{\ d\zb}{(1-\zb)^{1+2\ep}}
\PP(z)\PP(\zb) \big\langle\FLM(z\cdot 1,\zb \cdot 2)\big\rangle\\
-2\Cf  \int \limits_{0}^{1} \frac{\d z}{(1-z)^{1+2\ep}}\PP(z)\bigl<\FLM(z\cdot 1,2)+
\FLM(1,z\cdot 2)\bigr>
\int \limits_{z}^{1}\frac{\d\zb}{(1-\zb)^{1+2\ep}}\Bigg\}.
\end{split}
\label{eq:dc}
\ee
We can re-write Eq.~\eqref{eq:dc} to ensure that all singularities that 
appear in  $z$ and $\bar z$ integrals  are regulated with the 
plus-prescription.  This gives the final result for the double-collinear 
contribution 
\be
\begin{split}
-\big\langle
\big[\I-\SS\big]\big[\I-S_5\big]
\big[C_{41}C_{52}+C_{51}C_{42}\big]\dg4\dg5\FLM(1,2,4,5)
\big\rangle = \\
-\frac{[\as]^2 s^{-2\ep} }{\ep^2}\Bigg\{
\int \limits_0^1 \d z\; \d\bar z ~
\mathcal P_{qq,RR_2}(z)
\mathcal P_{qq,RR_2}(\bar z)~
\left\langle\frac{\FLM(z\cdot 1,\zb \cdot 2)}{z\bar z} \right\rangle
\\
-\int \limits_0^1 \d z\; \mathcal P_{qq,RR_6}(z)
\left\langle \frac{\FLM(z\cdot 1,2) + \FLM(1,\zb\cdot 2) }{z} \right\rangle\Bigg\}.
\end{split}
\label{eq:final_fourth}
\ee
The relevant splitting functions are given in Appendix~\ref{sect:defs}.

\subsection{The double-unresolved collinear limit: triple collinear}
\label{sect:tripcol}

In this section, we consider the 
contribution of the triple-collinear partitions to $\langle \FLM^{s_r c_t} \rangle $. 
It reads 
\bes
\sum_{i\in tc} 
\bigg\langle
\big[\I-\SS\big]\big[\I-S_5\big]
\bigg[
\theta^{(a)} \CC_i\big[\I-C_{5i}\big] + \theta^{(b)} \CC_i\big[\I-C_{45}\big] + \\
 + \theta^{(c)} \CC_i\big[\I-C_{4i}\big]+ \theta^{(d)} \CC_i\big[\I-C_{45}\big]
\bigg]\dg4 \dg5 w^{i4,i5}\FLM(1,2,4,5)
\bigg\rangle.
\label{eq4.148}
\end{split}
\ee
This contribution always contains the triple-collinear projection operator 
$\CC_i$ that acts on the 
hard matrix elements. For $i =1$, this gives, schematically, 
\be
\CC_1 \FLM(1,2,4,5) = \lp\frac{2}{s_{145}}\rp^2 P_{ggq}(1,4,5) \FLM(1-4-5,2),
\label{eq8.104}
\ee
where $s_{145} = (p_1-p_4-p_5)^2$ and $P_{ggq}(1,4,5) $  is the known  triple-collinear splitting 
function \cite{Campbell:1997hg,Catani:1998nv,Catani:1999ss}. The reduced matrix element 
in Eq.~\eqref{eq8.104}
has to be evaluated in the exact collinear limit, 
i.e. $p_{1-4-5}\equiv (E_1-E_4-E_5)/E_1\cdot p_1$. 
 Other projection 
operators  that appear in Eq.~\eqref{eq4.148} provide subtractions that are needed 
to make the triple-collinear splitting function integrable over the unresolved 
parts of the $(g_4,g_5)$ phase space.
For definiteness,  we focus here on the triple-collinear partition where 
gluons are emitted along the direction of the incoming quark with momentum 
$p_1$; this corresponds to taking  $i=1$ in Eq.~\eqref{eq4.148}.

To proceed further, we note that  the damping factors in Eq.~\eqref{eq4.148} 
can be removed  since the collinear projection 
operator $\CC_1$ acting on them yields 1. Next, we need to study the triple-collinear 
limit of the angular phase space. The generic phase space 
parametrization is described in Appendix~\ref{sect:phsp}
and we use it to compute the triple-collinear 
limits. We stress that since the phase space parametrization changes 
from sector to sector, we need 
to consider all the four sectors separately. 

Without going into further detail of the angular integration, it is clear 
that once this integration is performed,
each sector  in  Eq.~\eqref{eq4.148} provides the following 
contribution to the final integral over energies
\be
\int \theta^{(k)} \CC_1 \big[\I-C_{ij}\big] \d\Omega_{45}^{(k)} \FLM(1,2,4,5)
\equiv [\alpha_s]^2 \TC^{(k)}(E_1,E_4,E_5) \FLM(1-4-5,2),
\label{eq4.156}
\ee
where the auxiliary function $\TC^{(k)}$ in  Eq.~\eqref{eq4.156} is defined as 
\be
[\alpha_s]^2 \TC^{(k)}(E_1,E_4,E_5) = 
\int \theta^{(k)} \CC_1 \big[\I-C_{ij}\big]\; \d\Omega_{45}^{(k)} \;
\frac{4 P_{ggq}(z_4,z_5,z_1,s_{45},s_{41},s_{51}) }{s^2_{145}}.
\label{eq:tcdef}
\ee
We note that the reason that $\CC_1$ is  present 
in Eq.~\eqref{eq:tcdef} is that it still has to act on the phase space; its action on the 
matrix element has already been accounted for and resulted in the  factorized form 
of Eq.~\eqref{eq4.156} and the appearance of the triple-collinear splitting function 
in Eq.~\eqref{eq:tcdef}.  We use Eq.~\eqref{eq4.156} to write 
\bes
\!\!
& \big\langle
\big[\I-\SS\big]\big[\I-S_5\big]
\theta^{(k)} \CC_1\big[\I-C_{ij}\big] \dg4 \dg5 w^{i4,i5}\FLM(1,2,4,5)\big\rangle\bigg.
 \equiv [\alpha_s]^2\times\\
&\times
\int \limits_{0}^{\Em} \d E_4\; E_4^{1-2\ep}
\int \limits_{0}^{E_4} 
\d E_5\; E_5^{1-2\ep} 
\big[\I-\SS\big]
\big[\I-S_5\big] \TC^{(k)}(E_1,E_4,E_5) \big\langle  \FLM(145,2)\big\rangle,
\end{split}\label{eq:itc}
\ee
where $145 \equiv  1 - 4 -5$ in the collinear approximation.

In what follows, we discuss the integration over energies in  Eq.~\eqref{eq:itc}.
Our goal is to change variables in such a way that the argument of the 
hard matrix element becomes  $z \cdot 1$; once this happens, 
Eq.~\eqref{eq:itc} becomes a convolution of a hard matrix element 
with a splitting function.  Although, in principle, 
changing variables in an integral is straightforward, it turns 
out that it is beneficial to do so in 
different ways in the four terms that appear in Eq.~\eqref{eq:itc};  for this 
reason, we consider  them separately. We emphasize that since the individual 
contributions to Eq.~\eqref{eq:itc} 
 diverge, it is important to keep dimensional regularization in place until
the end of the computation.

We begin with the term that contains the  
identity operator $\I$ and    change the variables as follows 
\be
E_4 = E_1(1-z)\lp 1-\frac{r}{2}\rp,
~~~~~
E_5 = E_1(1-z)\frac{r}{2},
\label{eq4.159}
\ee
with $r\in(0,1)$ and $z\in (0,1)$. We note that the lower integration 
boundary for $z$ can be taken to be $z = 0$ because $\FLM(z\cdot1,2)$ 
always appears in this contribution. As we already discussed several times, 
this automatically cuts off the integral over $z$ at a proper minimal value. 
With this in mind, we write\footnote{For ease of notation, we will drop the 
sector index in the intermediate calculations and restore it at the end.}
\begin{align}
& \int \limits_{0}^{\Em} 
\d E_4\; E_4^{1-2\ep} 
\int \limits_{0}^{E_4} 
\d E_5\; E_5^{1-2\ep}
\TC(E_1,E_4,E_5)\FLM(145,2) =
\nonumber\\
& =\frac{E_1^{4-4\ep}}{2^{-2\ep}}
\int\limits_0^1\frac{\d z}{(1-z)^{1+4\ep}}
\frac{\d r}{r^{1+2\ep}}
\frac{1}{\lp1-\frac{r}{2}\rp^{1+2\ep}}
\bigg[
(1-z)^4 \lp\frac{r}{2}\rp^2\lp 1-\frac{r}{2}\rp^2\times\\
& \times
\TC\lp E_1,E_1(1-z)\lp 1-\frac{r}{2}\rp,
E_1(1-z)\frac{r}{2}\rp
\bigg]
\FLM(z\cdot1,2).\nonumber
\end{align}

Next, we consider the  $S_5$ operator. It describes the limit 
where $E_5 \to 0$ at fixed $E_4$. Calculating this limit with the parametrization 
in Eq.~\eqref{eq4.159} mixes $z$ and $r$ and, therefore, is inconvenient. 
A better way is to change the parametrization. We choose 
\be
E_4 = E_1(1-z),
~~~~
E_5 = E_1(1-z)r, 
\label{eq4.161}
\ee
where $r\in(0,1)$. In principle, we should use  $z>z_{\rm min}$ but,  since  $z$ enters the 
hard matrix element,  we can extend all the integrals to $z \in(0,1)$. We find 
\bes
& \int \limits_{0}^{\Em} 
\d E_4\; E_4^{1-2\ep}
\int \limits_{0}^{E_4} 
 \d E_5\; E_5^{1-2\ep}
S_5\TC(E_1,E_4,E_5)\FLM(1-4-5,2)=
E_1^{4-4\ep}\times\\
&
\int\limits_0^1\frac{\d z}{(1-z)^{1+4\ep}}
\frac{\d r}{r^{1+2\ep}}\; \FLM(z\cdot1,2) 
\bigg[(1-z)^4 r^2
\TC(E_1, E_1(1-z),E_1(1-z)r)\bigg]_{r\to 0}.
\end{split}
\ee
Note that the $r^2$ prefactor ensures that the $r\to 0$ limit of the square bracket
exists.

We can also use the change of variables in Eq.~\eqref{eq4.161} 
for terms with operators $\SS$ and $\SS S_5$.  The only difference is that since 
in those terms  $z$ does not appear in  $\FLM$,  we  have to keep
the lower integration boundary at   $z = z_{\rm min}$. We write  
\begin{align}
& \int \limits_{0}^{\Em} 
\d E_4\; E_4^{1-2\ep}
\int \limits_{0}^{E_4} 
 \d E_5\; E_5^{1-2\ep}
\SS\TC(E_1,E_4,E_5)\FLM(1-4-5,2)=
E_1^{4-4\ep}
\times
\nonumber
\\
&
\times\int\limits_{z_{min}}^1\frac{\d z}{(1-z)^{1+4\ep}}
\frac{\d r}{r^{1+2\ep}}\;  \FLM(1,2) 
\bigg[(1-z)^4 r^2
\TC(E_1,E_1(1-z),E_1(1-z)r)\bigg]_{z\to 1}.
\label{eq:tcss}
\end{align}
Also in this case, the $(1-z)^4$ prefactor ensures the existence of the $z\to 1$
limit of the term in the square bracket. 
The term  with an 
operator $\SS S_5$ in Eq.~\eqref{eq:itc} is obtained from Eq.~\eqref{eq:tcss} by taking 
the $r \to 0$ limit in the expression in square brackets.

To proceed further, it is convenient to define
two  $z$-dependent functions and a constant 
\begin{align}
& A_1(z) \equiv \frac{z}{2^{-2\ep}}
\int\limits_0^1 \frac{\d r}{r^{1+2\ep}}
\frac{1}{\lp 1-\frac{r}{2}\rp^{1+2\ep}} 
\nonumber
\\
&
~~~~~~
\times
\Bigg\{
\bigg[
(1-z)^4 \lp\frac{r}{2}\rp^2\lp 1-\frac{r}{2}\rp^2
\TC\lp E_1,E_1(1-z)\lp 1-\frac{r}{2}\rp,
E_1(1-z)\frac{r}{2}\rp
\bigg]
\nonumber
\\
&
~~~~~~
-\bigg[
(1-z)^4 \lp\frac{r}{2}\rp^2\lp 1-\frac{r}{2}\rp^2
\TC\lp E_1,E_1(1-z)\lp 1-\frac{r}{2}\rp,
E_1(1-z)\frac{r}{2}\rp
\bigg]_{r\to0}\Bigg\},
\nonumber
\\
& A_2(z) \equiv \frac{z}{2^{-2\ep}}
\int\limits_0^1 \frac{\d r}{r^{1+2\ep}}
\frac{1}{\lp 1-\frac{r}{2}\rp^{1+2\ep}} 
\nonumber
\\
&
~~~~~~
\times\bigg[
(1-z)^4 \lp\frac{r}{2}\rp^2\lp 1-\frac{r}{2}\rp^2
\TC\lp E_1,E_1(1-z)\lp 1-\frac{r}{2}\rp,
E_1(1-z)\frac{r}{2}\rp
\bigg]_{r\to0} 
\label{eq:adef}
\\
&
~~~~~~
-z\int\limits_0^1 \frac{\d r}{r^{1+2\ep}}
\bigg[(1-z)^4 r^2
\TC(E_1,E_1(1-z),E_1(1-z)r)\bigg]_{r\to 0},
\nonumber
\\
& A_3 \equiv \int\limits_0^1 \frac{\d r}{r^{1+2\ep}}
\Bigg\{
\bigg[(1-z)^4 r^2
\TC(E_1,E_1(1-z),E_1(1-z)r)\bigg]+
\nonumber
\\
&
~~~~~~
-\bigg[(1-z)^4 r^2
\TC(E_1,E_1(1-z),E_1(1-z)r)\bigg]_{r\to 0}
\Bigg\}_{z\to 1}.
\nonumber
\end{align}
We can further simplify the function 
$A_2$ if we realize that  the $S_5$ limit of the triple-collinear splitting function 
is homogeneous in 
$E_5$. This implies that the $r \to 0$ limit of the two $T_C$ functions 
in the formula for $A_2$ in Eq.~\eqref{eq:adef} are related and can be combined.
Changing variables $r \to r/2$ in the first term  
on the r.h.s. of the integral  for $A_2$, we obtain 
\bes
A_2(z) = z \bigg[
\int\limits_0^{1/2} \frac{\d r}{r^{1+2\ep}}\frac{1}{(1-r)^{1+2\ep}}
-\int\limits_0^1 \frac{\d r}{r^{1+2\ep}}\bigg]\\
\times
\bigg[(1-z)^4 r^2
\TC(E_1,E_1(1-z),E_1(1-z)r)\bigg]_{r\to0}.
\end{split}
\ee
Since the term in the square bracket no longer depends on $r$ after the $r\to 0$ limit
is taken, we can perform the $r$ integrations in the first line to get
\be
A_2(z) = \frac{z}{2\ep}
\left[1-\frac{\Gamma^2(1-2\ep)}{\Gamma(1-4\ep)}\right]
\bigg[(1-z)^4 r^2
\TC(E_1,E_1(1-z),E_1(1-z)r)\bigg]_{r\to0}.
\label{eq:a2def}
\ee

We can now write the result for the integral that we are interested in using 
$A_{1,2}(z)$ and $A_3$. We find\footnote{At this point, we restore the 
sector label.}
\bes
& \big\langle
\big[\I-\SS\big]\big[\I-S_5\big]
\theta^{(k)} \CC_1\big[\I-C_{ij}\big] \dg4 \dg5 w^{i4,i5}\FLM(1,2,4,5)\big\rangle
=   [\alpha_s]^2 E_1^{4-4\ep} \times \bigg.\\
&\times \Bigg\{ \int\limits_0^1\d z\frac{ A_1^{(k)}(z)+A_2^{(k)}(z)}{(1-z)^{1+4\ep}}
\left\langle \frac{\FLM(z\cdot1,2)}{z}\right\rangle
-   
\int \limits_{z_{\rm min}}^1 \frac{A_3^{(k)} \d z} {(1-z)^{1+4\ep}}
\big\langle\FLM(1,2)\big\rangle\Bigg\}
.
\end{split}
\ee
This integral can be re-written in such a way that all the $z \to 1$ singularities 
are regulated by plus-prescriptions. We have already discussed how this can be done 
 several times; for this reason, 
we do  not repeat this discussion  again and only present the result.  It reads 
\bes
& \big\langle
\big[\I-\SS\big]\big[\I-S_5\big]
\theta^{(k)} \CC_1\big[\I-C_{ij}\big] \dg4 \dg5 w^{i4,i5}\FLM(1,2,4,5) 
\big \rangle \bigg.
\\
=  & [\alpha_s]^2 E_1^{4-4\ep} \int \limits_{0}^{1} \d z 
\left [ R^{(k)}(z) + \frac{R^{(k)}_+}{[(1-z)^{1+4\ep}]_+} 
+ R^{(k)}_\delta \delta(1-z) 
\right ] \left\langle\frac{\FLM(z \cdot 1, 2)}{z}\right\rangle,
\end{split} 
\ee
where 
\bes
R^{(k)}(z) &= \frac{A_1^{(k)}(z)+A_2^{(k)}(z)-A_1^{(k)}(1)-A_2^{(k)}(1)}{(1-z)^{1+4\ep}},
\;\;\;\;
R^{(k)}_+ = A_1^{(k)}(1)+A_2^{(k)}(1),\\
R^{(k)}_\delta & = \frac{(\Em/E_1)^{-4\ep}-1}{4\ep} A_3^{(k)} 
-\int \limits_0^1 \frac{\d r}{r^{1+2\ep}}\frac{(1+r)^{4\ep}-1}{4\ep}\; F^{(k)}(r).
\end{split}
\ee
The functions  $A^{(k)}_{1,2}(z)$ and the constant 
$A^{(k)}_3$ are given in Eqs.~(\ref{eq:adef}, \ref{eq:a2def}). 
We have also used the  following notation 
\be
\begin{split}
& \frac{1}{[(1-z)^{1+4\ep}]_+} 
= \sum \limits_{n=0}^{\infty} \frac{(-4 \ep )^n}{n!} {\cal D}_n(z),
\;\;\;\;\;\\
& F^{(k)}(r) =  \bigg[(1-z)^4 r^2
\TC^{(k)}(E_1,E_1(1-z),E_1(1-z)r)\bigg]_{z\to1}. 
\end{split}
\ee
We are now in position to write the contribution 
of this term in the final form 
\be
\begin{split}
&\sum\limits_{k} \big\langle
\big[\I-\SS\big]\big[\I-S_5\big]
\theta^{(k)}\CC_1\big[\I-C_{ij}\big]
\dg4 \dg5 w^{14,15}\FLM(1,2,4,5)
\big\rangle = \frac{[\alpha_s]^2}{\ep} \times\\
&\times
\left[\frac{s}{4}\right]^{-2\ep}
\sum \limits_{k}^{}
\int \limits_0^1 \d z \bigg[
R^{(k)}(z) + \frac{R_+^{(k)}}{\big[(1-z)^{1+4\ep}\big]_+}
+R_\delta^{(k)}
\delta(1-z)\bigg]
\left\langle\frac{\FLM(z\cdot1,2)}{z}\right\rangle.
\label{eq7.158}
\end{split}
\ee
All the terms in  Eq.~\eqref{eq7.158}  can be expanded in power series in  
the dimensional regularization parameter $\ep$.  The functions $R^{(k)}(z)$ 
and the constants $R_+^{(k)}$ and $R^{(k)}_{\delta}$ are calculated numerically.

\section{Pole cancellation and finite remainders}
\label{sect:fres}

We are now in  position to discuss the final result for the NNLO QCD 
contribution to the cross section.  We consider 
\be
\d\hat\sigma^{\rm NNLO} = \d\sigma^{\rm RR}
+ \d\sigma^{\rm RV}
+ \d\sigma^{\rm VV}
+ \d\sigma^{\rm ren}
+ \d\sigma^{\rm CV}.
\label{eq258}
\ee
All the different contributions to Eq.~\eqref{eq258} were considered 
in the previous sections. It should be clear from these discussions 
that  the result for the NNLO cross section is given by a linear combination 
of integrated matrix elements with different multiplicities, which
may or may not be convoluted with generalized splitting 
functions. Since, 
for well-defined observables, the cancellation of soft
and collinear  divergences  occurs point-by-point in the phase space,  
contributions proportional to  $\FLM(1,2,4,5)$, $\FLM(1,2,4)$, $\FLM(z \cdot 1, 2, 4)$ 
etc. must be separately finite.  For this reason, it is convenient to present 
the result for the NNLO QCD contribution to the cross section as a sum of seven terms 
\be 
\begin{split}
{\rm d} \hat\sigma^{\rm NNLO} & =  \d  \hat\sigma^{\rm NNLO}_{\FLM(1,2,4,5)} 
+ \d  \hat\sigma^{\rm NNLO}_{\FLM(1,2,4)}
+ \d  \hat\sigma^{\rm NNLO}_{\FLM(z \cdot 1, \bar z \cdot 2) }
+ \d  \hat\sigma^{\rm NNLO}_{\FLM(z \cdot 1,2)}
\\
& + \d \hat \sigma^{\rm NNLO}_{\FLM(1, z \cdot 2)}
+ \d \hat \sigma^{\rm NNLO}_{\FV^{\rm fin}(1,2)}
+ \d  \hat \sigma^{\rm NNLO}_{\FLM(1,2)},
\end{split}
\label{eq259}
\ee
which are individually finite.
Each of the individual terms in Eq.~\eqref{eq259} has a subscript that 
indicates  the highest multiplicity 
matrix element that it contains. 
Below we collect all the different 
contributions to $\d \hat\sigma^{\rm NNLO}$ and present 
finite  remainders for terms with different multiplicities. 
For simplicity, we fix the arbitrary parameter $\Em=\sqrt{s}/2$. 

\subsection{Terms involving $ \hat {\cal O}_{NLO}\FLM(1,2,4,5)$ }
This contribution is the only one that involves the 
 matrix element for $q \bar q \to V+gg$. 
We repeat here the result, already given in Eq.~\eqref{eq4.12}
\begin{align}
&  \d \hat \sigma^{\rm NNLO}_{\FLM(1,2,4,5)} =
\langle \FLM^{s_rc_r} \rangle  = 
\sum_{(ij)\in dc}\bigg\langle
\big[\I-\SS\big]\big[\I-S_5\big]
\bigg[(\I- C_{5j})(\I-C_{4i})\bigg]\times
\nonumber
\\
&\quad\quad\quad\quad\quad\quad
\times\dg4\dg5 w^{i4,j5}\FLM(1,2,4,5)
\bigg\rangle
\nonumber
\\
&\quad\quad
+\sum_{i\in tc} 
\bigg\langle
\big[\I-\SS\big]\big[\I-S_5\big]
\bigg[
\theta^{(a)} \big[\I-\CC_i\big]\big[\I-C_{5i}\big] + 
\theta^{(b)} \big[\I-\CC_i\big]\big[\I-C_{45}\big] 
\\
&\quad\quad\quad\quad~~
 + \theta^{(c)} \big[\I-\CC_i\big]\big[\I-C_{4i}\big]+ 
\theta^{(d)} \big[\I-\CC_i\big]\big[\I-C_{45}\big]
\bigg]
\nonumber
\\
&\quad\quad\quad\quad~~
\times\dg4 \dg5 w^{i4,i5}\FLM(1,2,4,5)
\bigg\rangle.\nonumber
\end{align}
It follows that $ \d \hat \sigma^{\rm NNLO}_{\FLM(1,2,4,5)}$ is expressed 
through a  combination of  nested soft and collinear 
subtractions and can be directly computed in four dimensions.

\subsection{Terms involving $ \hat {\cal O}_{NLO}\FLM(1,2,4)$ }

We continue  with terms that involve $\FLM(1,2,4)$. 
They  are present in the double-real 
contribution, Eqs.~(\ref{eq:final_second},\ref{eq:final_third}) and in the 
real-virtual contribution,  Eq.~\eqref{eq:final_rv}; they are also found in terms 
that appear due to ultraviolet 
Eq.~\eqref{eq5.1} and collinear Eq.~\eqref{eq:final_pnlo}
renormalizations  of the next-to-leading order cross section.  Extracting these 
terms, we observe that all the $1/\ep$ singularities cancel out. The finite 
remainder reads 
\begin{align}
\!\!\!& {\rm d} \hat\sigma^{\rm NNLO}_{124} = 
\asontwopimu\Bigg\{
\int\limits_0^1 \d z 
\bigg[4\Cf\Dt_1(z) - \hat P^{(0)}_{qq}(z) \LMu - \hat P^{(\ep)}_{qq}(z)
\bigg]
\nonumber
 \\
\!\!\!& \times
\Bigg\langle\ONLO\left[
\frac{\FLM(z\cdot1,2,4)+\FLM(1,z\cdot2,4)}{z}\right]
\Bigg\rangle  
+
2\Cf
\int\limits_0^1 \d z\;
\Dt_0(z) 
\nonumber
\\
\!\!\!&  
\times\Bigg\langle
\ln\frac{\rho_{41}}{4}\hat{\mathcal O}_{\rm NLO}
\frac{\tilde w_{5||1}^{15,14}\FLM(z\cdot1,2,4)}{z}
+
\ln\frac{\rho_{42}}{4}\hat{\mathcal O}_{\rm NLO}
\frac{\tilde w_{5||2}^{25,24}\FLM(1,z\cdot2,4)}{z}
\Bigg\rangle
\nonumber
\\
\!\!\!& +\Cf\Bigg\langle\hat{\mathcal O}_{\rm NLO}
\bigg[
\frac{2}{3}\pi^2 
-2 \ln\frac{2E_4}{\sqrt{s}}\ln\frac{\rho_{41}}{4} \tilde w_{5||1}^{14,15}
-2 \ln\frac{2E_4}{\sqrt{s}}\ln\frac{\rho_{42}}{4} \tilde w_{5||2}^{24,25}
\bigg]\FLM(1,2,4)
\Bigg\rangle 
\\
\!\!\!& +\Ca\Bigg\langle\hat{\mathcal O}_{\rm NLO}
\bigg[
\frac{137}{18}-\frac{4}{3}\pi^2
+\frac{11}{6} \ln \frac{\mu^2}{E_4^2}
-\frac{11}{6}\lp
\ln\frac{\rho_{14}}{\rho_{24}} \tilde w_{4||5}^{14,15}+
\ln\frac{\rho_{24}}{\rho_{14}} \tilde w_{4||5}^{24,25}
\rp
\nonumber
 \\
\!\!\!& +\frac{3}{2}\ln\frac{2E_4}{\sqrt{s}}
+\ln^2\frac{2E_4}{\sqrt{s}}+
\frac{3}{4}\ln\frac{\rho_{14}\rho_{24}}{4}
+\ln\frac{2 E_4}{\sqrt{s}}\ln\frac{\rho_{14}\rho_{24}}{4}
+\Li_2\lp 1-\frac{\rho_{14}}{2}\rp  
\nonumber
\\
\!\!\!& +\Li_2\lp 1-\frac{\rho_{24}}{2}\rp\bigg]
\FLM(1,2,4)\Bigg\rangle
\nonumber
-\frac{\Ca}{3}\big\langle \hat{\mathcal O}_{\rm NLO} 
r_\mu r_\nu \FLM^{\mu\nu}(1,2,4)\big\rangle
\Bigg\}
 \\
\!\!\!& + \big\langle \hat{\mathcal O}_{\rm NLO}
\FVF(1,2,4)\big\rangle,
\nonumber
\end{align}
where 
\be
\Dt_i(z) = \D_i(z) - \frac{1+z}{2} \ln^i(1-z).
\ee

\subsection{Terms involving $\FVF(1,2)$ and $\FVVF(1,2)$}

Next,  we collect the finite remainders of the one-loop and two-loop virtual contributions 
to the $q \bar q \to V$ process.  These contributions appear  in  the real-virtual, 
the double-virtual, the 
collinear  subtraction  and the ultraviolet  renormalization. Upon combining them and 
expanding the resulting contributions in $\ep$, we obtain 
\bes
&\d\hat\sigma^{\rm NNLO}_{\FVF(1,2)}
=
\asontwopimu\Bigg\{\frac{2\pi^2}{3}\Cf
\big\langle \FVF(1,2)\big\rangle+
\big\langle \FVsqF(1,2)\big\rangle+
\big\langle \FVVF(1,2)\big\rangle
\\
&\!+
\int\limits_0^1 \d z \left[4\Cf \Dt_1(z) - \LMu \hat P^{(0)}_{qq}(z)
-P^{(\ep)}_{qq}(z)\right]
\bigg\langle
\frac{\FVF(z\cdot1,2)+\FVF(1,z\cdot2)}{z}
\bigg\rangle\Bigg\} .
\end{split}
\ee

\subsection{Terms of the form $\mathcal P_1 \otimes \d\sigma \otimes \mathcal P_2$}

Terms of the type $\mathcal P_1 \otimes \d\sigma \otimes \mathcal P_2$, 
where $\mathcal P_{1,2}$ are some splitting functions,
appear in the double-real contribution as well as  
in the collinear renormalization. Combining all the relevant terms, we find
\be
\begin{split}
\d\hat\sigma^{\rm NNLO}_{\FLM(z \cdot 1,\bar z \cdot 2)}=
\lp\asontwopimu\rp^2 \Cf^2
\int \limits_{0}^{1} \d z\; \d\bar z  
 \left [ 2\Dt_0(z) \ln\lp\frac{\mu^2}{s}\rp - 4\Dt_1(z) -(1-z)\right ]
 \\
\times
\Bigg\langle \frac{\FLM(z\cdot 1,\bar z\cdot 1)}{z\bar z} \Bigg\rangle
\left[2\Dt_0(\bar z) \ln\lp\frac{\mu^2}{s}\rp - 4\Dt_1(\bar z) -(1-\bar z) \right].
\end{split} 
\ee

\subsection{Terms of the form $\mathcal P \otimes \d\sigma$}

These terms appear  in  the double-real, real-virtual,  collinear subtraction 
and ultraviolet renormalization contributions.  We note that starting 
from ${\cal O}(1/\ep)$,  the part of the double-real contribution related to the 
integral of the triple-collinear  splitting function is only known 
numerically, see the discussion  in Section~\ref{sect:tripcol}.

Combining all the terms, we observe analytic cancellation of the poles up to $1/\ep^2$.
For the $1/\ep$ poles and the finite part, it is useful to split the contribution into
a scale-independent and a scale-dependent term
\be
\d\hat\sigma^{\rm NNLO}_{\FLM(z \cdot 1,2)} \equiv 
\d\hat\sigma^{\rm NNLO}_{\FLM(z \cdot 1,2)}(\mu^2=s)+
\Delta^{\rm NNLO}_{\FLM(z\cdot 1,2)}(\mu).
\ee
We also introduce an expansion  of the 
functions $R^{(k)}(z)$ and the constants $R^{(k)}_+$, which were introduced  in Section~\ref{sect:tripcol}, in powers of $\ep$
\be
\sum \limits_{k\in {\rm sectors}} 
R^{(k)}(z) = R^{(0)}(z) + \ep R^{(\ep)}(z)+ {\cal O}(\ep^2),\;\;
\sum \limits_{k\in {\rm sectors}} R_+^{(k)} = R_+^{(0)} + \ep R_+^{(\ep)} + {\cal O}(\ep^2).
\ee
The scale-independent term reads 
\begin{align}
&\d\hat\sigma^{\rm NNLO}_{\FLM(z \cdot 1,2)}(\mu^2=s) = 
\nonumber
\\
&\left[\asontwopimu\right]^2 \int\limits_0^1 \d z
\Bigg\{
\Cf^2\Bigg[
8\Dt_3(z) + 4\Dt_1(z)(1+\ln2)+4\Dt_0(z)\bigg[\frac{\pi^2}{3}\ln2 +4\zeta_3\bigg]
 \nonumber\\
& +\frac{5z-7}{2} + \frac{5-11z}{2}\ln z
+ (1-3z)\ln2 \ln z+\ln(1-z)\bigg[\frac{3}{2}z-(5+11z)\ln z\bigg]
 \nonumber\\
&\quad
+2(1-3z)\Li_2(1-z)
 \nonumber\\
& +(1-z)\bigg[\frac{4}{3}\pi^2+\frac{7}{2}\ln^2 2-2\ln^2(1-z)
+\ln2\big[4\ln(1-z)-6\big]+\ln^2 z 
\nonumber
 \\
&\quad
+\Li_2(1-z)\bigg]
+(1+z)\bigg[-\frac{\pi^2}{3}\ln z - \frac{7}{4}\ln^2 2 \ln z 
-2\ln 2 \ln(1-z)\ln z 
  \nonumber\\
&\quad
+4 \ln^2(1-z)\ln z-\frac{\ln^3 z}{3}+\big[4\ln(1-z)-2\ln2\big]\Li_2(1-z)
\bigg]
\nonumber\\
&\quad
+\left[\frac{1+z^2}{1-z}\right]\ln(1-z)
\big[3\Li_2(1-z)-2\ln^2 z\big]-
\frac{5-3z^2}{1-z}\Li_3(1-z)
\label{eq9.10}
\\
&\quad
+\frac{\ln z}{(1-z)}\bigg[
12 \ln(1-z)-\frac{3-5z^2}{2}\ln^2(1-z)
-\frac{7+z^2}{2}\ln2 \ln z\bigg]
\Bigg] \nonumber\\
&\quad
+\Ca\Cf\Bigg[
-\frac{22}{3}\Dt_2(z)+\lp\frac{134}{9}-\frac{2}{3}\pi^2\rp \Dt_1(z)+
\bigg[-\frac{802}{27}+\frac{11}{18}\pi^2   \nonumber\\
&\quad
+(2\pi^2-1)\frac{\ln2}{3}
+11\ln^2 2 + 16 \zeta_3\bigg]\Dt_0(z)+\frac{37-28z}{9}+
\frac{1-4z}{3}\ln2 \nonumber\\
&\quad
-\lp\frac{61}{9}+\frac{161}{18}z\rp\ln(1-z)+
(1+z)\ln(1-z)\bigg[\frac{\pi^2}{3}-\frac{22}{3}\ln2\bigg] \nonumber\\
&\quad
-(1-z)\bigg[\frac{\pi^2}{6}+\Li_2(1-z)\bigg]
-\frac{2+11z^2}{3(1-z)}\ln2\ln z
-\frac{1+z^2}{1-z}\Li_2(1-z)\times \nonumber\\
&\quad
\times\big[2\ln2+3\ln(1-z)\big]\Bigg] 
+R^{(\ep)}_{+} \D_0(z) + R^{(\ep)}(z)
\Bigg\}\Bigg\langle\frac{\FLM(z\cdot1,2)}{z}
\Bigg\rangle\nonumber.
\end{align}
The scale-dependent term reads
\begin{align}
&\Delta^{\rm NNLO}_{\FLM(z \cdot 1,2)}(\mu) = 
\left[\asontwopimu\right]^2\int\limits_0^1 \d z \Bigg\{
\Cf^2\Bigg[
-12\Dt_1(z)-12\Dt_2(z) -6+5z   \nonumber\\
&\quad
+2 (1-z)\ln(1-z)
-2\ln z \frac{1+z+z^2}{1-z}
-(1+z)\bigg[2\ln z\ln(1-z) + \frac{\ln^2 z}{2}+ \nonumber\\
&\quad
+2\Li_2(1-z)\bigg]
+2\frac{1+z^2}{1-z}\big[\ln(1-z)\ln z-\Li_2(1-z)\big]\Bigg] \nonumber\\
&\quad
+\Ca\Cf\Bigg[
\frac{22}{3}\Dt_1(z)-\lp\frac{67}{9}-\frac{\pi^2}{3}\rp\Dt_0(z)
-\frac{5-8z}{6} - \frac{2+11z^2}{6(1-z)}\ln z \\
&\quad
+\frac{1+z^2}{1-z}\Li_2(1-z)\Bigg]\Bigg\}
\Bigg\langle\frac{\FLM(z\cdot1,2)}{z}\Bigg\rangle
\times \ln\lp\frac{\mu^2}{s}\rp \nonumber\\
&+\lp\asontwopimu\rp^2\int\limits_0^1 dz \Bigg\{
\Cf^2\bigg[
4\Dt_1(z)+6\Dt_0(z)-(1-z)-\frac{1+3z^2}{2(1-z)}\ln z\bigg] \nonumber\\
&\quad
-\frac{11}{6}\Ca\Cf \Dt_0(z)\Bigg\}
\Bigg\langle\frac{\FLM(z\cdot1,2)}{z}\Bigg\rangle
\times \ln^2\lp\frac{\mu^2}{s}\rp. \nonumber
\end{align}
To arrive at these results, we check the cancellation of $1/\ep$ poles in 
$\d\hat\sigma^{\rm NNLO}_{\FLM(z \cdot 1,2)} $ and then,
assuming that the cancellation is exact, deduce the analytic
form of $R^{(0)}(z)$ and $R_+^{(0)}$. These analytic results are then used 
in the scale-dependent term.
Thus the only numerical contributions needed for the finite part are $R^{(\ep)}(z)$ and $R_+^{(\ep)}$.

\subsection{Terms involving $\FLM(1,2)$}

All the different  contributions to the NNLO cross section produce terms
proportional to $\FLM(1,2)$. These include constants $R_\delta^{(k)}$ originating
from the triple-collinear
splitting function, which, as mentioned in the previous subsection,
are only known numerically. As before, we introduce an expansion in $\ep$ for
these constants.
\be
\sum \limits_{k\in{\rm sectors}} 
R_\delta^{(k)} = R_\delta^{(0)} + \ep R_\delta^{(\ep)} + {\cal O}(\ep^2).
\ee
Furthermore, for the double-soft contribution we know the abelian constants
of Eq.~\eqref{eq4.19} analytically, but only  have numerical results
for the non-abelian constants, which are reported in Table~\ref{tab:ds}.
Thus, for each order in $1/\ep$, we check the cancellation of terms numerically and then, assuming that the
cancellation is actually exact, we deduce an analytic form of the
triple-collinear splitting and double-soft constants at this order.
This form is then used in determining the cancellation at lower orders in $1/\ep$.
Thus, the only numerical constants appearing in our final formula
are $R_\delta^{(\ep)}$ and $c^{\SS}_{0,C_A C_F}$. The final formula reads
\begin{align}
&\d\hat\sigma^{\rm NNLO}_{\FLM(1,2)} = \lp\asontwopimu\rp^2\Bigg[
2R^{(\ep)}_{\delta} + 
\Cf^2\bigg(
\frac{8}{45}\pi^4+\frac{7}{3}\pi^2\ln^2 2 - 16\zeta_3\ln 2\bigg)\nonumber \\
&
+\Ca\Cf\bigg(
\frac{4214}{81}
+\frac{403}{72}\pi^2
-\frac{17}{48}\pi^4
-\frac{671}{36} \zeta_3 + c^{\SS}_{0,\Ca\Cf}  
-\frac{445}{54}\ln 2 - \frac{22}{9}\pi^2 \ln 2 
\nonumber \\
&
-\frac{149}{18}\ln^2 2 
+ 4\pi^2 \ln^2 2+
\frac{22}{9}\ln^3 2 - \frac{16}{3}\ln^4 2-34\zeta_3\ln2 
\bigg)
\Bigg]\big\langle\FLM(1,2)\big\rangle 
\label{eq290}
\\
&+\lp\asontwopimu\rp^2
\Bigg\{
\Cf^2\bigg[\lp\frac{9}{2}-\frac{2}{3}\pi^2\rp \ln \lp\frac{\mu^2}{s}\rp
-\lp\frac{3}{4}+\pi^2+28\zeta_3\rp \bigg]
\nonumber
\\
&
+\Ca\Cf
\bigg[-\frac{11}{4} \ln \lp\frac{\mu^2}{s}\rp
+\lp-\frac{17}{12}+6\zeta_3\rp
\bigg]\Bigg\}
\ln \lp \frac{\mu^2}{s} \rp \big\langle \FLM(1,2)\big\rangle.
\nonumber 
\end{align}

\section{Numerical results}
\label{sect:numerics}

Having described the subtraction procedure in some detail, we will now
study how well it works in practice.  We have implemented it in a
partonic Monte Carlo program to compute NNLO QCD corrections to the
production of a vector boson $\gamma^*$ in proton-proton
collisions.\footnote{We remind the reader that we only consider the $q
  \bar q$ annihilation channel and restrict ourselves to gluonic
  corrections in this paper.}  The calculation is fully differential;
we consider decays of the virtual photons to massless leptons and
study NNLO QCD corrections to lepton observables. 
We extracted the relevant matrix elements from Refs.~\cite{Giele:1991vf,Bern:1997sc}
as implemented in~\cite{mcfm}, and from Refs.~\cite{Badger:2005jv,Gehrmann:2005pd}.
 For all
computations reported below we employ the NNLO parton distribution
functions from the NNPDF3.0 set~\cite{Ball:2014uwa}.

\begin{figure}[t]
\centering
\includegraphics[width=0.65\textwidth]{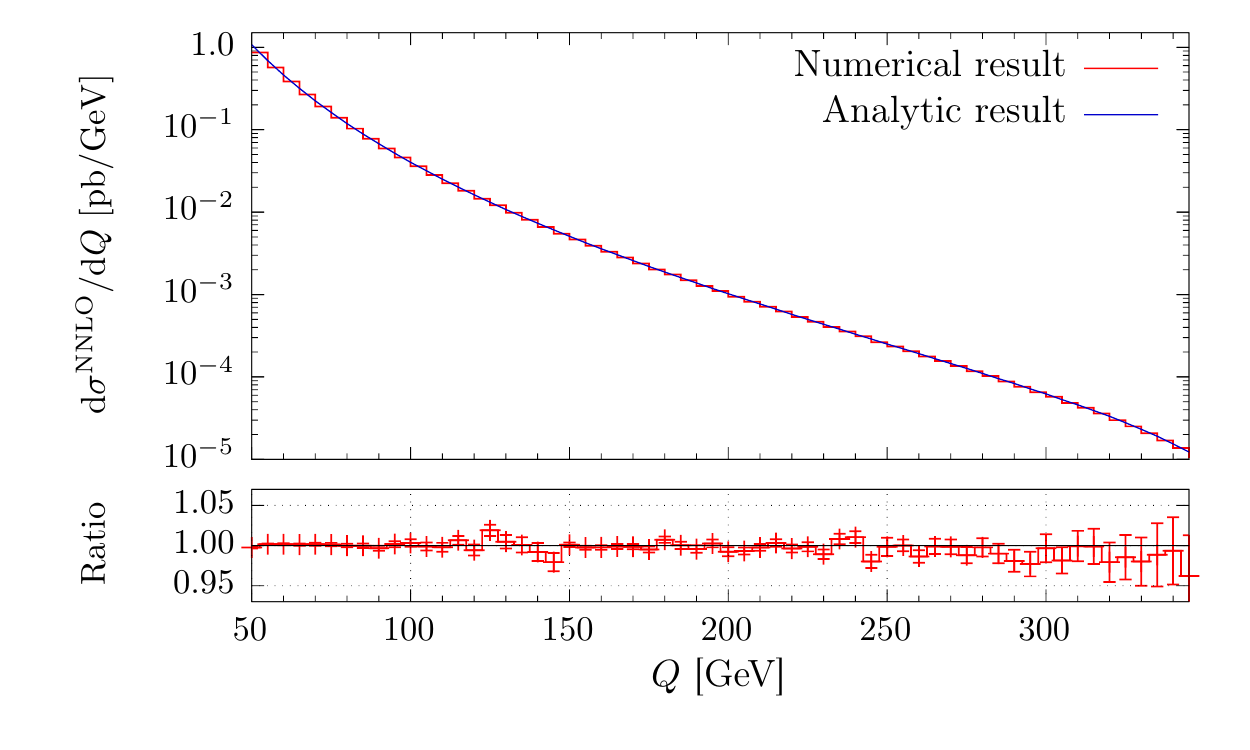}
\caption{Comparison of the NNLO QCD contribution ${\rm d}\sigma^{\rm NNLO}/{\rm d}Q$ 
computed in this paper with 
the analytic results in Ref.~\cite{Hamberg:1990np}.}
\label{fig:vNcomp}
\end{figure}

We begin by comparing the analytic result for the NNLO QCD correction
to the $pp \to \gamma^* \to e^+e^- + X$ cross section, which we
extract from Ref.~\cite{Hamberg:1990np}, and the result of the
numerical computation based on the formulas reported in the previous
section.  We wish to emphasize that this comparison is performed using
the NNLO \textit{contribution} to the cross section, and not the full
cross section at NNLO, which would have included LO and NLO contributions as well.  We take
$14~{\rm TeV}$ as the center-of-mass collision energy.  We include
lepton pairs with invariant masses $Q$ in the range $ 50~{\rm GeV} < Q
< 350~{\rm GeV}$ and take $\mu = 100~{\rm GeV}$ for the
renormalization and factorization scales.  We obtain the NNLO
{\it corrections} to the cross sections 
\be {\rm d} \sigma^{\rm NNLO} =
14.471(4)~{\rm pb},\;\;\;\; {\rm d} \sigma_{\rm analytic}^{\rm NNLO} =
14.470~{\rm pb},
\label{eq292}
\ee 
where the first result is ours and the second is extracted from
Ref.~\cite{Hamberg:1990np}.  The agreement between the two results is
quite impressive; it is significantly better than a permille.  To
further illustrate the degree of agreement, we repeat the comparison using the
kinematic distribution ${\rm d} \sigma^{\rm NNLO}/{\rm d}Q$, shown in
Fig.~\ref{fig:vNcomp}.  In the upper pane of Fig.~\ref{fig:vNcomp}, we
see a perfect agreement of analytic and numerical results for a range
of $Q$-values where the cross section changes by five orders of
magnitude.
The ratio of numerical and analytic cross sections is shown in the
lower pane of Fig.~\ref{fig:vNcomp}. We see that the agreement is
between a fraction of permille and a few percent for all values of $Q$ 
considered. 
We reiterate that we plot the NNLO correction to the differential cross section and not 
the full cross section at NNLO.
Given that the
NNLO contribution changes the NLO result by about 10\%, the
permille to percent precision on the NNLO {\it correction} leads to
almost absolute precision for physical {\it cross sections} and 
{\it simple kinematic distributions}. We will further illustrate this point
below.
Before doing so, we note that we found a similar level of agreement for individual
color structures and for individual contributions to the final result. We also note that although
we report results for a single scale choice here,  
using the results in the previous sections and the known amplitudes for $q\bar{q} \to e^+e^- + X$, it is easy to check {\it analytically} the scale dependence of our result against the one reported in
Ref.~\cite{Hamberg:1990np}. Full agreement is found.

\begin{figure}[!ht]
\centering
\includegraphics[width=0.48\textwidth]{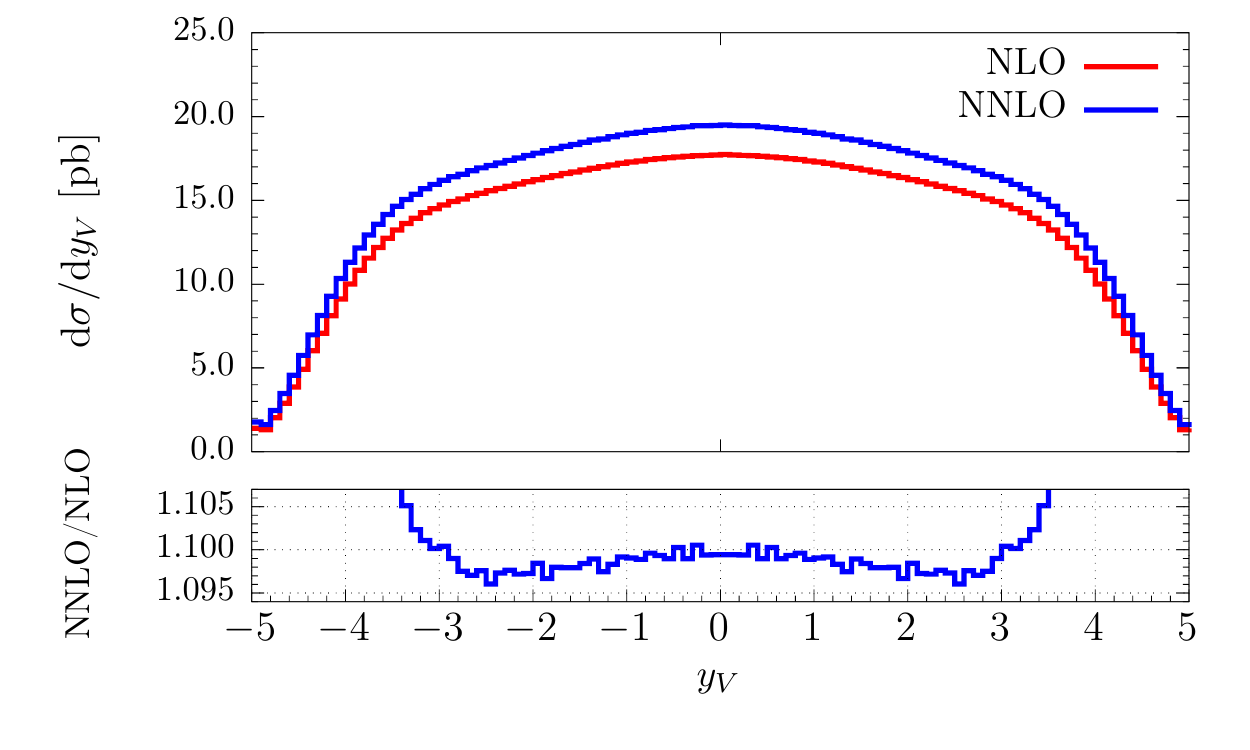}
\includegraphics[width=0.48\textwidth]{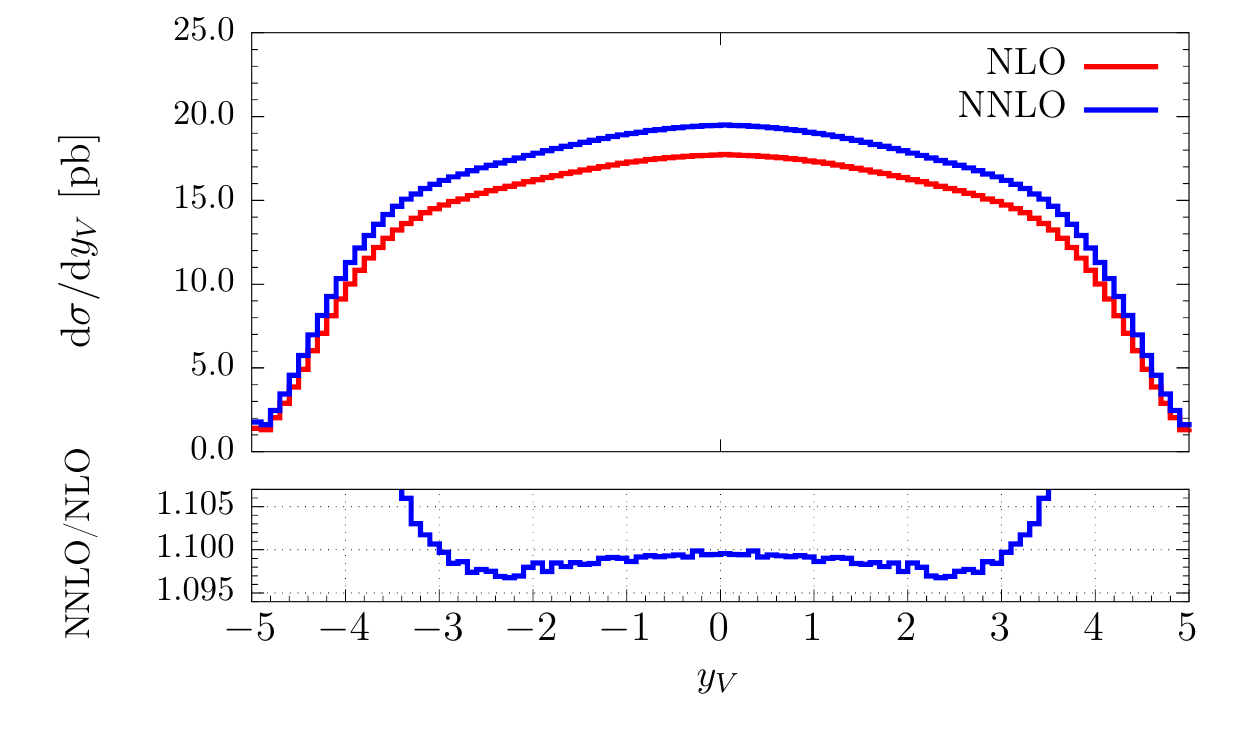}
\includegraphics[width=0.48\textwidth]{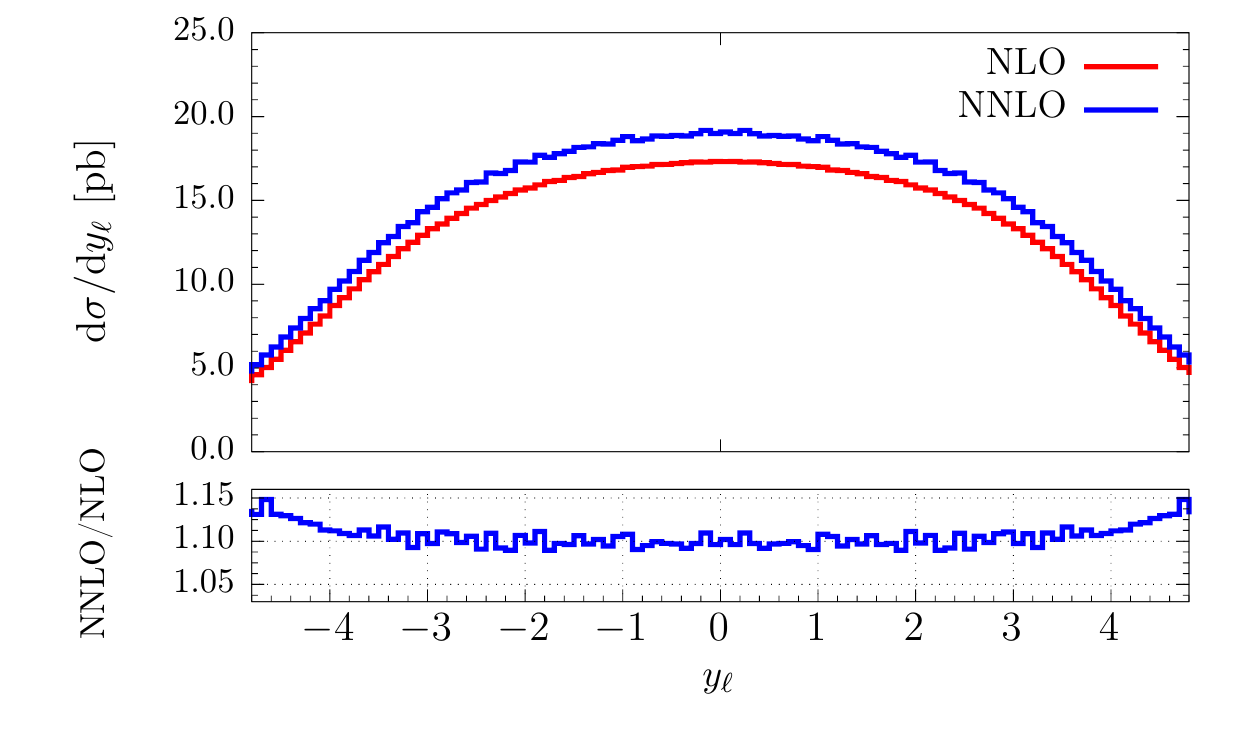}
\includegraphics[width=0.48\textwidth]{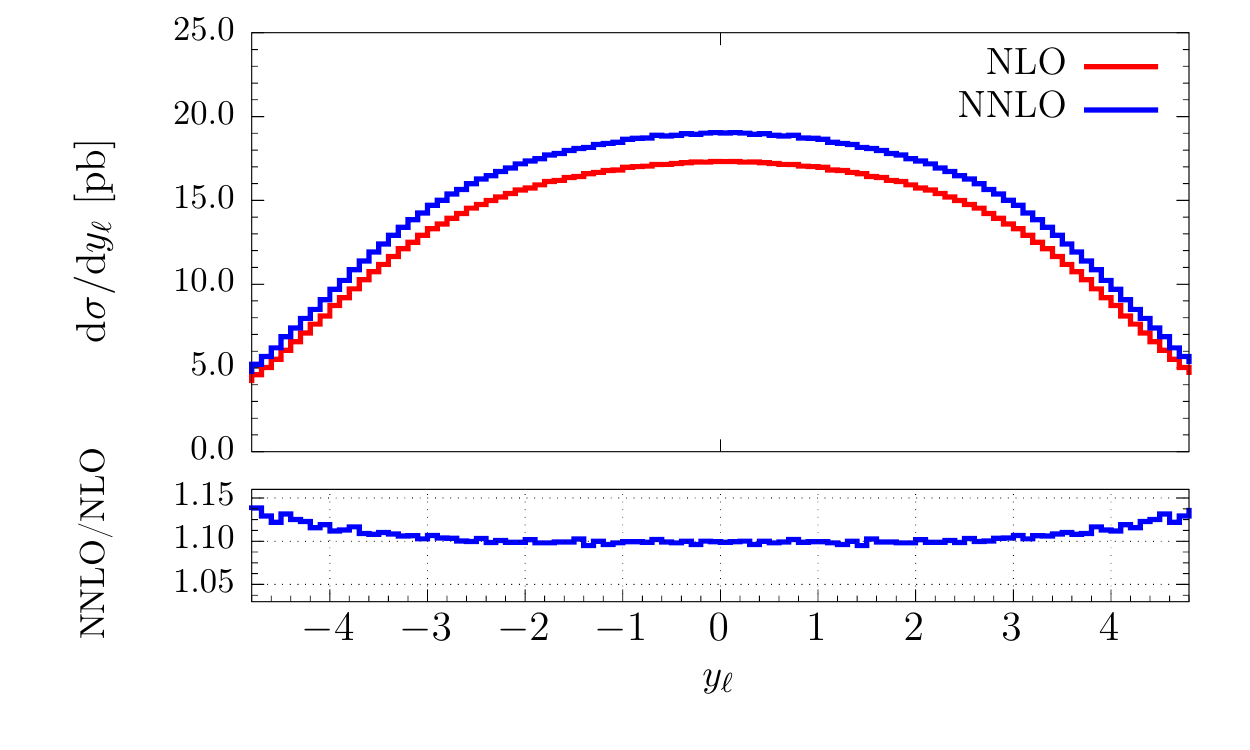}
\includegraphics[width=0.48\textwidth]{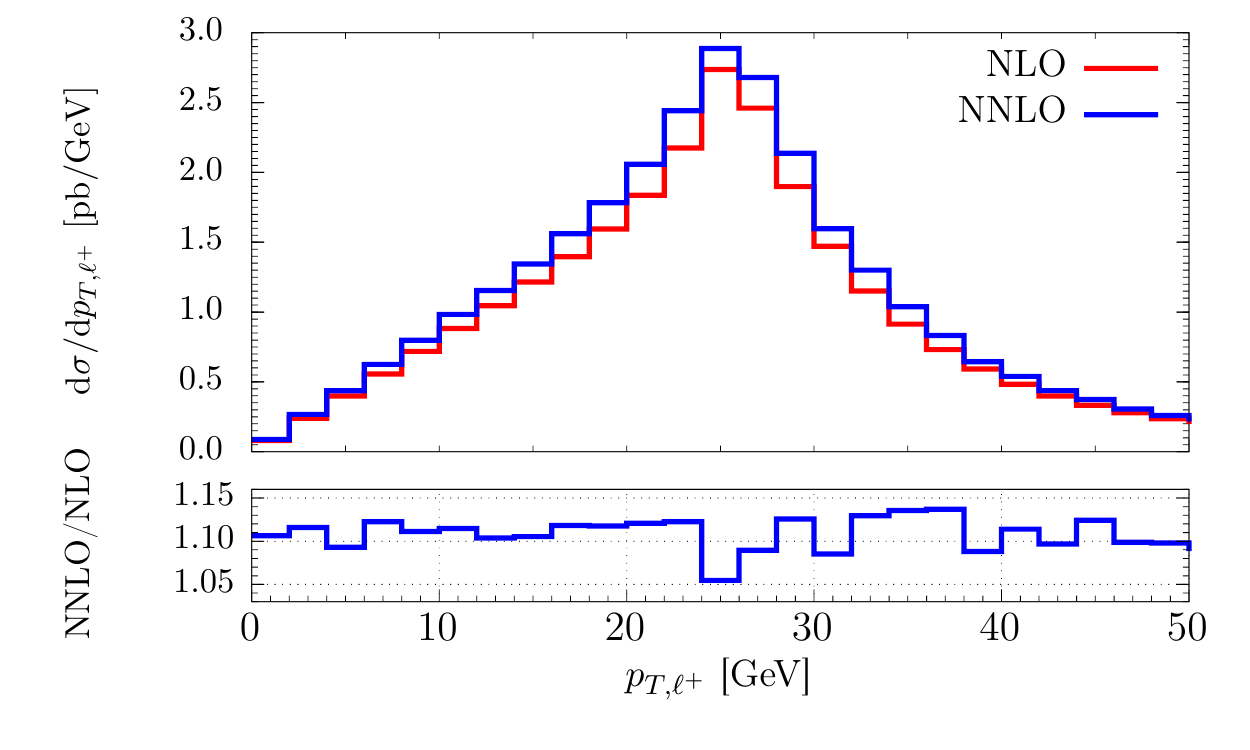}
\includegraphics[width=0.48\textwidth]{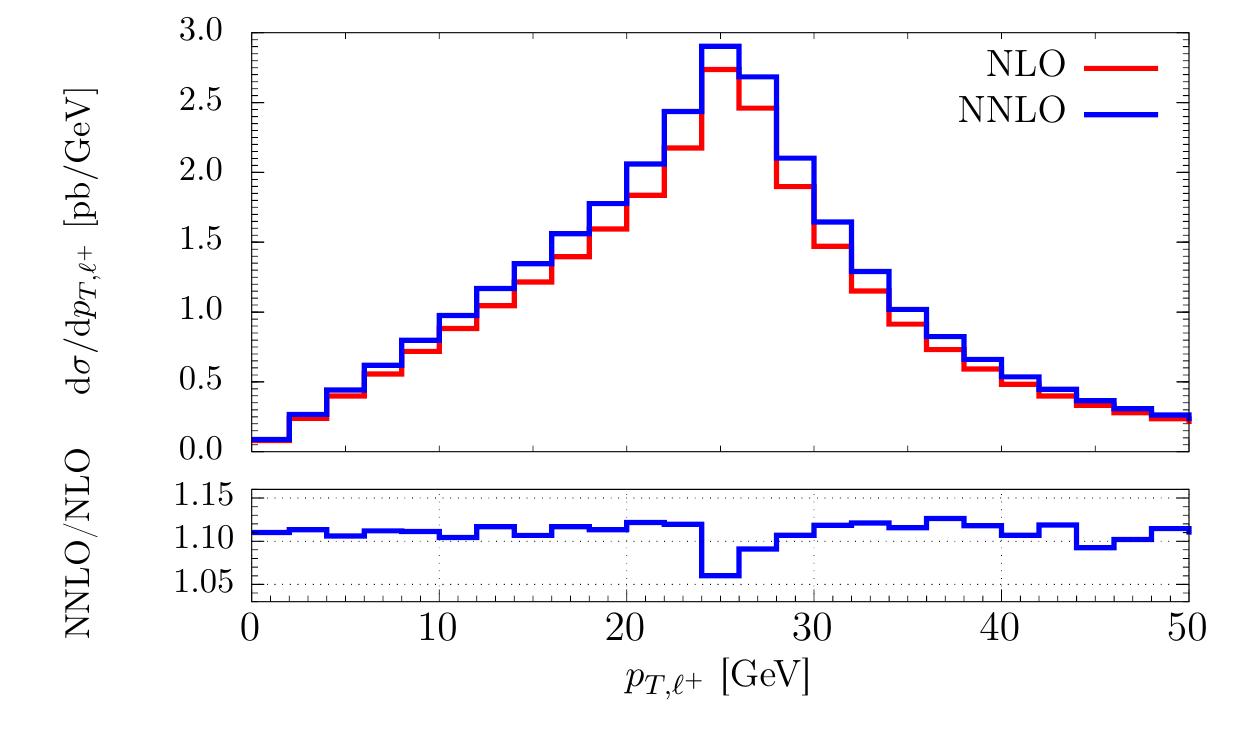}
\caption{Upper panes: Rapidity distribution of the vector boson, rapidity distribution of a lepton 
and $p_T$ distribution of a lepton at different orders of perturbation theory. 
Lower panes: the ratio of  NNLO/NLO prediction for a given observable. Plots on the left: the runtime 
of ${\cal O}(10)$ CPU hours; plots on the right: the runtime of ${\cal O}(100)$ CPU hours.
Note that the dip in the ratio of NNLO/NLO lepton $p_T$ distribution at $p_T\sim 25~{\rm GeV}$
is a physical feature and {\it not} a fluctuation.
  }
\label{fig:otherplots1}
\end{figure}

As we mentioned in the Introduction,  
one of the important issues for current   NNLO QCD computations 
is their practicality. 
For example, with the increasing precision of  Drell-Yan measurements, one may require 
very accurate theoretical predictions for fiducial volume cross sections. It is then 
important to clarify whether a given implementation of the NNLO QCD corrections can produce 
results that satisfy advanced stability requirements  and, if so,  how much CPU time is needed  to achieve them.  

To illustrate this aspect of our computational scheme, we show the
rapidity distribution of the dilepton pair, the rapidity distribution
of a lepton, and the lepton transverse momentum distribution in
Fig.~\ref{fig:otherplots1}.  The plots on the left and on the right
provide identical information: the upper panes show next-to-leading
and next-to-next-to-leading order predictions for the respective
observable, and the lower panes the ratio of the NNLO to NLO
distributions. The difference between the plots on the left and the
plots on the right is the CPU time required to obtain them; it changes
from $\mathcal{O}(10)$ CPU hours for the plots on the left, to
$\mathcal{O}(100)$ CPU hours for the plots on the right.  The
different run times are reflected in different bin-to-bin fluctuations
seen in both plots.  The bin-to-bin fluctuations for the two rapidity
distributions are at the percent-level or better in the plots on the
left, and they become practically unobservable in the plots on the
right. The situation is slightly worse for the transverse momentum of
the lepton.  However, this observable is rather delicate in the
$\gamma^*$ case, as each bin receives contributions from a large range
of invariant masses.  The introduction of a $Z$ boson propagator will
localize the bulk of the cross section in a much smaller invariant mass
window, and lead to improved stability in this case.\footnote{
The state-of-the-art comparison of this and other
observables in Drell-Yan production between different NNLO codes 
was presented in Ref.~\cite{Alioli:2016fum}.}
Nevertheless, the results shown in Fig.~\ref{fig:otherplots1} imply
that the numerical implementation of our subtraction scheme allows for
high precision computations, while also delivering results that are
acceptable for phenomenology even after relatively short run times.

\section{Conclusions}
\label{sect:conclusion}

In this paper we described a modification of the residue-improved subtraction 
scheme \cite{czakonsub}
that allows us to remove one of the five sectors that are traditionally  
used to fully factorize  singularities of the double-real emission matrix elements squared. 
The redundant sector includes correlated soft-collinear limits where energies  of emitted gluons 
and their angles vanish in a correlated fashion. Once this sector is removed, the  physical 
picture of  independent soft and collinear emissions leading  to singularities 
in scattering amplitudes is recovered and the bookkeeping simplifies considerably.

Using these simplifications,  we reformulated a NNLO subtraction scheme,  based on nested subtractions 
of soft and collinear singularities that, in a straightforward way,   leads to an integrable remainder 
for the double-real emission cross section. 
The subtraction terms are related  to cross sections of reduced multiplicity; they can be 
rewritten in a way that allows us to prove the cancellation of $1/\ep$ singularities 
independent of the hard matrix elements. Once singularities cancel,  
the  NNLO QCD corrected cross section is written in terms of quantities that 
can be computed in four dimensions. 

Although we believe that this framework is applicable for generic NNLO QCD
computations, in this paper, for the sake of simplicity, we
studied dilepton pair production in quark-antiquark
annihilation and computed gluonic contributions to
NNLO corrections. We implemented our formulas in a numerical program
and used it to calculate NNLO QCD {\it corrections} to the
production cross section of a vector boson in hadron collisions with a
sub-percent precision.  We also showed that kinematic distributions,
including the lepton rapidity and transverse momentum distributions,
can be computed precisely and efficiently.  We look forward to the
application of the computational framework discussed in this paper to more complex processes,
relevant for the LHC phenomenology.

\vskip 1cm
  
{\bf Acknowledgements:} We are grateful to Bernhard Mistlberger for
providing a cross-check on the analytic formula for the NNLO Drell-Yan
coefficient function.  We would like to thank KIT and CERN for
hospitality at various stages of this project.  The work of F.C. was
supported in part by the ERC starting grant 637019 ``MathAm'' and by the ERC
advanced grant 291377 ``LHCtheory''.  
K.M. and R.R. are supported by the German Federal Ministry for
Education and Research (BMBF) under grant 05H15VKCCA.

\newpage

\appendix

\pagenumbering{roman}

\section{Generalized splitting functions}
\label{sect:defs}

In this Appendix, we summarize all the various splitting functions that appear in our calculation. 

We start by presenting the Altarelli-Parisi splitting functions that are relevant for
the calculation. 
At NLO, we only require the leading order splitting function
\be
\hat P^{(0)}_{qq}(z) = \Cf \left[2\D_0(z)-(1+z)+\frac{3}{2}\delta(1-z)\right].
\label{eq:Pqq_AP}
\ee
For NNLO computations, two more splitting functions
are needed, see Eqs.~(\ref{eq:ren},\ref{eq:rensplit}). The first one is 
the convolution of two leading order
splitting functions. It reads
\bes
\left[\hat P^{(0)}_{qq}\otimes \hat P^{(0)}_{qq}\right](z) = 
\Cf^2\bigg[ 8 \D_1(z)+6 \D_0(z) + \lp\frac{9}{4}-\frac{2}{3}\pi^2\rp
\delta(1-z)  \\
-\frac{(1+3z^2)}{1-z}\ln z - 4(1+z)\ln(1-z) -5 - z\bigg]. 
\end{split}\label{eq:PqqAPoPqqAP}
\ee
The second is the splitting function $\hat P_{qq}^{(1)}$. We
emphasize that for our case, we need the NLO splitting function for a
continuous quark line entering the hard matrix element after the
emission of two gluons. 
We then only have to consider the non-singlet
NLO splitting function from which the contribution of identical quarks
is subtracted.
We extract the relevant information from Ref.~\cite{Hamberg:1990np}. 
$\hat P_{qq}^{(1)}$ in
Eq.~\eqref{eq:rensplit} must then be identified with
\begin{align}
\hat P^{(1)}_{qq,\widetilde{\rm NS}}(x) = 
&\Ca\Cf \Bigg[
\frac{3\pi^2(1+x)-124 x -19}{18}
+\lp\frac{67}{9}-\frac{\pi^2}{3}\rp \D_0(x)
+\frac{2+11x^2}{6(1-x)}\ln x 
\nonumber
\\
&\quad\quad\quad\quad
-\frac{1+x^2}{1-x}\Li_2(1-x)
+\delta(1-x)\lp\frac{17}{24}+\frac{11}{18}\pi^2-3\zeta_3\rp
\Bigg]
\nonumber
\\
&+\Cf^2\Bigg[3-2x 
-2\frac{1+x^2}{1-x}\ln(1-x)\ln x +2\ln x +
\frac{1+3x^2}{2(1-x)}\ln^2 x
\label{eq:PAPqq_NLO}
\\
&\quad\quad\quad\quad
+2\frac{1+x^2}{1-x}\Li_2(1-x)
+\delta(1-x)\lp\frac{3}{8}-\frac{\pi^2}{2}+6\zeta_3\rp\Bigg].
\nonumber
\end{align}

We now list the definitions of the various splitting functions used in the text.
In these formulas, we use the definition
\be
L_1 = \ln\lp2\frac{\Em}{\sqrt{s}}\rp.
\ee

\begin{itemize} 

\item[a)] The tree-level splitting function used in the NLO computation is defined as
\begin{align}
\PqqRRtwodelta(z) &= \Cf \Bigg\{\bigg[
\frac{3}{2} \delta(1-z) + 2\D_0(z) - 4 \D_1(z)\ep + 4\D_2(z)\ep^2\bigg]
\nonumber
\\
&
-(1+z) + \bigg[2(1+z)\ln(1-z) -(1-z)\bigg]\ep  
\label{eq:PqqRRtwodelta}
\\
&+\bigg[ 2 (1-z) \ln(1-z) -2 (1+z) \ln^2(1-z) \bigg]\ep^2
\Bigg\} + \mathcal O(\ep^3).
\nonumber
\end{align}

\item[b)] The splitting function $\mathcal P_{qq,{\rm NLO_{CV}}}(z)$ reads
\bes
\mathcal P_{qq,{\rm NLO_{CV}}}(z)  
&
= \Cf\hat P^{(0)}_{qq} \left[\frac{2\pi^2}{3} + \ep (\pi^2-4\zeta_3)\right] 
\\
&  +\Cf^2\frac{1+z^2}{1-z} \ln z \left[ \ln z - \ep
\lp \frac{2 \pi^2}{3} - \frac{3}{2}\ln z + \ln^2 z\rp\right]
+\mathcal O(\ep^2).
\end{split}
\label{eq:PqqNLOCV}
\ee

\item[c)] The splitting 
function $\big[\mathcal P_{qq}\otimes\mathcal P_{qq}\big]_{\rm NLO_{CV}}(z)$
reads 
\begin{align}
&\big[\mathcal P_{qq}\otimes\mathcal P_{qq}\big]_{\rm NLO_{CV}}(z)
=\Cf^2\Bigg\{
\bigg[6\D_0(z)+8\D_1(z) + \lp\frac{9}{4}-\frac{2}{3}\pi^2\rp \delta(1-z)\bigg]
\nonumber
\\
&+\bigg[\frac{4}{3}\pi^2 \D_0(z)-6\D_1(z)-12\D_2(z)-8\zeta_3\delta(1-z)\bigg]\ep 
\nonumber
\\
&+\bigg[16\zeta_3\D_0(z)-\frac{8}{3}\pi^2\D_1(z)+6\D_2(z)+\frac{32}{3}\D_3(z)
 -\frac{8}{45}\pi^4\delta(1-z)\bigg]\ep^2 
\nonumber
\\
&+\bigg[-5-z-4(1+z)\ln(1-z) - \frac{(1+3z^2)\ln z}{1-z}\bigg]
\nonumber
\\
&+\bigg[-\frac{3}{2}(1-z)+(5+z)\ln(1-z)+2(3+z)\ln z
+\frac{4\ln^2 z-6\ln z}{1-z} 
\label{eq:pqqOpqqNLOcv}
\\
&\quad\quad
+(1+z)\big[2\Li_2(z)-3\ln^2 z+6\ln^2 (1-z)-\pi^2\big]\bigg]\ep 
\nonumber
\\
&+\bigg[\pi^2\lp z-\frac{5}{3}\rp + (1-z)\big[3\ln(1-z)-2\big] -2\ln z
-2(3+z)\ln^2 z  
\nonumber
\\
&\quad\quad
+\frac{\ln^2 z}{1-z}\lp6-\frac{8}{3}\ln z\rp +
(6-2z)\Li_2(z)+
(1+z)\bigg(-\frac{16}{3}\ln^3(1-z)
\nonumber
\\
&\quad\quad\quad
-\ln^2(1-z)\big[3+2\ln z\big]
+\frac{2}{3}\pi^2\ln z + 2 \ln^3 z 
+\ln(1-z) 
\nonumber 
\\
&\quad\quad\quad
\times\big[2\pi^2-4\Li_2(z)\big] -4\Li_3(z)-4\Li_3(1-z)-4\zeta_3\bigg)
\bigg]\ep^2\Bigg\} + \mathcal O(\ep^3).
\nonumber
\end{align}

\item[d)] The one-loop splitting functions used in the computation of the real-virtual 
limits reads
\begin{align}
&P_{qq}^{\rm loop,i}(z) = 
-(1-z) P_{qq}(z)\Bigg\{
\bigg[\frac{1}{\ep^2}\frac{\Gamma^2(1-\ep)\Gamma^2(1+\ep)}
{\Gamma(1-2\ep)\Gamma(1+2\ep)}
\frac{1}{(1-z)^{2\ep}}+2\Li_2(1-z)
\nonumber
\\
&-2\ep\Li_2(1-z)\ln(1-z)\bigg]\Ca
+2\bigg[\frac{\ln z}{\ep}
-\Li_2(1-z)-\ln z\ln(1-z)  
\label{eq:pqqloopi}
\\
&+\bigg(
\frac{1}{2}\ln^2(1-z)\ln z + \ln(1-z) \Li_2(1-z) - \Li_3(1-z)
\bigg)\ep\bigg]\Cf
\Bigg\}\nonumber
\\
&+(\Ca-\Cf) \mathcal P^{\rm RV,new}_{qq}(z)
+\mathcal O(\ep^2),
\nonumber
\end{align}
where 
\bes
\mathcal P_{qq}^{\rm RV,new}(z) = -\Cf(1-z)\bigg[z + (1+z-z \ln(1-z))\ep\bigg] + \mathcal O (\ep^2).
\end{split}
\ee

\item[e)] The tree-level splitting function used in the real-virtual contribution  is defined as 
\bes
& \mathcal P_{qq,RV_1}(z) = \Cf\Bigg\{
2\big[\delta(1-z)\LM1-\D_0(z)\big] 
+2\big[2\D_1(z)-\delta(1-z)\LM1^2\big]\ep  \\
& +\left[\frac{4}{3}\delta(1-z)\LM1^3-4\D_2(z)\right]\ep^2+
\left[-\frac{2}{3}\delta(1-z)\LM1^4 + \frac{8}{3}\D_3(z)\right]\ep^3
 +(1+z) 
\\
&+ \big[(1-z) - 2(1+z) \ln(1-z)\big]\ep
+2\bigg[(1+z)\ln^2(1-z)
 -(1-z)\times\\
&\times \ln(1-z)\bigg]\ep^2
+\Bigg[
2(1-z)\ln^2(1-z)
-\frac{4}{3}(1+z) \ln^3(1-z)
\Bigg ]\ep^3\Bigg \} + \mathcal O(\ep^4).
\end{split}\label{eq:PqqRV_1}
\ee

\item[f)]
The one-loop splitting function used in the real-virtual contribution  reads
\begin{align}
& \mathcal P_{RV,2}(z) = 
-\Ca\Cf\Bigg\{
\frac{1}{\ep^2}\left[\delta(1-z)\LM1 - \D_0(z)\right]+
\frac{2}{\ep}\left[2\D_1(z)-\delta(1-z)\LM1^2\right]
\nonumber
\\
& +
\left[
-\frac{1}{3}\delta(1-z)\LM1\lp\pi^2-8\LM1^2\rp
+\frac{\pi^2}{3} \D_0(z) - 8 \D_2(z)
\right]+
\bigg[\frac{2}{3}\delta(1-z)\LM1^2
\nonumber
\\
&\times\lp\pi^2-4\LM1^2\rp
-\frac{4}{3}\pi^2\D_1(z)+\frac{32}{3}\D_3(z)\bigg]\ep
+\frac{1+z}{2\ep^2}
+\frac{1}{2\ep}\bigg[(1-z)-4(1+z)
\nonumber
\\
&
\times\ln(1-z)\bigg]
 +\bigg[
-\frac{z}{2}-2(1-z)\ln(1-z) + (1+z)\bigg( 4\ln^2(1-z)-\frac{\pi^2}{6}\bigg)
\nonumber\\
&
-\frac{(1+z^2)\Li_2(1-z)}{(1-z)}\bigg]
 +\bigg[
\frac{3}{2} z\ln(1-z) + 
\frac{1+z}{6}\big[ -3 + 4 \pi^2\ln(1-z)
\nonumber
\\
&
-32\ln^3(1-z)\big]
+(1-z)\bigg(4\ln^2(1-z)-\frac{\pi^2}{6} + \Li_2(1-z)\bigg)
+3\left[\frac{1+z^2}{1-z}\right]
\label{eq:PqqRV_2}
\\
&\times \ln(1-z)\Li_2(1-z)
\bigg]\ep
\Bigg\}
-\Cf^2\Bigg\{
-\frac{1}{\ep}\left[ \frac{(1+z^2)\ln z}{1-z}\right]
+\bigg[
\frac{z}{2} + (1-z)\ln z 
\nonumber
\\
&
+\frac{1+z^2}{1-z} \big[ 3 \ln(1-z)\ln z 
+ \Li_2(1-z)\big]
\bigg]
+
\bigg[
\frac{1}{2}\big[1+z-3z\ln(1-z)\big]
\nonumber
\\
&-(1-z)\big[3\ln(1-z)\ln z 
+ \Li_2(1-z)\big]
-\frac{1}{2}\left[\frac{1+z^2}{(1-z)}\right]\big[9\ln^2(1-z)\ln z 
\nonumber
\\
&
+ 6 \ln(1-z)\Li_2(1-z)
 -2\Li_3(1-z)\big]
\bigg]\ep
\Bigg\} + \mathcal O(\ep^2).
\nonumber
\end{align}

\item[g)] The splitting function $\mathcal P_{qq,RR_1}$ reads 
\bes
& \mathcal P_{qq,RR_1}(z) = \Cf\Bigg\{\bigg[
2\big[\D_0(z)-\LM1\delta(1-z) \big]
-4\big[2\D_1(z) -\LM1^2\delta(1-z)\big]\ep\\
&+16\big[\D_2(z) - \frac{1}{3}\LM1^3\delta(1-z)\big]\ep^2
-\frac{16}{3}\big[4 \D_3(z) - \LM1^4\delta(1-z)\big]\ep^3
\bigg]\\
&+\bigg[-(1+z)
+ \big[4(1+z)\ln(1-z) -(1-z)\big]\ep  \\
&+\big[4(1-z)\ln(1-z) - 8 (1+z) \ln^2(1-z) \big]\ep^2  \\
&+\left[\frac{32}{3}(1+z)\ln^3(1-z)  - 8 (1-z)\ln^2(1-z) \right]\ep^3
\bigg]
\Bigg\}
+\mathcal O(\ep^4).
\end{split}\label{eq:pqqrrI_b}
\ee

\item[h)]

The splitting function $\mathcal P_{qq,RR_2}(z)$ reads  
\begin{align}
\mathcal P_{qq,RR_2}(z) &= \Cf \Bigg\{\bigg[
2\D_0(z) - 4 \D_1(z)\ep + 4\D_2(z)\ep^2 -\frac{8}{3}\D_3(z)\ep^3\bigg]
\nonumber
\\
&+\bigg[
-(1+z) + \big[2(1+z)\ln(1-z) -(1-z)\big]\ep  
\nonumber
\\
&+\big[ 2 (1-z) \ln(1-z) -2 (1+z) \ln^2(1-z) \big]\ep^2  
\label{eq:pqqrr2_b}
\\
&+\left[-2(1-z)\ln^2(1-z) +\frac{4}{3}(1+z)\ln^3(1-z) \right]\ep^3
\bigg]
\Bigg\} + \mathcal O(\ep^3).
\nonumber
\end{align}

\item[i)] The splitting function $\mathcal P_{qq,RR_3}(z) $ reads 
\begin{align}
& \mathcal P_{qq,RR_3}(z) = \Cf \Bigg\{
\left[ 2\LM1 \D_0(z) - 2\D_1(z) - \delta(1-z)\LM1^2\right]
\nonumber
\\
+&
\left[ 6\D_2(z) - 4 \LM1 \D_1(z) - 2 \LM1^2 \D_0(z) + 2 \delta(1-z)\LM1^3\right] \ep
\nonumber
\\
+&\left[
-\frac{28}{3}\D_3(z) + 4\LM1 \D_2(z) + 4 \LM1^2 \D_1(z) + 
\frac{4}{3}\LM1^3 \D_0(z) - \frac{7}{3}\delta(1-z) \LM1^4
\right]\ep^2
\nonumber
\\
+&
\left[(1+z)(\ln(1-z)-\LM1)\right] + 
\bigg[
(1-z)(\ln(1-z)-\LM1)
\label{eq:pqqrr3_b}
\\
+&(1+z) \lp
\LM1^2 + 2 \LM1 \ln(1-z) -3 \ln^2(1-z)
\rp
\bigg]\ep
\nonumber
\\
+&
\bigg[
(1-z)\lp
\LM1^2 + 2 \LM1 \ln(1-z) - 3\ln^2(1-z)\rp+(1+z)\times
\nonumber
\\
\times&\lp
-\frac{2}{3}\LM1^3 - 2 \LM1 \ln^2(1-z) + \frac{14}{3}\ln^3(1-z)
-2\ln(1-z)\LM1^2
\rp
\bigg]\ep^2
\Bigg\} + \mathcal O(\ep^3).
\nonumber
\end{align}

\item[j)] The splitting function $\mathcal P_{qq,RR_4}(z)$ reads  

\bes
\mathcal P_{qq,RR_4}(z)&=\Cf\Bigg\{
\left[
\delta(1-z)\LM1^2-2\D_1(z)
\right]+
\left[
-2\delta(1-z)\LM1^3 + 6\D_2(z)
\right]\ep\\
&+\left[
\frac{7}{3}\delta(1-z)\LM1^4 - \frac{28}{3}\D_3(z)
\right]\ep^2
+\left[(1+z)\ln(1-z)\right]\\
&+\left[(1-z)\ln(1-z) -3 (1+z) \ln^2(1-z)\right]\ep\\
&+\left[
\frac{14}{3}(1+z)\ln^3(1-z) -3(1-z)\ln^2(1-z)
\right]\ep^2\Bigg\} + \mathcal O(\ep^3).
\end{split}\label{eq:pqqrr4_b}
\ee

\item[k)]  The splitting function $\mathcal P_{qq,RR_5}(z)$ reads
\bes
\mathcal P_{qq,RR_5}(z) = \Cf \bigg[&
2\big[\D_0(z) - \LM1 \delta(1-z)\big]+
\big[4\LM1^2 \delta(1-z) -8 \D_1(z)\big]\ep  \\
&-\frac{3+z}{2} + 
2(3+z)\ln(1-z) \ep\bigg] + \mathcal O(\ep^2).
\end{split}\label{eq:pqqrr5_b}
\ee

\item[l)] The splitting function $\mathcal P_{qq,RR_6}(z)$ reads 
\bes
& \mathcal P_{qq,RR_6}(z)=\Cf^2\Bigg(
4\D_1(z) - 12\D_2(z)\ep + \frac{56}{3}\D_3(z)\ep^2
-2(1+z)\ln(1-z)\\
&\quad\quad\quad
+\left[-2(1-z)\ln(1-z) +6 (1+z) \ln^2(1-z)\right]\ep\\
&\quad\quad\quad
+\left[6(1-z)\ln^2(1-z)-\frac{28}{3}(1+z)\ln^3(1-z)\right]\ep^2
\Bigg)+\mathcal O(\ep^3).
\end{split}\label{eq:pqqrr6}
\ee

\item[m)] The splitting function $\big[\mathcal P_{qq}\otimes\mathcal P_{qq}\big]_{RR}(z) $ 
is defined as 
\begin{align}
& \big[\mathcal P_{qq}\otimes\mathcal P_{qq}\big]_{RR}(z) = 
\Cf^2  \Bigg\{
\bigg[
8\D_1(z)
-\frac{2}{3}\pi^2 \delta(1-z)
- 4\D_0(z) \LM1
\bigg]
\nonumber
\\
& 
+\bigg[
\lp\frac{8}{3}\pi^2 + 4 \LM1^2\rp \D_0(z)
+ 8 \LM1 \D_1(z)-24 \D_2(z) - 16 \zeta_3 \delta(1-z)
\bigg]\ep
\nonumber
\\ & 
+
\bigg[
\lp-\frac{8}{3}\LM1^3 + 64 \zeta_3\rp \D_0(z)
-\lp 8 \LM1^2 + \frac{32}{3} \pi^2\rp \D_1(z)
- 8 \LM1 \D_2(z)  
\nonumber
\\
&
+\frac{112}{3} \D_3(z) 
- \frac{2}{5}\pi^4 \delta(1-z)
\bigg]\ep^2\Bigg\} 
+\Cf^2 \Bigg\{
\bigg[
-2(1-z) + (1+z)
\big( \ln z 
\nonumber
\\
&
- 4 \ln (1-z)+2\LM1 \big)
\bigg]
+
\bigg[
2(1-z) \lp 2\ln(1-z)+\LM1\rp+
(1+z)
\big[
12\ln^2(1-z)
\nonumber
\\
&
+2\ln z - \ln^2 z + 4 \Li_2(z)-2\pi^2
 - 2 \LM1^2 - 4 \LM1 \ln(1-z)\big]
\bigg]\ep
+
\bigg[
-2(1-z)
\label{eq:PqqotimesPqqrr_b}
\\
&
\times
\big(2\ln^2(1-z)+2\LM1 \ln(1-z)+\LM1^2 + 5\big)
+(1+z)
\bigg(
4\ln^2(1-z)\LM1 
\nonumber
\\
&
+4 \ln(1-z)\LM1^2 +\frac{4}{3}\LM1^3 
-\frac{4}{3}\pi^2 + 8\pi^2 \ln(1-z) 
- \frac{56}{3} \ln^3(1-z)
-5 \ln z 
\nonumber
\\
&
+ \frac{2}{3}\pi^2 \ln z
- 8 \ln^2(1-z)\ln z 
- 2 \ln^2 z + 
\frac{2}{3}\ln^3 z
+8 \Li_2(z) 
\nonumber
\\
&- 16 \ln(1-z) \Li_2(z) 
-16 \Li_3(1-z)-8 \Li_3(z) - 24 \zeta_3
\bigg)
\bigg]\ep^2\Bigg\} + \mathcal O (\ep^3).
\nonumber
\end{align}

\end{itemize}

We also require the following constants
\bes
&\tilde\gamma_g(\ep) =
\frac{11}{6}\Ca + 
\Ca\lp \frac{137}{18}-\frac{2\pi^2}{3}\rp \ep + 
\Ca\lp \frac{823}{27}-\frac{11}{18}\pi^2 - 16\zeta_3 \rp\ep^2+
\mathcal O(\ep^3),\\
&\tilde\gamma_g(\ep,\kt)  = 
-\frac{\Ca}{3} - \frac{7\Ca}{9}\ep + \mathcal O(\ep^2),
\label{eq:gamma}
\end{split}
\ee
and
\bes
&\delta_g(\ep) =
\Ca\lp -\frac{131}{72}+\frac{\pi^2}{6} + \frac{11}{6}\ln2\rp
+\Ca\lp -\frac{1541}{216}+\frac{11}{18}\pi^2-\frac{\ln2}{6}+4\zeta_3\rp\ep\\
&+\Ca\lp -\frac{9607}{324}+\frac{125}{216}\pi^2+\frac{7}{45}\pi^4
+\ln2 + \frac{11}{18}\pi^2\ln2 + \frac{77}{6}\zeta_3\rp\ep^2+
\mathcal O(\ep^3).
\label{eq:delta}
\end{split}
\ee

\section{Phase space parametrization and partitioning }
\label{sect:phsp}

We consider the phase space element of the two gluons 
\be
\dg4 \dg5  \theta(E_{\rm max} - E_4) \theta( E_4 - E_5)
\label{eq:ap11}
\ee
and discuss its parametrization. 

We take $E_4 = E_{\rm max}\; x_1$, $E_5 = E_{\rm max}\; x_1 x_2$ with $ x_1\in(0,1)$
and $  x_2\in (0,1)$ and write the phase space in Eq.~\eqref{eq:ap11} as 
\be
\begin{split}
& [dg_4][dg_5] = \frac{ {\rm d} x_1 }{x_1^{1+4\ep}} \frac{ {\rm d} x_2 }{x_2^{1+2\ep}} \;
x_1^4 \;  x_2^2  \; E_{\rm max}^{4 - 4 \ep} \; {\rm d \Omega}_{45},
\;\;\;\;\;{\rm d}\Omega_{45} = 
\frac{{\rm d} \Omega_4^{(d-1)}}{ 2 ( 2\pi)^{d-1}} \frac{{\rm d} \Omega_5^{(d-1)}}{ 2 ( 2\pi)^{d-1}}.
\end{split} 
\ee

We will now introduce the parametrization of the angular phase space. This parametrization 
is tailored the  process that we are interested 
in 
\be
\bar q(p_1) + q(p_2) \to  V(p_V) + g(p_4) + g(p_5),
\ee
and, as we will see, allows us to expose all the collinear singularities in a straightforward manner. 
We assume that momenta of quarks in the initial state point along the $z$-axis
\be
p_{1,2} = E_{1,2} \left ( t^\mu  \pm e_3^\mu \right ),\;\;\;
t^\mu = (1,0,0,0;...),\;\;\;e_3^\mu = (0,0,0,1;0,0...), 
\ee
and parametrize  the gluon momenta as 
\be
\begin{split} 
& p_4 = E_4 \big ( t^\mu + \cos \theta_{41} e_3^\mu + \sin \theta_{41} b^\mu \big ),
\\
& p_5  =E_5 \big ( t^\mu + \cos \theta_{51} e_3^\mu  + \sin \theta_{51} \left ( \cos \varphi_{45}  b^\mu 
+ \sin \varphi_{45} a^\mu \big )
\right ),\label{eq:mom_spin}
\end{split} 
\ee
The ($d$-dimensional) unit vectors $b^\mu$ and $a^\mu$ are chosen in such a way that 
\be
t \cdot a = e_3 \cdot a = t \cdot b = e_3 \cdot b = a \cdot b = 0. 
\ee
Given this choice, the angular phase space is written as \cite{czakonsub}
\be
{\rm d}\Omega_{45} = 
\frac{{\rm d} \Omega_{b}^{(d-2)} {\rm d} \Omega_{a}^{(d-3)}}{ 2^{6\ep} (2 \pi)^{2 d -2} }
\left [ \eta_4 (1-\eta_4) \right ]^{-\ep} \left [ \eta_5 ( 1- \eta_5 ) \right ]^{-\ep} 
\frac{| \eta_4 - \eta_5|^{1-2\ep} }{D^{1-2\ep} }
 \frac{ {\rm d} \eta_4 {\rm d} \eta_5 {\rm d} \lambda}{[\lambda ( 1- \lambda)]^{1/2+\ep}}, 
\ee
where $\eta_{i} = \eta_{i1}$ and $\eta_{ij} = \rho_{ij}/2$, 
$ 
D = \eta_4 + \eta_5 - 2 \eta_4 \eta_5 + 2 ( 2\lambda - 1) \sqrt{\eta_4 \eta_5 ( 1- \eta_4) (1 - \eta_5) }$,
and 
\be
\begin{split} 
& \eta_{45} = \frac{|\eta_4 - \eta_5|^2}{D},\;\;\;\; \sin^2 \varphi_{45} 
= 4 \lambda ( 1- \lambda) \frac{|\eta_4 - \eta_5 |^2}{D^2}.
\end{split} 
\ee

The phase space can be split into four different sectors that we will refer to as 
$a,b,c,d$.  The following parametrization is choosen for each of the 
four sectors~\cite{czakonsub}
\begin{itemize} 
\item[ a)] $\eta_4 = x_3,\;\;\; \eta_5 = x_3 x_4/2$;
\;\;\;\;\;\;\;\;\;
 b) $\eta_4 = x_3,\;\;\; \eta_5 = x_3 ( 1-  x_4/2)$;
\item[ c)] $\eta_4 = x_3 x_4/2 ,\;\;\; \eta_5 = x_3 $;\;\;\;\;\;\;\;\;\;\;
d)  $\eta_4 = x_3 (1-x_4/2),\;\;\; \eta_5 = x_3$.
\end{itemize} 
Below we present the phase space  for each of the four sectors employing the above 
parametrization.  To this end, we will need  the following function 
\be
N(x_3,x_4, \lambda) = 1 + x_4 ( 1 - 2 x_3) - 2(1-2 \lambda) \sqrt{x_4(1-x_3) ( 1- x_3 x_4)}.
\ee
It turns out that the angular phase spaces  for sectors $a$ and $c$ and 
for sectors $b$ and $d$ are identical.  The results read 
\be
\begin{split}
& {\rm d} \Omega_{45}^{(a,c)} = 
\frac{{\rm d} \Omega_{b}^{(d-2)} {\rm d} \Omega_{a}^{(d-3)}}{ 2^{6\ep} (2 \pi)^{2 d -2} }
\frac{{\rm d} x_3 }{x_3^{1+2\ep}}\frac{{\rm d} x_4 }{ x_4^{1+ \ep} }
\frac{{\rm d} \lambda}{ (\lambda ( 1- \lambda) )^{1/2+\ep}}\; 
F_{\ep}^{-\ep}\; F_0 \; x_3^2 x_4 = \\
&\quad\quad\quad= \left[\frac{1}{8\pi^2}\frac{(4\pi)^{\ep}}{\Gamma(1-\ep)}\right]^2
\left[\frac{\Gamma^2(1-\ep)}{\Gamma(1-2\ep)}\right]
\left[
\frac{\d\Omega^{(b)}_{d-2}}{\Omega_{d-2}}
\frac{\d\Omega^{(a)}_{d-3}}{\Omega_{d-3}}\right]\times\\
&\quad\quad\quad\quad\times
\frac{{\rm d} x_3}{x_3^{1+2\ep}}\frac{{\rm d} x_4}{x_4^{1+ \ep} }
\frac{{\rm d} \lambda}{ \pi(\lambda ( 1- \lambda) )^{1/2+\ep}}\; 
\; (256  F_{\ep})^{-\ep} \; 4F_0\; x_3^2 x_4,\\
& F_\ep =\frac{(1-x_3) (1-x_3 x_4/2) (1-x_4/2)^2}{2 N(x_3,x_4/2,\lambda)^2},\;\;\;\;   F_0 = \frac{(1-x_4/2)}{2 N(x_3,x_4/2,\lambda)},\;\;\;\Bigg.
\end{split}
\label{eqa:11}
\ee
and
\be
\begin{split}
& {\rm d} \Omega_{45}^{(b,d)} = 
\frac{{\rm d} \Omega_{b}^{(d-2)} {\rm d} \Omega_{a}^{(d-3)}}{ 2^{6\ep} (2 \pi)^{2 d -2} }
\frac{{\rm d} x_3}{x_3^{1+2\ep}}\frac{{\rm d} x_4}{ x_4^{1+ 2 \ep} }
\frac{{\rm d} \lambda}{ (\lambda ( 1- \lambda) )^{1/2+\ep}}\; 
F_{\ep}^{-\ep}\; F_0 \; x_3^2 x_4^2=\\
&\quad\quad\quad= \left[\frac{1}{8\pi^2}\frac{(4\pi)^{\ep}}{\Gamma(1-\ep)}\right]^2
\left[\frac{\Gamma^2(1-\ep)}{\Gamma(1-2\ep)}\right]
\left[
\frac{\d\Omega^{(b)}_{d-2}}{\Omega_{d-2}}
\frac{\d\Omega^{(a)}_{d-3}}{\Omega_{d-3}}\right]\times\\
&\quad\quad\quad\quad\times
\frac{{\rm d} x_3}{x_3^{1+2\ep}}\frac{{\rm d} x_4 }{ x_4^{1+ 2 \ep} }
\frac{{\rm d} \lambda}{ \pi(\lambda ( 1- \lambda) )^{1/2+\ep}}\; 
(256  F_{\ep})^{-\ep} \;4F_0\; x_3^2 x_4^2,\\
& F_\ep =\frac{(1-x_3) (1-x_4/2) (1-x_3(1-x_4/2) ) }{4 N(x_3,1-x_4/2,\lambda)^2},\;\;\;\;   
F_0 = \frac{1}{4 N(x_3,1-x_4/2,\lambda)}.\;\;\;\Bigg.
\label{eqa:12}
\end{split}
\ee
Furthermore, it is beneficial to remove singular dependence on $\lambda$ 
in Eqs.~(\ref{eqa:11},\ref{eqa:12}) by changing variable $\lambda \to y$ as 
\be
\lambda = \sin^2\lp \frac{\pi}{2}y\rp,\;\;\;\;\;\;
\frac{\d\lambda}{\pi\sqrt{\lambda(1-\lambda)}}=\d y,~~~~~~~
y\in (0,1).
\ee

To deal with one singularity at a time, we have to partition the 
phase space. We write  
\be
1 = w^{14,15}  + w^{24,25} + w^{14,25} + w^{15,24},
\label{eqa14}
\ee
where 
\be
\begin{split} 
& w^{14,15} = \frac{\rho_{24} \rho_{25}}{d_4 d_5} \left ( 1 + \frac{\rho_{14}}{d_{4521}}
 + \frac{\rho_{15}}{d_{4512}} \right ),\;\;\;\;\;\;\;
\\
& w^{24,25} = \frac{\rho_{14} \rho_{15}}{d_4 d_5} \left ( 1 + \frac{\rho_{25}}{d_{4521}}
 + \frac{\rho_{24}}{d_{4512}} \right ),
\\
& w^{14,25} = \frac{\rho_{24} \rho_{15} \rho_{45}}{d_4 d_5 d_{4512}},\;\;\;\;\;
w^{24,15} = \frac{\rho_{14} \rho_{25} \rho_{45}}{d_4 d_5 d_{4521}}.
\end{split} 
\ee
We have introduced the following notation 
\be
d_{i =4,5} =\rho_{1i} + \rho_{2i} = 2,\;\;\;\; 
d_{4521} = \rho_{45} + \rho_{42}+\rho_{51},\;\;\;\;
d_{4512} = \rho_{45} + \rho_{41}+\rho_{52}.
\ee
In Eq.~\eqref{eqa14}, the term $w^{14,15}$ corresponds to the 
triple-collinear sector where singular radiation occurs 
along the direction of the incoming quark with momentum 
$p_1$, $w^{24,25}$ to the triple-collinear sector where singular radiation 
occurs along the direction of the incoming 
antiquark with momentum $p_2$, and $w^{14,25}$ and $w^{15,24}$ to the double-collinear 
sectors.

\end{document}